\documentclass[aps,prc,twocolumn,showpacs,amssymb,amsmath,amsfonts,floatfix,nofootinbib]{revtex4-2}
\usepackage{graphicx}% Include figure files
\usepackage{amsmath}
\usepackage[usenames]{color}
\usepackage{epsfig}
\usepackage{epstopdf}
\usepackage{amssymb,amsmath}
\usepackage{amsmath}
\usepackage{epstopdf}
\usepackage{sidecap}
\usepackage{cases}
\usepackage{enumitem}
\usepackage{bm}
\usepackage{overpic}
\usepackage{upgreek}
\usepackage{morefloats}
\usepackage{hyperref}

\definecolor{grey}{rgb}{0.9,0.9,0.9}
\definecolor{black}{rgb}{0,0,0}

\newcommand{\be}{\begin{eqnarray}}
\newcommand{\ee}{\end{eqnarray}}
\newcommand{\bc}{\begin{center}}
\newcommand{\ec}{\end{center}}
\newcommand{\beq}{\begin{eqnarray}}
\newcommand{\eea}{\end{eqnarray}}

\newcommand{\Ocal}{\mathcal{O}}
\newcommand{\OcalTilde}{\tilde{\mathcal{O}}}

\newcommand{\Real}{\text{Re}}
\newcommand{\Imag}{\text{Im}}

\newcommand{\HISKP}{Helmholtz-Institut f\"{u}r Strahlen- und Kernphysik, Universit\"{a}t Bonn, Germany}

\begin{document}

\allowdisplaybreaks

\title{New graphical criterion for the selection of complete sets of polarization observables and its application to single-meson photoproduction as well as electroproduction}
\author{Y.~Wunderlich}\email[Corresponding author: ]{wunderlich@hiskp.uni-bonn.de}\affiliation{\HISKP}  
%\author{...}
%\affiliation{\HISKP}
% \author{P.~Kroenert}
% \affiliation{\HISKP}
% \author{F.~Afzal}
% \affiliation{\HISKP} 
% \author{A.~Thiel}
% \affiliation{\HISKP} 
%\affiliation{...}

\date{\today}
\begin{abstract}
This paper combines the graph-theoretical ideas behind Moravcsik's theorem with a completely analytic derivation of discrete phase-ambiguities, recently published by Nakayama. The result is a new graphical procedure for the derivation of certain types of complete sets of observables for an amplitude-extraction problem with $N$ helicity-amplitudes. \\
The procedure is applied to pseudoscalar meson photoproduction ($N = 4$ amplitudes) and electroproduction ($N = 6$ amplitudes), yielding complete sets with minimal length of $2N$ observables. For the case of electroproduction, this is the first time an extensive list of minimal complete sets is published. Furthermore, the generalization of the proposed procedure to processes with a larger number of amplitudes, i.e. $N > 6$ amplitudes, is sketched. The generalized procedure is outlined for the next more complicated example of two-meson photoproduction ($N = 8$ amplitudes).
\end{abstract}

\maketitle

\section{Introduction} \label{sec:Intro}

Hadron spectroscopy is and has been a very important tool for the improvement of our understanding of non-perturbative QCD. Reactions among particles with spin have always been of central importance for spectroscopy. For the spectroscopy of baryons~\cite{Klempt:2009pi,Ireland:2019uwn} in particular, most experimental activities in the recent years have taken place at facilities capable of measuring reactions induced by electromagnetic probes. Well-known experiments, all capable of measuring the photoproduction of one or several pseudoscalar mesons (as well as vector mesons), are the CBELSA/TAPS experiment at Bonn~\cite{Sparks:2010vb, Thiel:2012yj, Gottschall:2013uha, Hartmann:2014mya,Afzal:2020geq,CBELSA/TAPS:2020yam}, CLAS at JLab (Newport News)~\cite{Dugger:2013, Strauch:2015zob, Senderovich:2015lek, Mattione:2017fxc, Ho:2017kca, Collins:2017sgu, Kunkel:2017src}, A2 at MAMI (Mainz)~\cite{Hornidge:2012ca, Sikora:2013vfa, Schumann:2015ypa, Adlarson:2015byy, Annand:2016ppc, Gardner:2016irh, Kashevarov:2017kqb, Dieterle:2017myg, Briscoe:2019cyo},  and LEPS at SPring-8 (Hy$\bar{\text{o}}$go Prefecture)~\cite{Kohri:2017kto,Kohri:2020ucd}. The GlueX Collaboration has started exploring completely new kinematic regimes for single-meson photoproduction recently~\cite{AlGhoul:2017nbp,Adhikari:2019gfa,Adhikari:2020cvz}. Furthermore, new datasets on electroproduction have or will become available, measured by the CLAS-collaboration~\cite{Burkert:2016kyi,Burkert:2019kxy,Burkert:2020akg,Markov:2019fjy}.

The currently accepted canonical method to determine physical properties of resonances (i.e. masses, widths and quantum numbers) from data are analyses using so-called energy-dependent (ED) partial-wave analysis (PWA-) models. Elaborate reaction-theoretic models are constructed in order to obtain the amplitude as a function of energy. Then, after fitting the data, the resulting amplitude is analytically continued into the complex energy-plane in order to search for the resonance-poles. Well-known examples for such approaches are the SAID-analysis~\cite{SAID,Arndt:1994bu,Arndt:1995bj,Arndt:2006bf,Workman:2012hx}, the Bonn-Gatchina model~\cite{BnGa,BnGaFormalismBook,Anisovich:2011fc,Sarantsev:2014jba}, the J\"{u}lich-Bonn model~\cite{Ronchen:2012eg,Ronchen:2014cna,Ronchen:2015vfa,Ronchen:2018ury,Mai:2021vsw} and the MAID-analysis~\cite{MAID,Drechsel:2007if,Tiator:2018heh,Tiator:2018pjq}, among others.

In an approach that is complementary to the above-mentioned ED fits, one can ask for the maximal amount of information on the underlying reaction-amplitudes that can be extracted from the data without introducing any model-assumptions. One thus considers a generic {\it amplitude-extraction problem}, which is concerned with the extraction of $N$ so-called spin-amplitudes (often specified as helicity-amplitudes~$H_{i}$ or transversity-amplitudes~$b_{i}$~\cite{Chiang:1996em}) out of a set of $N^{2}$ polarization ob\-serva\-bles. Such an amplitude-extraction problem takes place at each point in the kinematical phase-space individually. For a $2 \rightarrow 2$ reaction, this means at each point in energy and angle. For a $2 \rightarrow n$ reaction with $n \geq 3$ particles in the final state, the amplitude-extraction problem has to be solved in each higher-dimensional 'bin' of phase-space, where the phase-space is spanned by $(3 [2 + n] - 10)$ independent kinematical variables~\cite{Eden:1966dnq}. In any case, the unknown overall phase can in principle have an arbitrary dependence on the full reaction-kinematics.

For an ordinary scattering-experiment such as the ones discussed in this work, the determination of the overall phase from data for a single reaction alone is a mathematical impossibility, due to the fact that observables are always bilinear hermitean forms of the $N$ amplitudes~\cite{MyDiploma,MyPhD}. Alternative experiments have been proposed in the literature in order to remedy this problem: Goldberger and collaborators suggested a Hanbury-Brown and Twiss experiment to measure the overall phase~\cite{Goldberger:1963}, while Ivanov proposed to use Vortex beams in order to access information on the angular dependence of the overall phase~\cite{Ivanov:2012na}. However, both of these proposed methods cannot be realized experimentally at the time of this writing. The only alternative consists of the introduction of additional theoretical constraints. As many past studies on the mathematical physics of inverse scattering-problems have shown~\cite{Newton:1968zs,Martin:1969xs,Atkinson:1972hr,Itzykson:1973sz,Atkinson:1973wt,Atkinson:1975rv,Bowcock:1976ax,Chadan:1977pq,Martin:2020jlu}, the unitarity of the $S$-matrix is a very powerful constraint for restricting the overall phase. Unitarity-constraints are of course inherent to many of the ED fit-approaches mentioned above, since almost all of them are formulated for a simultaneous analysis of multiple coupled-channels. However, within the context of an amplitude-extraction problem for {\it one} individual reaction, the overall phase cannot be determined, at least as long as the discussion remains fully model-independent. 

When confronted with a general amplitude-extraction problem, the question of minimizing the measurement effort leads one in a natural way to the search for so-called {\it complete experiments}~\cite{Barker:1975bp,Chiang:1996em} (or {\it complete sets of observables}). These are minimal subsets of the full set of $N^{2}$ ob\-serva\-bles that allow for an unambiguous extraction of the $N$ amplitudes up to one overall phase. 
For an amplitude-extraction problem with an arbitrary number of amplitudes $N$, one can find the following compelling heuristic argument for the fact that at least $2N$ ob\-serva\-bles are required in order to determine the $N$ amplitudes up to one unknown overall phase (cf. the introductions of references~\cite{Wunderlich:2020umg,Moravcsik:1984uf}, as well as footnote~1 in reference~\cite{Keaton:1996pe}). At least $2 N - 1$ observables are needed in order to fix $N$ moduli and $N - 1$ relative-phases. However, with $2 N - 1$ observables, there generally still remain so-called {\it discrete ambiguities}~\cite{Keaton:1995pw,Chiang:1996em}. The resolution of these discrete ambiguities requires at least one additional observable. In this way, one obtains a minimum number of $2N$ observables. This heuristic argument of course tells nothing about {\it how} these $2N$ observables have to be selected. This is then the subject of works like the present one.

The minimal number of $2N$ has indeed turned out to be true for the specific reactions we found treated in the literature. For Pion-Nucleon scattering ($N = 2$), the argument demonstrating that indeed all four accessible observables have to be measured is still quite simple (cf. reference~\cite{Wunderlich:2020umg} as well as the introduction of reference~\cite{Anisovich:2013tij}). The process of pseoduscalar meson photoproduction ($N = 4$) has been treated at length in the literature. Based on earlier results by Keaton and Workman~\cite{Keaton:1995pw, Keaton:1996pe}, Chiang and Tabakin found in a seminal work~\cite{Chiang:1996em} that $8$ carefully selected observables can constitute a minimal complete set for this process. The result by Chiang and Tabakin has recently been substantiated in a rigorous algebraic proof by Nakayama~\cite{Nakayama:2018yzw}, where all the discrete phase-ambiguities implied by quite arbitrary selection-patterns of observables were derived and the conditions for the resolution of these ambiguities were clearly stated. Some of Nakayama's derivations will also turn out to be important for this work. For pseudoscalar meson electroproduction ($N = 6$), the construction of some complete sets with $12$ observables has been outlined by Tiator and collaborators~\cite{Tiator:2017cde}, but an extensive list of complete sets has not been given in the latter reference. This is something that will be improved upon in the present work. Finally, the process of two-meson photoproduction ($N = 8$) has been treated as well in some quite explicit works~\cite{Arenhoevel:2014dwa,Kroenert:2020ahf}. The complete sets for this process indeed have a minimal length of $16$~\cite{Kroenert:2020ahf}.

In mathematical treatments of complete experiments such as the ones mentioned up to this point, one always assumes the ob\-serva\-bles to have vanishing measurement uncertainty. Once the mathematically 'exact' complete sets have been established in this way, one can study the influence of non-vanishing measurement-uncertainties using high-level statistical methods. This has been done in a number of recent works by Ireland~\cite{Ireland:2010bi} and the Ghent-group~\cite{Vrancx:2013pza, Vrancx:2014yja, Nys:2015kqa}.

An interesting alternative approach for the deduction of complete sets of observables is given by Moravcsik's theorem~\cite{Moravcsik:1984uf}. This theorem has been reexamined in a recent work~\cite{Wunderlich:2020umg}, where it has received slight corrections for the case of an odd number of amplitudes~$N$. The theorem is formulated in the language of a 'geometrical analog'~\cite{Moravcsik:1984uf}, which yields a useful representation of complete sets in the shape of graphs. The advantages of the theorem are that it can be applied directly to any amplitude-extraction problem irrespective of~$N$. Furthermore, it can be fully automated on a computer. However, the approach also has it's drawbacks: for larger $N$ (i.e. $N > 6$)~\cite{Roberts:2004mn,Arenhoevel:2014dwa,Kroenert:2020ahf,Pichowsky:1994gh}, the number of relevant graph-topologies grows very rapidly, as $(N - 1)!/2$~\cite{Wunderlich:2020umg}. This alone makes the computations quite expensive for more involved amplitude-extraction problems. Another drawback of (the modified form of) Moravcsik's theorem consists of the fact that for $N \geq 4$, the derived complete sets do not have the minimal length~$2N$ any more, but are rather slightly over-complete (see in particular section~VII of reference~\cite{Wunderlich:2020umg}). The reason for the latter fact is that Moravcsik directly considered just the bilinear products $b_{i}^{\ast} b_{j}$ of amplitudes. However, polarization ob\-serva\-bles for $N \geq 4$ generally are invertible linear combinations of such bilinear products. 

The present work is an attempt to devise an approach similar to (the modified form of) Moravcsik's theorem~\cite{Wunderlich:2020umg}, but which can get the length of the derived complete sets down to~$2N$, for amplitude-extraction problems with~$N \geq 4$. Generally, the proposed approach can be applied to any amplitude-extraction problem with an {\it even} number of amplitudes~$N$. In order to achieve this, we combine the graph-theoretical ideas according to references~\cite{Moravcsik:1984uf,Wunderlich:2020umg} with the results derived by Nakayama~\cite{Nakayama:2018yzw} for the discrete phase-ambiguities implied by selections of pairs of observables. Although these ambiguities have been originally derived by Nakayama for photoproduction~\cite{Nakayama:2018yzw}, we get a criterion that directly facilitates deriving minimal complete sets for electroproduction as well. A crucial new ingredient for the procedure proposed in this work is that the graphs have to be embued with additional directional information. The graphical criterion formulated in this work, together with the types of graphs needed for it, are to our knowledge new. 

This paper is organized as follows. The new graphical criterion is motivated and deduced in section~\ref{sec:NewCriterion}, as a result of the combination of the ideas behind Moravcsik's theorem and the phase-ambiguities as derived by Nakayama~\cite{Nakayama:2018yzw}. We then illustrate the new criterion in applications to single-meson photo- and electroproduction in sections~\ref{sec:Photoproduction} and~\ref{sec:Electroproduction}. Some ideas on the generalization of the proposed procedure to problems with a larger number of $N > 6$ amplitudes are stated in section~\ref{sec:Generalization}, which is followed by the conclusion in section~\ref{sec:ConclusionsAndOutlook}. Three appendices collect a review of the recently published modified form of Moravcsik's theorem~\cite{Wunderlich:2020umg}, as well as lengthy calculations which are however of vital importance for the present work. Elaborate lists of the newly derived complete sets for electroproduction can be found in the supplemental material~\cite{Supplement}.

%\clearpage

%\section{The modified form of Moravcsik's theorem} \label{sec:MoravcsikTheorem}

\section{The new graphical criterion} \label{sec:NewCriterion}

In this section, the new graphical criterion for complete sets of observables is derived deductively. It is based on a combination of the graph-theoretical ideas from Moravcsik's theorem~\cite{Moravcsik:1984uf,Wunderlich:2020umg}, where each complete sets of observables has a lucid representation in terms of a graph, and recent derivations of discrete phase-ambiguities given in full detail by Nakayama~\cite{Nakayama:2018yzw}. Since Moravcsik's theorem serves as a useful reference point to the new ideas developed in this section, and also to keep this work self-contained, a review of a recently published slightly modified version of the theorem is given in appendix~\ref{sec:ReviewMoravcsik}. In this appendix, also some pictorial examples for complete graphs according to Moravcsik can be found. 

We start with the standard assumption that the moduli $\left| b_{i} \right|$ of the~$N$ amplitudes~$b_{1},\ldots,b_{N}$ have already been determined from a set of $N$ diagonal observables (cf. appendix~\ref{sec:ReviewMoravcsik} and references~\cite{Moravcsik:1984uf,Chiang:1996em,Nakayama:2018yzw,Wunderlich:2020umg}). Consider now a so-called (non-diagonal) {\it shape-class} composed of four observables, which is a mathematical structure that repeatedly appears in the problems of single-meson photoproduction and electroproduction (cf. Table~\ref{tab:PhotoObservables} in section~\ref{sec:Photoproduction} and Table~\ref{tab:ElectroObservables} of section~\ref{sec:Electroproduction}). The four observables belonging to the shape-class, which we denote by the super-script '$n$', are given by the following linear combinations of bilinear amplitude-products (the notation for the observables is taken over from reference~\cite{Nakayama:2018yzw}):
%
%\begin{widetext}
\begin{align}
 \Ocal^{n}_{1+} &= \Imag \left[  b_{j}^{\ast} b_{i} + b_{l}^{\ast} b_{k}  \right]  \nonumber \\
  &= \left| b_{i} \right| \left| b_{j} \right| \sin \phi_{ij} + \left| b_{k} \right| \left| b_{l} \right| \sin \phi_{kl}  , \label{eq:NonTrivialShapeClassObsI} \\
 \Ocal^{n}_{1-}  &= \Imag \left[  b_{j}^{\ast} b_{i} - b_{l}^{\ast} b_{k}  \right]  \nonumber \\
 &= \left| b_{i} \right| \left| b_{j} \right| \sin \phi_{ij} - \left| b_{k} \right| \left| b_{l} \right| \sin \phi_{kl}   , \label{eq:NonTrivialShapeClassObsII} \\
 \Ocal^{n}_{2+}  &= \Real \left[  b_{j}^{\ast} b_{i} + b_{l}^{\ast} b_{k}  \right]  \nonumber \\
 &=  \left| b_{i} \right| \left| b_{j} \right| \cos \phi_{ij} + \left| b_{k} \right| \left| b_{l} \right| \cos \phi_{kl}  , \label{eq:NonTrivialShapeClassObsIII} \\
 \Ocal^{n}_{2-}   &= \Real \left[  b_{j}^{\ast} b_{i} - b_{l}^{\ast} b_{k}  \right]  \nonumber \\
 &=  \left| b_{i} \right| \left| b_{j} \right| \cos \phi_{ij} - \left| b_{k} \right| \left| b_{l} \right| \cos \phi_{kl}  . \label{eq:NonTrivialShapeClassObsIV}
\end{align}
%\end{widetext}
%
The four indices $i,j,k,l \in 1,\ldots,N$ (for either $N = 4$ in case of photoproduction, or $N = 6$ for electroproduction) have to be all pairwise distinct. In this way, every shape-class composed of four observables, which has the above-given structure, is in one-to-one correspondence to a particular pair of relative phases~$\left\{ \phi_{ij}, \phi_{kl} \right\}$. In the case of photoproduction ($N = 4$, section~\ref{sec:Photoproduction}), one has three shape-classes of this type, while for electroproduction ($N=6$, section~\ref{sec:Electroproduction}), one encounters seven such shape-classes, containing four observables each. 

A shape-class composed of four observables such as in equations~\eqref{eq:NonTrivialShapeClassObsI} to~\eqref{eq:NonTrivialShapeClassObsIV} really represents the {\it simplest non-trivial example} of such a class, since any simpler combination of bilinear amplitude-products would just amount to the real- and imaginary parts of the products $b_{i}^{\ast} b_{j}$ themselves, without any additional linear combination (cf. discussions in appendix~\ref{sec:ReviewMoravcsik}). For problems with $N > 6$ amplitudes, one generally encounters more involved shape-classes (cf. section~\ref{sec:Generalization}).

\begin{figure}
 \begin{center}
\includegraphics[width = 0.595 \textwidth,trim={3.7cm 4.35cm 2.0cm 1.0cm},clip]{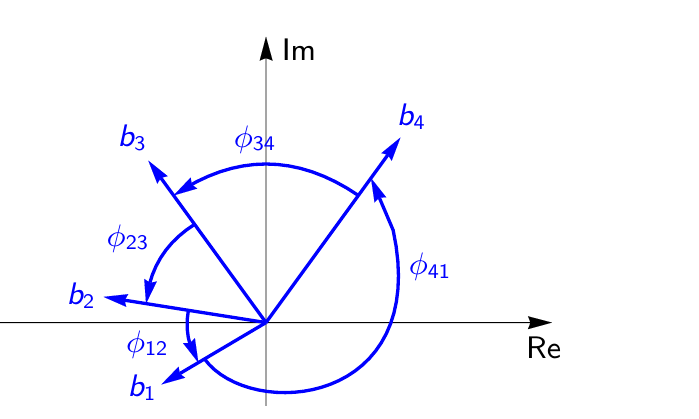} 
\end{center}
\vspace*{-5pt}
\caption{The general consistency relation~\eqref{eq:GeneralConsistencyRelation} is illustrated for the example of an amplitude-extraction problem with~$N = 4$ amplitudes~$b_{1}, \ldots, b_{4}$. The relation one deduces geometrically from the given diagram is~$\phi_{12} + \phi_{23} + \phi_{34} + \phi_{41} = 2 \pi$, which up to addition of~$2 \pi$ is equivalent to~$\phi_{12} + \phi_{23} + \phi_{34} + \phi_{41} = 0$.}
\label{fig:ConsistencyRelationFigure}
\end{figure}

Before discussing the discrete phase-ambiguities implied by different selections of observables picked from the shape-class given in equations~\eqref{eq:NonTrivialShapeClassObsI} to~\eqref{eq:NonTrivialShapeClassObsIV}, we need to introduce another important part of the proofs yet to be presented, which is given by so-called {\it consistency relations}~\cite{Nakayama:2018yzw, Wunderlich:2020umg}. In case the {\it connectedness-criterion} is fulfilled by the graphs that represent potentially complete sets of observables~(cf. discussions further below in this section and in appendix~\ref{sec:ReviewMoravcsik}), one can establish a consistency relation among all the occurring relative-phases, which generally takes the shape\footnote{The consistency relation~\eqref{eq:GeneralConsistencyRelation}, as well as all other relations among phases appearing in this work, is only valid up to addition of multiples of $2 \pi$.}:
\begin{equation}
%\centering
 \phi_{1i} + \phi_{ij} + \ldots + \phi_{k1} = 0 . \label{eq:GeneralConsistencyRelation}
\end{equation}
The pairings of indices in relative-phases occurring in this relation is in one-to-one correspondence to the considered graph-topology. The consistency-relation~\eqref{eq:GeneralConsistencyRelation} is a natural constraint for an arrangement of $N$ amplitudes in the complex plane (cf. the illustration given in Figure~\ref{fig:ConsistencyRelationFigure}) and any valid solution of the considered amplitude-extraction problem has to satisfy it. It will turn out to be important for this work to fix a standard-convention for writing down consistency-relations: we want to isolate all relative-phases on one side of the equation-sign (such as in equation~\eqref{eq:GeneralConsistencyRelation}), want all relative-phases to have a positive sign and want the index-pairings in the appearing relative-phases to correspond to a {\it definite direction of translation} (or just short: a {\it direction}) for the graph. The direction of translation is fixed by starting at amplitude-point '1', then stepping through the graph along direct connections of amplitudes which have to be in one-to-one correspondence to the sequence of indices appearing in equation~\eqref{eq:GeneralConsistencyRelation}, until ending up again at amplitude-point '1'. This convention will turn out to be important for the discussion from here on. 

Consistency-relations such as~\eqref{eq:GeneralConsistencyRelation} are crucial for the derivation of fully complete sets. A selection of observables picked from several copies of the above-given shape-class, with the selection corresponding to a particular considered graph, leads to a set of $\mathcal{N}_{\text{amb.}}$ potentially ambiguous solutions\footnote{For the ambiguities of real- and imaginary parts of bilinear products~$b_{j}^{\ast} b_{i}$ considered in case of Theorem~2 in appendix~\ref{sec:ReviewMoravcsik}, the discrete phase-ambiguities are always two-fold for each relative-phase individually and thus one always has~$\mathcal{N}_{\text{amb.}} = 2^{N}$. For the selections of observables from the non-diagonal shape-class considered in this section, $\mathcal{N}_{\text{amb.}}$ may differ from~$2^{N}$.}. For each of these $\mathcal{N}_{\text{amb.}}$ discrete phase-ambiguities, one can write down a consistency relation, where the respective ambiguous solutions are labelled by a corresponding superscript-$\lambda$ on the relative-phases:
\begin{equation}
%\centering
 \phi^{\lambda}_{1i} + \phi^{\lambda'}_{ij} + \ldots + \phi^{\lambda^{(N)}}_{k1} = 0 . \label{eq:GeneralConsistencyRelationAmbiguityCases}
\end{equation}
The criteria stated in Theorem~1 derived in this section, as well as Theorem~2 from appendix~\ref{sec:ReviewMoravcsik}, are now the results of a careful analysis of all possible cases where no degeneracies\footnote{Two equations from the $\mathcal{N}_{\text{amb.}}$ possibilities~\eqref{eq:GeneralConsistencyRelationAmbiguityCases} are called {\it degenerate} in case they can be transformed into each other using the following two operations~\cite{Wunderlich:2020umg, Kroenert:2020ahf}: \begin{itemize} \item[$\diamond$] multiplication of the whole equation by $(-1)$, \item[$\diamond$] addition (and/or subtraction) of multiples of $2 \pi$. \end{itemize}} occur any more among the $\mathcal{N}_{\text{amb.}}$ possible relations~\eqref{eq:GeneralConsistencyRelationAmbiguityCases} (see also appendix~A of reference~\cite{Wunderlich:2020umg} for a more detailed derivation of Theorem~2). The only difference is that Theorem~2 is only valid in the basis of fully {\it decoupled} bilinear products~$b_{j}^{\ast} b_{i}$, while Theorem~1 to be deduced below holds for selections of observables from the non-decoupled shape-class given in equations~\eqref{eq:NonTrivialShapeClassObsI} to~\eqref{eq:NonTrivialShapeClassObsIV}. In case all degeneracies are indeed resolved (in case of either Theorem~1 or Theorem~2), full completeness is obtained and the solution of the amplitude-extraction problem is thus unique.

We now proceed to enumerate the discrete ambiguities for the relative phases implied by the selection of any pair of observables from the four quantities~\eqref{eq:NonTrivialShapeClassObsI} to~\eqref{eq:NonTrivialShapeClassObsIV}. A full derivation of these ambiguities has been given by Nakayama~\cite{Nakayama:2018yzw}, based on earlier ideas by Chiang and Tabakin~\cite{Chiang:1996em}. In the following, we only cite the results. A full derivation according to Nakayama is outlined in more detail in appendix~\ref{sec:NakayamaDerivation}, in order to keep the present work self-contained.

\begin{figure}
 \begin{center}
\includegraphics[width = 0.45 \textwidth,trim={2.0cm 1.0cm 2.0cm 0},clip]{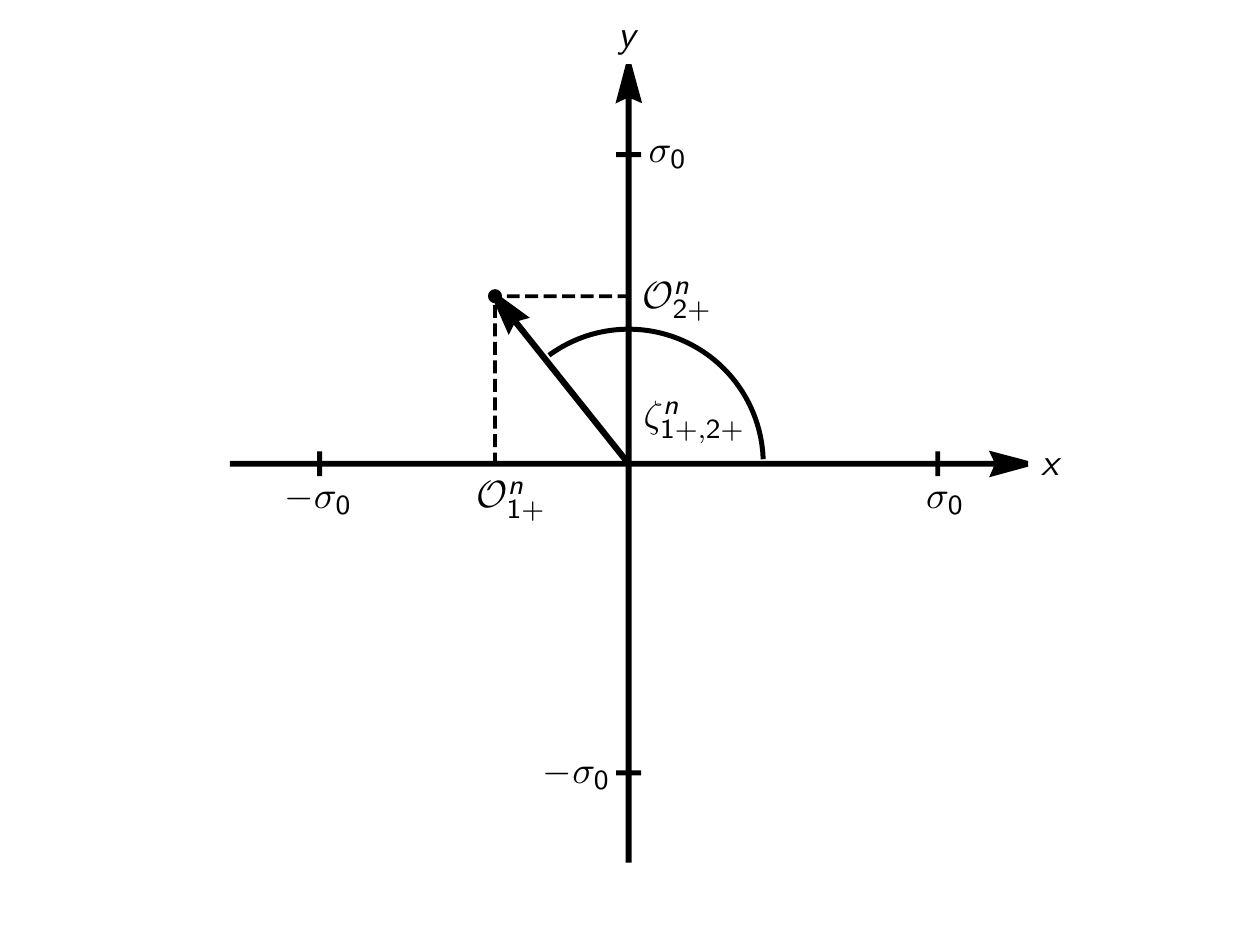} 
\end{center}
\vspace*{-5pt}
\caption{The meaning of the angle $\zeta \equiv \zeta^{n}_{1+,2+}$ is illustrated. This picture directly corresponds to the definitions~\eqref{eq:DefLengthN} and~\eqref{eq:ZetaAngleFormalDefinition} in appendix~\ref{sec:NakayamaDerivation}. The modulus of the value for each observable $\left( \Ocal^{n}_{1+}, \Ocal^{n}_{2+} \right)$ is limited by the unpolarized differential cross-section $\sigma_{0}$ of the considered process.}
\label{fig:ZetaAngleDefinition}
\end{figure}

One does not need any elaborate additional derivations in case the pair of observables is selected in such a way that the bilinear amplitude-products fully {\it decouple}. This is also the case in which Theorem~2 from appendix~\ref{sec:ReviewMoravcsik} can be directly used, i.e. the case of the two following possible selections (see also reference~\cite{Nakayama:2018yzw}):
\begin{itemize}
 \item[A.1)] $\left( \Ocal^{n}_{1+}, \Ocal^{n}_{1-} \right)$:
 
 This particular selection of observables allows for the isolation of both sines of the relative-phases $\phi_{ij}$ and $\phi_{kl}$, according to the following linear combinations of observables:
 \begin{equation}
  \sin \phi_{ij} = \frac{\Ocal^{n}_{1+} + \Ocal^{n}_{1-}}{2 \left| b_{i} \right| \left| b_{j} \right|} \text{, } \sin \phi_{kl} = \frac{\Ocal^{n}_{1+} - \Ocal^{n}_{1-}}{2 \left| b_{k} \right| \left| b_{l} \right|}     . \label{eq:SineIsolationEquations}
 \end{equation}
 In this way, one obtains a discrete sine-type ambiguity for the two relative phases (cf. equation~\eqref{eq:SinTypeAmbiguity}):
 \begin{equation}
  \phi_{ij}^{\lambda} = \phi_{ij}^{\pm} = \begin{cases}  + \alpha_{ij}, \\ \pi - \alpha_{ij} ,   \end{cases}    \phi_{kl}^{\lambda'} = \phi_{kl}^{\pm} = \begin{cases}  + \alpha_{kl}, \\ \pi - \alpha_{kl} ,   \end{cases} \label{eq:SineIsolationDiscreteAmbiguities}
 \end{equation}
 where the values of the two selected observables uniquely fix both $\alpha_{ij}$ and $\alpha_{kl}$ on the interval $\left[ - \pi / 2, \pi / 2 \right]$. Since both $\lambda$ and $\lambda'$ in equation~\eqref{eq:SineIsolationDiscreteAmbiguities} can take their values $\pm$ independently, the discrete ambiguity is four-fold.
 %\newpage
 \item[A.2)] $\left( \Ocal^{n}_{2+}, \Ocal^{n}_{2-} \right)$:
 
 For this particular selection of observables, one obtains an isolation of the cosines according to:
 \begin{equation}
  \cos \phi_{ij} = \frac{\Ocal^{n}_{2+} + \Ocal^{n}_{2-}}{2 \left| b_{i} \right| \left| b_{j} \right|} \text{, } \cos \phi_{kl} = \frac{\Ocal^{n}_{2+} - \Ocal^{n}_{2-}}{2 \left| b_{k} \right| \left| b_{l} \right|}     . \label{eq:CosineIsolationEquations}
 \end{equation}
 This leads to discrete cosine-type ambiguities for the two relative phases $\phi_{ij}$ and $\phi_{kl}$ (cf. equation~\eqref{eq:CosTypeAmbiguity}):
 \begin{equation}
  \phi_{ij}^{\lambda} = \phi_{ij}^{\pm} = \begin{cases}  + \alpha_{ij}, \\ - \alpha_{ij} ,   \end{cases}   \phi_{kl}^{\lambda'} = \phi_{kl}^{\pm} =  \begin{cases}  + \alpha_{kl}, \\ - \alpha_{kl} ,   \end{cases} \label{eq:CosineIsolationDiscreteAmbiguities}
 \end{equation}
 with $\alpha_{ij}$ and $\alpha_{kl}$ both fixed uniquely on the interval $\left[ 0, \pi \right]$, from the values of the two selected observables. The discrete phase-ambiguity is again four-fold (due to $\lambda, \lambda' = \pm$).
\end{itemize}
Once a 'crossed' pair of observables, i.e with one observable chosen from~$\Ocal^{n}_{1 \pm}$ and the other one from~$\Ocal^{n}_{2 \pm}$ is selected, the elaborate derivations outlined in appendix~\ref{sec:NakayamaDerivation} become necessary. These are however also the selections which are much more interesting and important for the graphical criterion proposed in this work. One has to distinguish the following four cases~\cite{Nakayama:2018yzw}: 
\begin{itemize}
 \item[B.1.)] $\left( \Ocal^{n}_{1+}, \Ocal^{n}_{2+} \right)$:
 
 For this selection of observables, one only obtains a two-fold discrete phase-ambiguity. Only the following two possible pairs of values are allowed for the relative-phases $\phi_{ij}$ and $\phi_{kl}$ (see reference~\cite{Nakayama:2018yzw} and appendix~\ref{sec:NakayamaDerivation})\footnote{The expressions for the ambiguities~\eqref{eq:TwoFoldPhaseAmbiguityII1} to~\eqref{eq:TwoFoldPhaseAmbiguityII4}, as derived in appendix~\ref{sec:NakayamaDerivation}, are formally a bit different compared to those of reference~\cite{Nakayama:2018yzw}. However, the most important features of the derived ambiguities (i.e. the signs of the $\zeta$-angles) remain the same and therefore the statements of Theorem~1 developed in this section do not change, no matter which formulas one uses. In order to keep the present work self-contained, we stick to the expressions for the ambiguities as derived in appendix~\ref{sec:NakayamaDerivation}.}:
 \begin{equation}
   \begin{cases} \phi_{ij} =  - \zeta + \alpha_{ij}  , \\  \phi_{kl} = - \zeta - \alpha_{kl} + \pi   , \end{cases} \hspace*{-7pt} \text{or} \begin{cases} \phi_{ij} = - \zeta - \alpha_{ij} + \pi  , \\  \phi_{kl} =  - \zeta + \alpha_{kl}  , \end{cases}  \label{eq:TwoFoldPhaseAmbiguityII1}
 \end{equation}
 where both $\alpha_{ij}$ and $\alpha_{kl}$ are uniquely fixed on the interval $\left[ - \pi / 2, \pi / 2 \right]$ via the values of the selected pair of observables (cf. equations~\eqref{eq:ExampleDerivationStepIII} and~\eqref{eq:ExampleDerivationStepIV} in appendix~\ref{sec:NakayamaDerivation}). \\ The quantity $\zeta$ in the definition of this two-fold ambiguity~\eqref{eq:TwoFoldPhaseAmbiguityII1} is the new ingredient which appears in case of a selection of a crossed pair of observables. As defined in reference~\cite{Nakayama:2018yzw}, this quantity $\zeta$ is equal to the polar angle in a $2$-dimensional coordinate system, where $\Ocal^{n}_{1+}$ defines the $x$-coordinate and $\Ocal^{n}_{2+}$ the $y$-coordinate (see Figure~\ref{fig:ZetaAngleDefinition}). We therefore call it a 'transitional angle'. This angle should actually be denoted as '$\zeta^{n}_{1+,2+}$', since it depends on the values of the selected pair of observables (cf. appendix ~\ref{sec:NakayamaDerivation} and reference~\cite{Nakayama:2018yzw}). However, in order to keep the notation as simple as possible, we only write $\zeta$ (and $\zeta', \zeta'', \ldots$ for any additional transitional angles that appear in an equation). The transitional angles are of vital importance for the resolution of degenerate consistency-relations\footnote{We note here that special values for the $\zeta$-angle exist where it may generally loose its ability to resolve degenerate consistency-relations, namely $\zeta = 0, \frac{\pi}{2}, \pi, \frac{3 \pi}{2}, 2 \pi$ and multiples thereof. Considering Figure~\ref{fig:ZetaAngleDefinition}, we see that these values occur when at least one observable in the pair $\left( \Ocal^{n}_{1+}, \Ocal^{n}_{2+} \right)$ vanishes. These special configurations belong to the surfaces of vanishing measure in the parameter-space, on which Theorem~1 can loose its validity (cf. comments made at the end of section~\ref{sec:NewCriterion}, as well as similar discussions in reference~\cite{Nakayama:2018yzw}). In the present work, we disregard such special cases.} and therefore also for the removal of phase-ambiguities (cf. reference~\cite{Nakayama:2018yzw}). They are therefore the {\it central objects of interest} for our proposed graphical criterion.
 \item[B.2.)] $\left( \Ocal^{n}_{1+}, \Ocal^{n}_{2-} \right)$:
  
In this case, one obtains the two-fold discrete phase-ambiguity (cf. appendix~\ref{sec:NakayamaDerivation})
 \begin{equation}
   \begin{cases}  \phi_{ij} = - \zeta + \alpha_{ij}  ,  \\ \phi_{kl} = \zeta - \alpha_{kl}  , \end{cases} \hspace*{-2.5pt} \text{or } \begin{cases}  \phi_{ij} = - \zeta - \alpha_{ij} + \pi ,  \\ \phi_{kl} = \zeta + \alpha_{kl} - \pi  , \end{cases}  \label{eq:TwoFoldPhaseAmbiguityII2}
 \end{equation}
 where the values of the selected pair of observables uniquely fix both $\alpha_{ij}$ and $\alpha_{kl}$ on the interval $\left[ - \pi / 2, \pi / 2 \right]$  (see equations~\eqref{eq:Example3DerivationStepIII} and~\eqref{eq:Example3DerivationStepIV} in appendix~\ref{sec:NakayamaDerivation}), as well as the value of the transitional angle $\zeta \equiv \zeta^{n}_{1+,2-}$.
 \item[B.3.)] $\left( \Ocal^{n}_{1-}, \Ocal^{n}_{2+} \right)$:
 
 For this selection of observables, one obtains the two-fold discrete phase-ambiguity  (see appendix~\ref{sec:NakayamaDerivation})
 \begin{equation}
   \begin{cases}  \phi_{ij} = - \zeta + \alpha_{ij}  ,  \\ \phi_{kl} = \zeta + \alpha_{kl} - \pi   , \end{cases} \vspace*{-10pt} \text{or} \begin{cases}  \phi_{ij} = - \zeta - \alpha_{ij} + \pi ,  \\ \phi_{kl} = \zeta - \alpha_{kl}  , \end{cases}   \label{eq:TwoFoldPhaseAmbiguityII3}
 \end{equation}
 where both $\alpha_{ij}$ and $\alpha_{kl}$ are uniquely fixed on the interval $\left[ - \pi / 2, \pi / 2 \right]$ from the values of the selected pair of observables (cf. equations~\eqref{eq:Example4DerivationStepIII} and~\eqref{eq:Example4DerivationStepIV} in appendix~\ref{sec:NakayamaDerivation}). The selected observables also fix the transitional angle $\zeta \equiv \zeta^{n}_{1-,2+}$.
 \item[B.4.)] $\left( \Ocal^{n}_{1-}, \Ocal^{n}_{2-} \right)$:
 
 This selection of observables implies the two-fold discrete phase-ambiguity (cf. appendix~\ref{sec:NakayamaDerivation})
 \begin{equation}
   \begin{cases} \phi_{ij} =  - \zeta + \alpha_{ij}   , \\  \phi_{kl} = - \zeta + \alpha_{kl}  , \end{cases} \hspace*{-7pt} \text{or} \begin{cases} \phi_{ij} = - \zeta - \alpha_{ij} + \pi  , \\  \phi_{kl} = - \zeta - \alpha_{kl} + \pi   , \end{cases}  \label{eq:TwoFoldPhaseAmbiguityII4}
 \end{equation}
 where the values of the selected pair of observables uniquely fix both $\alpha_{ij}$ and $\alpha_{kl}$ on the interval $\left[ - \pi / 2, \pi / 2 \right]$  (see equations~\eqref{eq:Example2DerivationStepIII} and~\eqref{eq:Example2DerivationStepIV} in appendix~\ref{sec:NakayamaDerivation}) and furthermore also define the value of the transitional angle $\zeta \equiv \zeta^{n}_{1-,2-}$.
\end{itemize}

The key is now to observe that the sign of the transitional angle $\zeta$ does {\it not} change for each of the two relative phases, i.e. $\phi_{ij}$ or $\phi_{kl}$, individually when passing from one ambiguous solution to the other one. This is true in all of the cases 'B.1', $\ldots$, 'B.4'. The sign of $\zeta$ may however vary in a comparison {\it between} $\phi_{ij}$ and $\phi_{kl}$. 

Still, when evaluating all the different cases possible for a particular consistency-relation~\eqref{eq:GeneralConsistencyRelationAmbiguityCases}, the $\zeta$-angles have a great power for resolving discrete ambiguities, or equivalently for removing degenerate pairs of equations. Thus, one has to carefully keep track of the signs of the $\zeta$'s appearing in equations~\eqref{eq:TwoFoldPhaseAmbiguityII1} to~\eqref{eq:TwoFoldPhaseAmbiguityII4}, when devising a graphical criterion.

There is another sign which is important: this has to do with the index-structure of the relative-phases, as they appear in our standard-convention for the consistency relation~\eqref{eq:GeneralConsistencyRelation} (or~\eqref{eq:GeneralConsistencyRelationAmbiguityCases}). For instance, in case $\phi_{ij}$ appears in this equation with reversed placement $\phi_{ji}$, there appears yet another sign one has to carefully keep track of.

Our proposed graphical criterion is now, in essence, a way to keep track of both the above-mentioned signs, in such a way that at least one transitional $\zeta$-angle survives in all the possible cases for the consistency relation~\eqref{eq:GeneralConsistencyRelationAmbiguityCases} (compare this to expressions given in section~III of reference~\cite{Nakayama:2018yzw}).

We always start with the standard-assumption that $N$ suitable observables have been measured in order to uniquely fix the $N$ moduli $\left| b_{1} \right|, \ldots, \left| b_{N} \right|$. For the selection of the remaining observables, which are supposed to uniquely fix all relative-phases between the $N$ amplitudes, the criterion reads as follows: \\

\textbf{\underline{Theorem 1 (Proposed graphical criterion)}} \\ 

Start with one possible topology for a connected graph with $N$ vertices of order two, i.e. with exactly two edges attached to each vertex. The vertices, or points, again represent the $N$ amplitudes of the problem. The chosen graph has to have exactly $N$ edges (or link-lines) and furthermore has to satisfy in addition the following constraint:
\begin{itemize}
 \item[$\diamond$] The graph should be chosen in such a way that only those connections of amplitude-points appear which are in direct correspondence to any pair of relative-phases $\left\{ \phi_{ij}, \phi_{kl} \right\}$ from a particular shape-class of four observables. The graph is thus constructed to exactly match selections of observables from classes with the structure~\eqref{eq:NonTrivialShapeClassObsI} to~\eqref{eq:NonTrivialShapeClassObsIV}. 
\end{itemize}
Now, select $N / 2$ pairs of observables from the shape-classes implied by the considered graph. The selection of $N/2$ pairs is the reason why the proposed approach can only be directly applied to problems with an {\it even} number of amplitudes~$N$. Draw the following connections of points, based on the selection made:
\begin{itemize}
 \item[$\diamond$] In case the selection 'A.1' has been made, draw a single dashed line which connects the respective amplitude-points (these are then two link-lines, in this case). In case a pair of observables has been chosen according to 'A.2', draw the corresponding pair of link-lines as single solid lines.
 \item[$\diamond$] In case any of the selections 'B.1', $\ldots$, 'B.4' has been made, draw a double-line for both connections of the corresponding amplitude-points. In case multiple such pairs of double-lines appear in the graph, draw a different style of double-line for each different shape-class (i.e. normal double-line, wavy double-line, dashed double-line, dotted double-line, $\ldots$). This has to be done in order to keep track of relative-phases fixed by different shape-classes of observables. 
\end{itemize}
Now, draw arrows into the $N$ link-lines which {\it indicate the direction of translation} through the graph, according to our standard-convention of writing the consistency relation~\eqref{eq:GeneralConsistencyRelation} (cf. comments below equation~\eqref{eq:GeneralConsistencyRelation}). We call these arrows 'directional arrows'. The standard-form of the consistency-relation~\eqref{eq:GeneralConsistencyRelation} would imply directional arrows pointing as follows: $1 \rightarrow i$, $i \rightarrow j$, $\ldots$, $k \rightarrow 1$. \\
Then, draw an additional '$\zeta$-sign arrow' next to each double-line or, depending on the graphic layout, {\it into} the double-line (cf. Figures in sections~\ref{sec:Photoproduction} and~\ref{sec:Electroproduction}). In case the considered double-line corresponds to an arbitrary relative-phase $\phi_{ab}$, the $\zeta$-sign arrow has to point from $a \rightarrow b$ in case the $\zeta$-angle appears with a {\it positive} sign in the ambiguity written in the corresponding case from 'B.1', $\ldots$, 'B.4' (cf. the descriptions of the cases above). The $\zeta$-sign arrow has to point from $b \rightarrow a$ in case the $\zeta$-angle appears with a {\it negative} sign in the equations defining the corresponding discrete phase-ambiguity (cases 'B.1', $\ldots$, 'B.4'). \\
The graph constructed in this lengthy procedure, and therefore also the corresponding set of observables, allows for a unique solution of the amplitude-extraction problem if it contains {\it at least one pair of double-lines} and furthermore satisfies the following criterion:
\begin{itemize}
 \item[(C1)] For {\it at least one of the pairs of double-lines} in the thus constructed graph, one of the following two conditions has to be fulfilled for the graph to be fully complete (note that both conditions cannot be satisfied at the same time):
 \begin{itemize}
 \item[$\diamond$] for {\it both} double-lines, the directional arrows have to point into the {\it same} direction as the corresponding $\zeta$-sign arrows,
 \item[$\diamond$] for {\it both} double-lines, the directional arrows have to point into the direction {\it opposite} to the direction of the respective $\zeta$-sign arrows.
 \end{itemize}
 The single dashed- and solid lines are not really important any more for this criterion\footnote{ The $\zeta$-angles have now taken the role of the 'residual summands of $\pi$' needed in the proof of Theorem~2 (see appendix~A of reference~\cite{Wunderlich:2020umg}).}, as opposed to Theorem~2 from appendix~\ref{sec:ReviewMoravcsik}.
\end{itemize}

As in the case of the modified form of Moravcsik's theorem (Theorem~2 in appendix~\ref{sec:ReviewMoravcsik}), the connectedness-condition imposed on the graphs considered in our new graphical criterion directly removes any possibilities for continuous ambiguities. The remaining conditions stated in Theorem~1 above then are included solely for the purpose of resolving all possible remaining discrete phase-ambiguities.

Both of the possible conditions stated in the criterion (C1) above make sure that the transitional $\zeta$-angle belonging to the corresponding pair of observables appears in all cases for the consistency relation~\eqref{eq:GeneralConsistencyRelationAmbiguityCases} with always the {\it same sign}. This is a plus-sign in case of the first condition mentioned in (C1), or a minus-sign in case of the second condition. Therefore, in exactly these cases the transitional $\zeta$-angles do {\it not} cancel out! This automatically removes all possible degeneracies among the possible cases for the consistency relation~\eqref{eq:GeneralConsistencyRelationAmbiguityCases}. We will illustrate in more detail how this works in our treatment of the example-case of single-meson photoproduction, in section~\ref{sec:Photoproduction}.

We note that the above-stated criterion is only valid for the special case of a selection of exactly two observables from each shape-class of four (cf. equations~\eqref{eq:NonTrivialShapeClassObsI} to~\eqref{eq:NonTrivialShapeClassObsIV}). In the case of Nakayama's work~\cite{Nakayama:2018yzw}, which treated single-meson photoproduction, this was called the '{\bf (2+2)}-case'. Certainly this specific assumption of choosing only pairs of observables from each shape-class restricts the complete sets which we can derive to this certain particular sub-set and the full set of possible complete experiments is certainly larger. We do not want to exclude the possibility that the graphical criterion stated above may in the future be generalized to more general selections of observables (such as the '{\bf (2+1+1)}-case' in Nakayama's work~\cite{Nakayama:2018yzw}), but at present it does not cover such more general possibilities.

As in the case of Moravcsik's theorem in its modified form (Theorem~2 from appendix~\ref{sec:ReviewMoravcsik}), there do exist singular sub-surfaces in the parameter-space composed of the relative-phases, on which Theorem~1 as stated above looses its validity. Nakayama also mentioned such configurations in his treatment of photoproduction~\cite{Nakayama:2018yzw}. However, such singular surfaces again have negligible measure and therefore we do not further consider such special cases in the present work.

Theorem~1 stated above allows for the graphical derivation of minimal complete sets of $2 N$ observables for the cases of single-meson photoproduction ($N = 4$) and electroproduction ($N = 6$), which has not been possible using the modified form of Moravcsik's theorem as stated in appendix~\ref{sec:ReviewMoravcsik} (see reference~\cite{Wunderlich:2020umg}). This fact will be illustrated in sections~\ref{sec:Photoproduction} and~\ref{sec:Electroproduction}. In case one wishes to consider problems with a larger number of $N > 6$ amplitudes, new obstacles appear which mainly are connected to the fact that the shape-classes encountered in these cases are more involved. We will comment on these issues in section~\ref{sec:Generalization}.

%\vspace*{5pt}

\section{Application to pseudoscalar meson photoproduction ($N = 4$)} \label{sec:Photoproduction}

Pseudoscalar meson photoproduction is generally described by $N = 4$ complex amplitudes, which are accompanied by $16$ polarization ob\-serva\-bles~\cite{Chiang:1996em, Nakayama:2018yzw}. The definitions of these observables in terms of transversity amplitudes $b_{1} , \ldots, b_{4}$ are given in Table~\ref{tab:PhotoObservables}. There exist $4$ shape-classes of diagonal ('D'), right-parallelogram ('PR'), anti-diagonal ('AD') and left-parallelogram ('PL') type (the importance of such shape-classes was originally pointed out in ref.~\cite{Chiang:1996em}). Every shape-class except for the class of diagonal observables ('D') has the generic form given in equations~\eqref{eq:NonTrivialShapeClassObsI} to~\eqref{eq:NonTrivialShapeClassObsIV} and thus contains $4$ ob\-serva\-bles. The diagonal shape-class 'D' contains the unpolarized differential cross section and the $3$ single-spin ob\-serva\-bles $\check{\Sigma}$, $\check{T}$ and $\check{P}$. Each of the $3$ non-diagonal shape-classes is in exact correspondence to one of the three groups of Beam-Target ($\mathcal{BT}$), Beam-Recoil ($\mathcal{BR}$),  and Target-Recoil ($\mathcal{TR}$) experiments, as indicated in Table~\ref{tab:PhotoObservables}.

\begin{table*}[ht]
 \begin{center}
 \begin{tabular}{lcr}
 \hline
 \hline
  Observable & \hspace*{5pt} Relative-phases  & \hspace*{10pt} Shape-class \\
  \hline 
  $\sigma_{0} = \frac{1}{2} \left( \left| b_{1} \right|^{2} + \left| b_{2} \right|^{2} + \left| b_{3} \right|^{2} + \left| b_{4} \right|^{2} \right)$ &  &     \\
  $- \check{\Sigma} = \frac{1}{2} \left( \left| b_{1} \right|^{2} + \left| b_{2} \right|^{2} - \left| b_{3} \right|^{2} - \left| b_{4} \right|^{2} \right)$  &  &   $\mathcal{S} = \mathrm{D}$ \\
  $- \check{T} = \frac{1}{2} \left( - \left| b_{1} \right|^{2} + \left| b_{2} \right|^{2} + \left| b_{3} \right|^{2} - \left| b_{4} \right|^{2} \right)$  &  &    \\
  $\check{P} = \frac{1}{2} \left( - \left| b_{1} \right|^{2} + \left| b_{2} \right|^{2} - \left| b_{3} \right|^{2} + \left| b_{4} \right|^{2} \right)$  &   &    \\
  \hline
   $\Ocal^{a}_{1+} = \left| b_{1} \right| \left| b_{3} \right| \sin \phi_{13} + \left| b_{2} \right| \left| b_{4} \right| \sin \phi_{24} = \mathrm{Im} \left[ b_{3}^{\ast} b_{1} + b_{4}^{\ast} b_{2} \right] = - \check{G}$  &  &  \\
   $\Ocal^{a}_{1-} = \left| b_{1} \right| \left| b_{3} \right| \sin \phi_{13} - \left| b_{2} \right| \left| b_{4} \right| \sin \phi_{24}  = \mathrm{Im} \left[ b_{3}^{\ast} b_{1} - b_{4}^{\ast} b_{2} \right] = \check{F}$  &  $\left\{ \phi_{13}, \phi_{24} \right\}$  & $a = \mathcal{BT} = \mathrm{PR}$ \\
   $\Ocal^{a}_{2+} = \left| b_{1} \right| \left| b_{3} \right| \cos \phi_{13} + \left| b_{2} \right| \left| b_{4} \right| \cos \phi_{24}  = \mathrm{Re} \left[ b_{3}^{\ast} b_{1} + b_{4}^{\ast} b_{2} \right] = - \check{E}$  &   &  \\
   $\Ocal^{a}_{2-} = \left| b_{1} \right| \left| b_{3} \right| \cos \phi_{13} - \left| b_{2} \right| \left| b_{4} \right| \cos \phi_{24} = \mathrm{Re} \left[ b_{3}^{\ast} b_{1} - b_{4}^{\ast} b_{2} \right] =  \check{H}$  &   &  \\
   \hline
   $\Ocal^{b}_{1+} = \left| b_{1} \right| \left| b_{4} \right| \sin \phi_{14} + \left| b_{2} \right| \left| b_{3} \right| \sin \phi_{23} = \mathrm{Im} \left[ b_{4}^{\ast} b_{1} + b_{3}^{\ast} b_{2} \right] = \check{O}_{z'}$  &   &  \\
   $\Ocal^{b}_{1-} = \left| b_{1} \right| \left| b_{4} \right| \sin \phi_{14} - \left| b_{2} \right| \left| b_{3} \right| \sin \phi_{23}  = \mathrm{Im} \left[ b_{4}^{\ast} b_{1} - b_{3}^{\ast} b_{2} \right] = - \check{C}_{x'}$  &  $\left\{ \phi_{14}, \phi_{23} \right\}$  & $b = \mathcal{BR} = \mathrm{AD}$ \\
   $\Ocal^{b}_{2+} = \left| b_{1} \right| \left| b_{4} \right| \cos \phi_{14} + \left| b_{2} \right| \left| b_{3} \right| \cos \phi_{23}  = \mathrm{Re} \left[ b_{4}^{\ast} b_{1} + b_{3}^{\ast} b_{2} \right] = - \check{C}_{z'}$  &   &  \\
   $\Ocal^{b}_{2-} = \left| b_{1} \right| \left| b_{4} \right| \cos \phi_{14} - \left| b_{2} \right| \left| b_{3} \right| \cos \phi_{23} = \mathrm{Re} \left[ b_{4}^{\ast} b_{1} - b_{3}^{\ast} b_{2} \right] = - \check{O}_{x'}$  &   &  \\
   \hline
   $\Ocal^{c}_{1+} = \left| b_{1} \right| \left| b_{2} \right| \sin \phi_{12} + \left| b_{3} \right| \left| b_{4} \right| \sin \phi_{34} = \mathrm{Im} \left[ b_{2}^{\ast} b_{1} + b_{4}^{\ast} b_{3} \right] = - \check{L}_{x'}$  &   &  \\
   $\Ocal^{c}_{1-} = \left| b_{1} \right| \left| b_{2} \right| \sin \phi_{12} - \left| b_{3} \right| \left| b_{4} \right| \sin \phi_{34}  = \mathrm{Im} \left[ b_{2}^{\ast} b_{1} - b_{4}^{\ast} b_{3} \right] = - \check{T}_{z'}$  &  $\left\{ \phi_{12}, \phi_{34} \right\}$  & $c = \mathcal{TR} = \mathrm{PL}$ \\
   $\Ocal^{c}_{2+} = \left| b_{1} \right| \left| b_{2} \right| \cos \phi_{12} + \left| b_{3} \right| \left| b_{4} \right| \cos \phi_{34}  = \mathrm{Re} \left[ b_{2}^{\ast} b_{1} + b_{4}^{\ast} b_{3} \right] = - \check{L}_{z'}$  &   &  \\
   $\Ocal^{c}_{2-} = \left| b_{1} \right| \left| b_{2} \right| \cos \phi_{12} - \left| b_{3} \right| \left| b_{4} \right| \cos \phi_{34} = \mathrm{Re} \left[ b_{2}^{\ast} b_{1} - b_{4}^{\ast} b_{3} \right] = \check{T}_{x'}$   &   &  \\
   \hline
   \hline
 \end{tabular}
 \end{center}
 \caption{The definitions of the $16$ polarization ob\-serva\-bles in pseudoscalar meson photoproduction (cf. ref.~\cite{Chiang:1996em}) are collected here. The observables are written in terms of transversity-amplitudes~$b_{1}, \ldots, b_{4}$. The non-diagonal observables are given in Nakayama's symbolic notation $\Ocal^{n}_{\nu \pm}$ (cf. reference~\cite{Nakayama:2018yzw}), but the ordinary names of the observables are given as well. The subdivision of the 16 observables into 4 shape-classes is explicitly shown. Furthermore, for the three non-diagonal shape-classes $a$, $b$ and $c$, the corresponding pairs of relative phases are indicated. The definitions and sign-conventions are chosen to be consistent with reference~\cite{MyPhD}.}
 \label{tab:PhotoObservables}
\end{table*}

For the ob\-serva\-bles in the non-diagonal shape-classes, we use the notation introduced by Nakayama~\cite{Nakayama:2018yzw}, which has also been used already in section~\ref{sec:NewCriterion}.

We mention here the fact that the $16$ observables can be written as bilinear hermitean forms defined in terms of a basis of $4 \times 4$ Dirac-matrices $\tilde{\Gamma}^{\alpha}$, which have been introduced in reference~\cite{Chiang:1996em} (see also reference~\cite{Wunderlich:2020umg}). We however do not list these Dirac-matrices explicitly in this work, although their internal structure is of course contained implicitly in all the mathematical facts leading to Theorem~1 of section~\ref{sec:NewCriterion}. Furthermore, the name 'shape-class' actually stems from the shapes of these Dirac-matrices~\cite{Chiang:1996em,MyPhD}.

We begin with the standard-assumption that all four observables from the diagonal shape-class ('D') have been measured in order to uniquely fix the four moduli $\left| b_{1} \right| , \ldots, \left| b_{4} \right| $. Therefore, the task is now to select four more observables from the remaining non-diagonal shape-classes $a$, $b$ and $c$, which corresponds to the determination of complete sets with minimal length $2 N = 8$, in order to uniquely specify the relative-phases. This is where the criterion formulated in Theorem~1 of section~\ref{sec:NewCriterion} becomes useful.

\begin{figure*}
 \begin{center}
\includegraphics[width = 0.95 \textwidth,trim={0.0cm 3.25cm 0.0cm 1.0cm},clip]{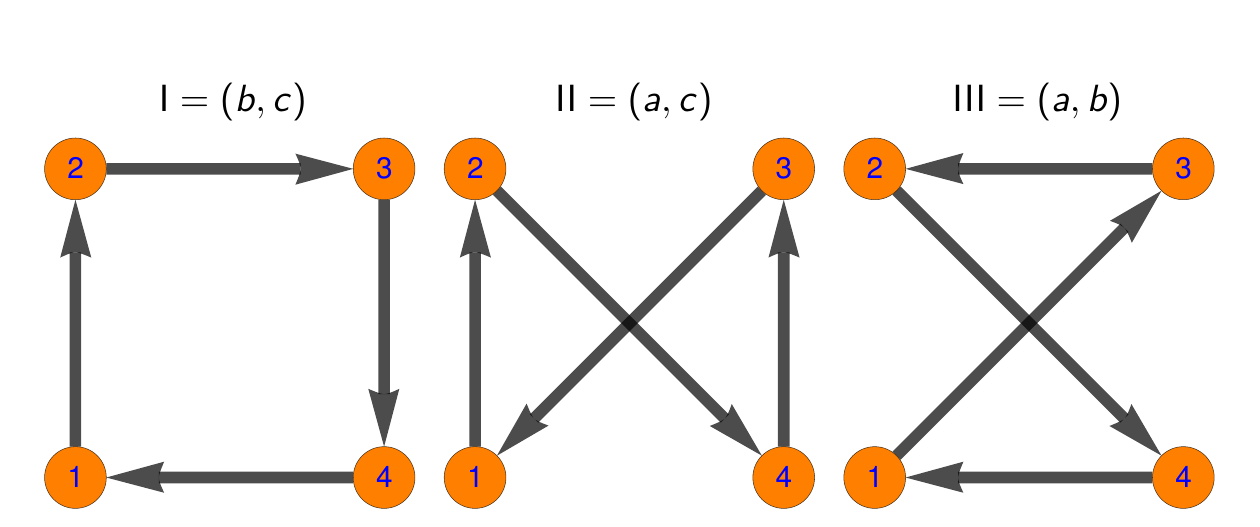} 
\end{center}
\vspace*{-5pt}
\caption{The three possible start-topologies for pseudoscalar meson photoproduction ($N = 4$ amplitudes) are drawn here. Each graph is drawn with a particular {\it direction}, which is intimately connected with to our way of writing the corresponding consistency relation, i.e. equations~\eqref{eq:ConsistencyRelPhotoproductionI},~\eqref{eq:ConsistencyRelPhotoproductionII} and~\eqref{eq:ConsistencyRelPhotoproductionIII} (cf. comments made below equation~\eqref{eq:GeneralConsistencyRelation} in section~\ref{sec:NewCriterion}). Each topology corresponds to the relative phases from a particular combination of two shape-classes (cf. Table~\ref{tab:PhotoObservables}), as is indicated above the graphs (cf. discussion in the main text).}
\label{fig:PhotoproductionStartTopologies}
\end{figure*}

The problem with $N = 4$ amplitudes allows for three non-trivial basic topologies for a connected graph (or 'closed loop') as demanded at the beginning of Theorem~1. The three topologies are shown in Figure~\ref{fig:PhotoproductionStartTopologies}, where also a definite {\it direction of translation} is indicated for each graph. If we consider for example the first box-like topology shown in Figure~\ref{fig:PhotoproductionStartTopologies}, we see that the indicated direction of the graph is in one-to-one correspondence to the following standard-convention for writing the consistency relation (see equation~\eqref{eq:GeneralConsistencyRelation}, as well as comments below that equation): 
\begin{equation}
 \text{I} = (b,c) \text{: } \phi_{12} + \phi_{23} + \phi_{34} + \phi_{41} = 0    .  \label{eq:ConsistencyRelPhotoproductionI}
\end{equation}
In exactly the same way, one can write a uniquely specified consistency-relation for each of the remaining two topologies, with their respective direction of translation. For the second and third topology shown in Figure~\ref{fig:PhotoproductionStartTopologies}, we have the expressions: 
\begin{align}
 \text{II} = (a,c) \text{: } \phi_{12} + \phi_{24} + \phi_{43} + \phi_{31} &= 0     ,  \label{eq:ConsistencyRelPhotoproductionII} \\
 \text{III} = (a,b) \text{: } \phi_{13} + \phi_{32} + \phi_{24} + \phi_{41} &= 0     .  \label{eq:ConsistencyRelPhotoproductionIII}
\end{align}

We again stress the fact that the sign-choices fixed in the standard-conventions~\eqref{eq:ConsistencyRelPhotoproductionI} to~\eqref{eq:ConsistencyRelPhotoproductionIII} are crucial for the applicability of the graphical criterion formulated in Theorem~1 of section~\ref{sec:NewCriterion}.

As already indicated in the equations~\eqref{eq:ConsistencyRelPhotoproductionI} to~\eqref{eq:ConsistencyRelPhotoproductionIII} written above, one should note that each of the topologies shown in Figure~\ref{fig:PhotoproductionStartTopologies}, as well as each of the consistency-relations~\eqref{eq:ConsistencyRelPhotoproductionI} to~\eqref{eq:ConsistencyRelPhotoproductionIII} is in direct correspondence to a particular combination of shape-classes from which the pairs of observables are to be picked (cf. Table~\ref{tab:PhotoObservables}). To be more precise, the first topology in Fig.~\ref{fig:PhotoproductionExampleLoops} corresponds to the combination of shape-classes $(b,c)$, the second topology relates to the combination $(a,c)$ and the third topology corresponds to $(a,b)$.

\begin{figure}
 \begin{center}
\includegraphics[width = 0.45 \textwidth,trim={0.0cm 0.5cm 0.0cm 0.5cm},clip]{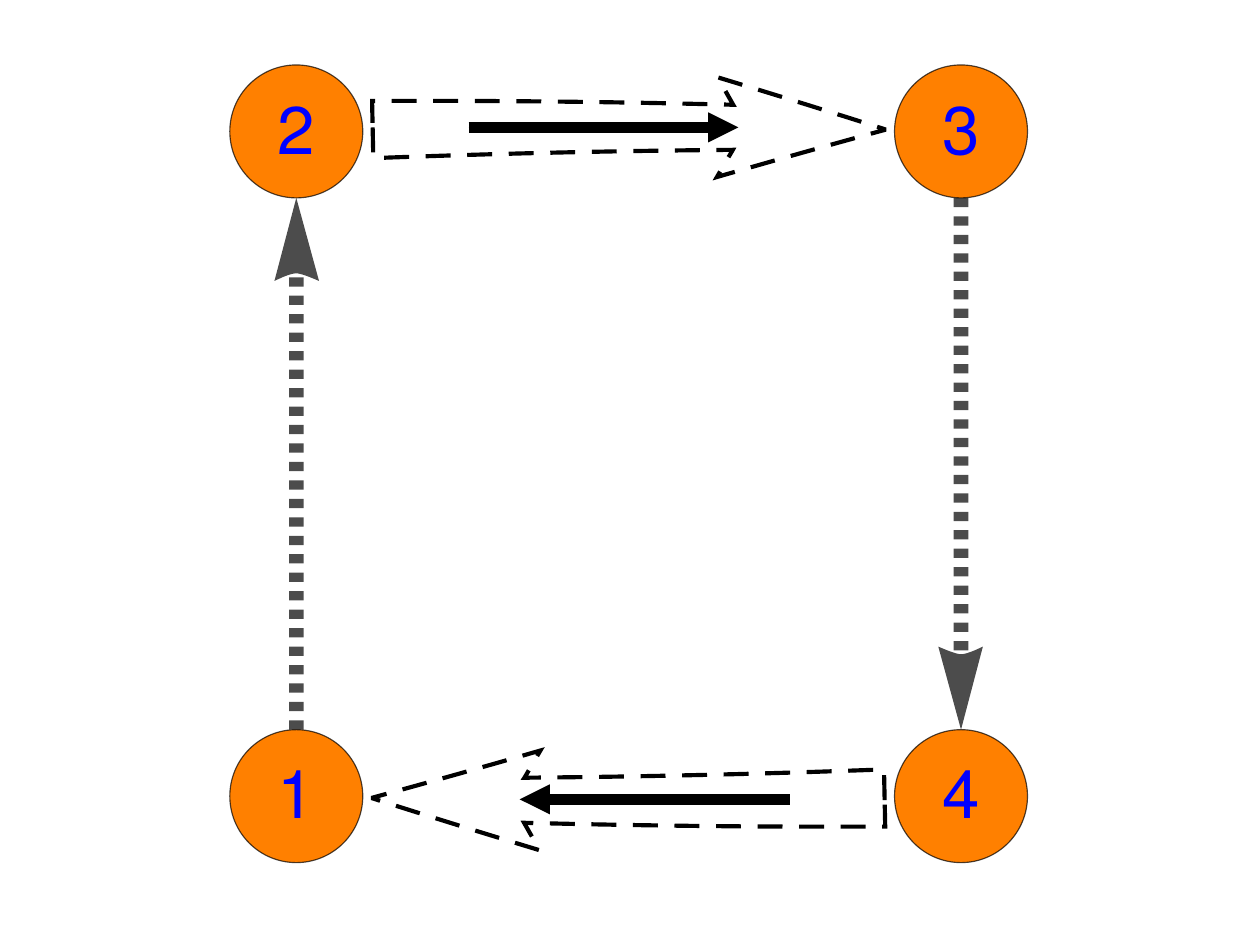} 
\end{center}
\vspace*{-5pt}
\caption{The first example for a fully complete graph according to Theorem~1 from section~\ref{sec:NewCriterion} is shown. This graph can be inferred from the selection of observables~\eqref{eq:PhotoproductionFirstExampleSet}. The dashed single-lined arrows indicate the fact that the selection 'A.1' has been applied for the two observables belonging to shape-class $c$, with corresponding relative-phases $\left\{ \phi_{12}, \phi_{34} \right\}$. The dashed double-lined arrows indicate the fact that a selection of type 'B' has been applied for the two observables from the shape-class $b$, with corresponding relative-phases $\left\{ \phi_{14}, \phi_{23} \right\}$. The $\zeta$-sign arrows have been drawn into the dashed double-lined arrows, according to the selection 'B.2' taken from shape-class~$b$ (cf. equation~\eqref{eq:TwoFoldPhaseAmbiguityII2}).}
\label{fig:PhotoproductionFirstExampleGraph}
\end{figure}

We now consider some examples for (in-) complete graphs, in order to illustrate how Theorem~1 (section~\ref{sec:NewCriterion}) works. Consider for instance the set of observables
\begin{equation}
  \left\{ \Ocal^{b}_{1+}, \Ocal^{b}_{2-}, \Ocal^{c}_{1+}, \Ocal^{c}_{1-} \right\}  . \label{eq:PhotoproductionFirstExampleSet}
\end{equation}
From this set of observables, one constructs the graph with box-like topology shown in Figure~\ref{fig:PhotoproductionFirstExampleGraph}, which satisfies the graphical criterion posed in Theorem~1.
The set~\eqref{eq:PhotoproductionFirstExampleSet} is composed of selection of type 'B.2' taken from the shape-class $b$ and a selection of type 'A.1' taken from shape-class $c$ (cf. discussion in section~\ref{sec:NewCriterion}). We now write explicitly all the possible cases for the consistency relation~\eqref{eq:ConsistencyRelPhotoproductionI} which follow from this particular selection of observables, and thus also correspond to the graph shown in Figure~\ref{fig:PhotoproductionFirstExampleGraph}. Since the discrete ambiguity in case 'A.1' is four-fold and for 'B.2' it is two-fold, we get the following eight relations (some multiples of $2 \pi$ have already been removed by hand in the following equations):
\begin{align}
  \alpha_{12} + \zeta - \alpha_{23}  + \alpha_{34} + \zeta - \alpha_{41}    &=  0  , \label{eq:FirstExampleConsistencyRelI} \\
  \pi - \alpha_{12} + \zeta - \alpha_{23}  + \alpha_{34} + \zeta - \alpha_{41}    &=  0  , \label{eq:FirstExampleConsistencyRelII} \\
  \alpha_{12} + \zeta - \alpha_{23}  + \pi - \alpha_{34} + \zeta - \alpha_{41}    &=  0  , \label{eq:FirstExampleConsistencyRelIII} \\
 - \alpha_{12} + \zeta - \alpha_{23}  - \alpha_{34} + \zeta - \alpha_{41}    &=  0  , \label{eq:FirstExampleConsistencyRelIV} \\
 \alpha_{12} + \zeta + \alpha_{23} - \pi  + \alpha_{34} + \zeta + \alpha_{41} - \pi    &=  0  , \label{eq:FirstExampleConsistencyRelV} \\
 \pi - \alpha_{12} + \zeta + \alpha_{23} - \pi  + \alpha_{34} + \zeta + \alpha_{41} - \pi    &=  0  , \label{eq:FirstExampleConsistencyRelVI} \\
 \alpha_{12} + \zeta + \alpha_{23} - \pi  + \pi - \alpha_{34} + \zeta + \alpha_{41} - \pi     &=  0  , \label{eq:FirstExampleConsistencyRelVII} \\
 - \alpha_{12} + \zeta + \alpha_{23} - \pi  - \alpha_{34} + \zeta + \alpha_{41} - \pi   &=  0  . \label{eq:FirstExampleConsistencyRelVIII} 
\end{align}
It can be seen that no degenerate pair of equations exists in this case. This is true due to the fact that the respective transitional $\zeta$-angle ($\zeta \equiv \zeta^{b}_{1+,2-}$ in this case) always appears with the {\it same sign} in each equation. One always obtains a term '$2 \zeta$' in each equation, since the $\zeta$-angles belonging to the two different relative-phases $\phi_{23}$ and $\phi_{14}$ do not cancel out. The graph shown in Figure~\ref{fig:PhotoproductionFirstExampleGraph} is just right for such a cancellation {\it not} to occur. Furthermore, we recognize the graphs constructed in Theorem~1 from section~\ref{sec:NewCriterion} to be in principle just graphical summaries of all cases for a particular consistency relation, corresponding to a specific set of observables.

\begin{figure}
 \begin{center}
\includegraphics[width = 0.45 \textwidth,trim={0.0cm 0.5cm 0.0cm 0.5cm},clip]{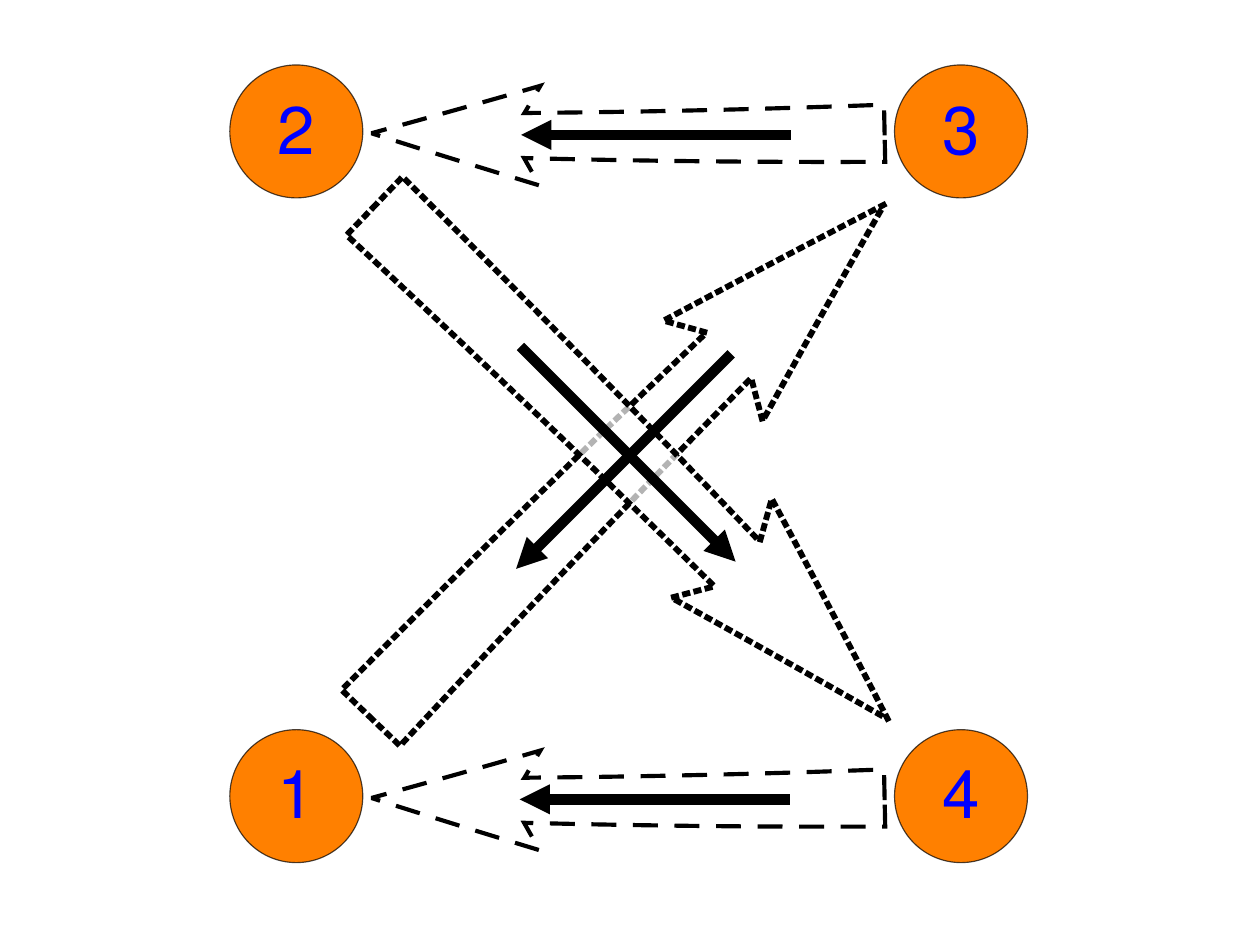} 
\end{center}
\vspace*{-5pt}
\caption{The second example for a fully complete graph according to Theorem~1 from section~\ref{sec:NewCriterion} is shown, which can be inferred from the selection of observables~\eqref{eq:PhotoproductionSecondExampleSet}. The dashed double-lined arrows indicate a selection of type 'B' from the shape-class $b$ (relative phases $\left\{ \phi_{14}, \phi_{23} \right\}$), while the dotted double-lined arrows represent a 'B'-type selection from shape-class $a$ (relative phases $\left\{ \phi_{13}, \phi_{24} \right\}$). The $\zeta$-sign arrows have been drawn into the respective double-lined arrows according to the selection~\eqref{eq:PhotoproductionSecondExampleSet}.}
\label{fig:PhotoproductionSecondExampleGraph}
\end{figure}

As a next example for a fully complete set, the following selection of observables is considered: 
\begin{equation}
  \left\{ \Ocal^{a}_{1+}, \Ocal^{a}_{2-}, \Ocal^{b}_{1+}, \Ocal^{b}_{2+} \right\}  . \label{eq:PhotoproductionSecondExampleSet}
\end{equation}
This set implies the graph shown in Figure~\ref{fig:PhotoproductionSecondExampleGraph}. 
This graph satisfies all the criteria posed by Theorem~1. The observables from shape-class $a$ are picked according to the case 'B.2' from section~\ref{sec:NewCriterion}, while the pair of observables from the class $b$ has been selected according to case 'B.1'. In both these cases, the discrete phase-ambiguity is two-fold. Therefore, we have to consider the following four cases for the consistency relation~\eqref{eq:ConsistencyRelPhotoproductionIII} (again removing possible summands of $2 \pi$ by hand)
\begin{align}
&- \zeta + \alpha_{13} + \zeta' + \alpha_{32} - \pi + \zeta  - \alpha_{24} + \zeta' - \alpha_{41}  \nonumber \\
&= 2 \zeta' + \alpha_{13} + \alpha_{32} - \alpha_{24} - \alpha_{41} - \pi   = 0  , \label{eq:SecondExampleConsistencyRelI} \\
& - \zeta + \alpha_{13} + \zeta' - \alpha_{32}  + \zeta  - \alpha_{24} + \zeta' + \alpha_{41} - \pi  \nonumber \\
&= 2 \zeta' + \alpha_{13} - \alpha_{32} - \alpha_{24} + \alpha_{41} - \pi   = 0  , \label{eq:SecondExampleConsistencyRelII} \\
& - \zeta - \alpha_{13} + \pi + \zeta' + \alpha_{32} - \pi + \zeta  + \alpha_{24} - \pi + \zeta' - \alpha_{41} \nonumber \\ &= 2 \zeta' - \alpha_{13} + \alpha_{32} + \alpha_{24} - \alpha_{41} - \pi   = 0  , \label{eq:SecondExampleConsistencyRelIII} \\
&  - \zeta - \alpha_{13} + \pi + \zeta' - \alpha_{32} + \zeta  + \alpha_{24} - \pi + \zeta' + \alpha_{41} - \pi \nonumber \\ & = 2 \zeta' - \alpha_{13} - \alpha_{32} + \alpha_{24} + \alpha_{41} - \pi   = 0  , \label{eq:SecondExampleConsistencyRelIV}
\end{align}
where the angle $\zeta = \zeta^{a}_{1+,2-}$ is defined from the pair $\left( \Ocal^{a}_{1+}, \Ocal^{a}_{2-}  \right)$, while the second angle $\zeta' \equiv \zeta^{b}_{1+,2+}$ belongs to the observables $\left( \Ocal^{b}_{1+}, \Ocal^{b}_{2+}  \right)$. Again, no pair of degenerate consistency-relations exists, since the $\zeta'$-angles remain in the equations, while the $\zeta$-angles cancel each other out.  

\begin{figure}
 \begin{center}
\includegraphics[width = 0.45 \textwidth,trim={0.0cm 0.5cm 0.0cm 0.5cm},clip]{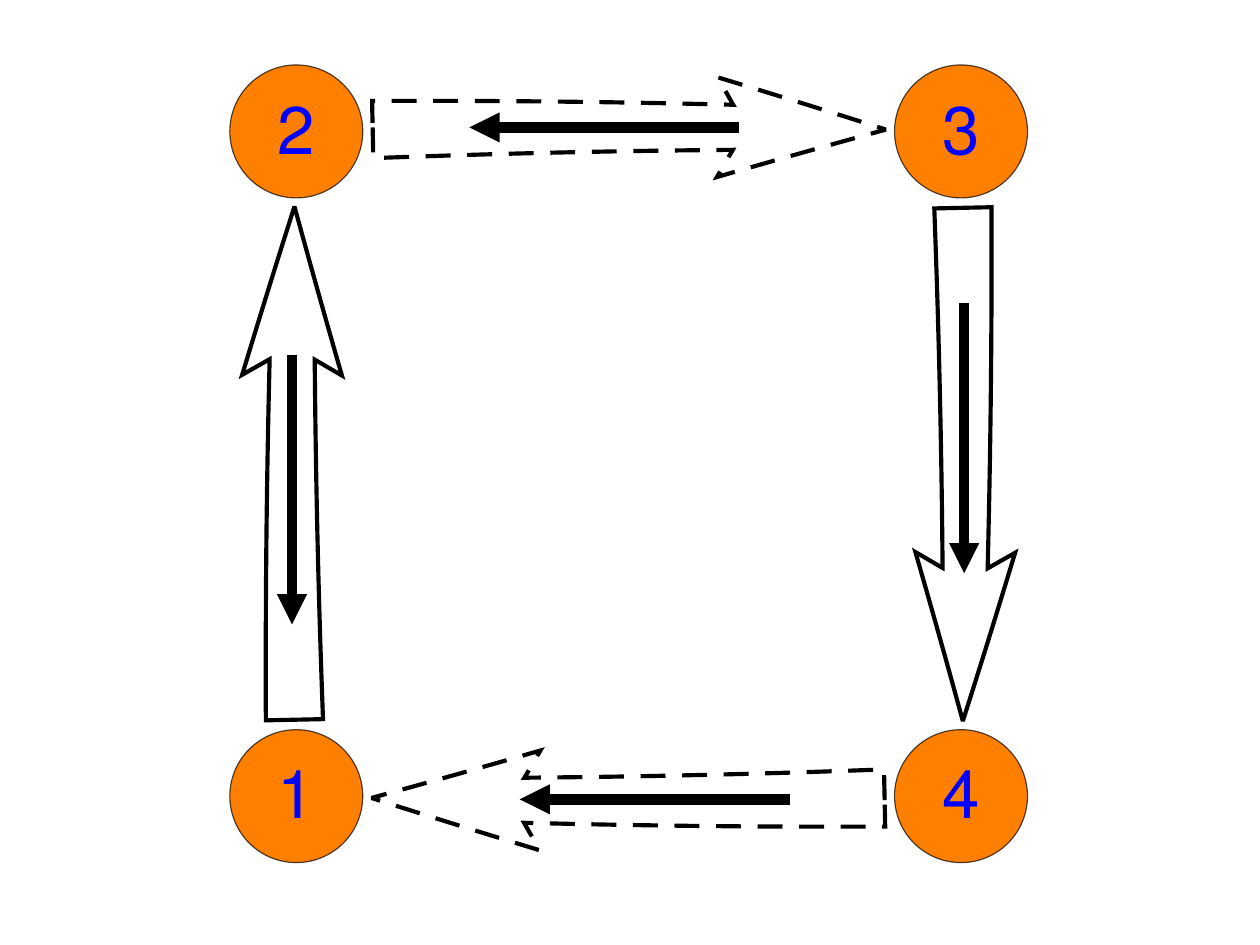} 
\end{center}
\vspace*{-5pt}
\caption{The third example-graph mentioned in the main text is shown here, which can be inferred from the set of observables~\eqref{eq:PhotoproductionThirdExampleSet}. The solid and dashed double-lined arrows have the same meaning as in Figure~\ref{fig:PhotoproductionFirstExampleGraph}. The $\zeta$-sign arrows, which have been drawn inside of the double-lined arrows, correspond to the selection~\eqref{eq:PhotoproductionThirdExampleSet}. This graph is {\it not} fully complete according to Theorem~1 from section~\ref{sec:NewCriterion}.}
\label{fig:PhotoproductionThirdExampleGraph}
\end{figure}

As a third example, we consider the following set:
\begin{equation}
  \left\{ \Ocal^{b}_{1+}, \Ocal^{b}_{2+}, \Ocal^{c}_{1+}, \Ocal^{c}_{2-} \right\}  . \label{eq:PhotoproductionThirdExampleSet}
\end{equation}
This set leads to the graph shown in Figure~\ref{fig:PhotoproductionThirdExampleGraph}, which does not satisfy the criteria of Theorem~1.
The selection 'B.1' has been picked from the shape-class $b$, while the combination 'B.2' has been picked from shape-class $c$. In both cases, the discrete phase-ambiguity is again two-fold, which implies the following four cases for the consistency relation:
\begin{align}
& - \zeta + \alpha_{12} - \zeta' - \alpha_{23} + \pi + \zeta  - \alpha_{34} + \zeta' - \alpha_{41}  \nonumber \\
&=  \alpha_{12} - \alpha_{23} - \alpha_{34} - \alpha_{41} + \pi   =  0  , \label{eq:ThirdExampleConsistencyRelI} \\
& - \zeta + \alpha_{12} - \zeta' + \alpha_{23} + \zeta - \alpha_{34} + \zeta' + \alpha_{41} - \pi \nonumber \\
&=  \alpha_{12} + \alpha_{23} - \alpha_{34}  + \alpha_{41} - \pi  =  0  , \label{eq:ThirdExampleConsistencyRelII} \\
& - \zeta - \alpha_{12} + \pi - \zeta' - \alpha_{23} + \pi + \zeta  + \alpha_{34} - \pi + \zeta' - \alpha_{41}  \nonumber \\
&= - \alpha_{12} - \alpha_{23} + \alpha_{34} - \alpha_{41} + \pi   =  0  , \label{eq:ThirdExampleConsistencyRelIII} \\
& - \zeta - \alpha_{12} + \pi - \zeta' + \alpha_{23}  + \zeta  + \alpha_{34} - \pi + \zeta' + \alpha_{41} - \pi   \nonumber \\
&=  - \alpha_{12} + \alpha_{23} + \alpha_{34} + \alpha_{41} - \pi   =  0  . \label{eq:ThirdExampleConsistencyRelIV}
\end{align}
The angle $\zeta = \zeta^{c}_{1+,2-}$ belongs here to the pair $\left( \Ocal^{c}_{1+}, \Ocal^{c}_{2-}  \right)$, while the second angle $\zeta' \equiv \zeta^{b}_{1+,2+}$ is defined by the observables $\left( \Ocal^{b}_{1+}, \Ocal^{b}_{2+}  \right)$. However, now we observe that all pairs of $\zeta$-angles cancel each other out in all the above-given equations. Then, degenerate pairs of relations emerge. For instance, it is possible to transform the equation~\eqref{eq:ThirdExampleConsistencyRelII} into equation~\eqref{eq:ThirdExampleConsistencyRelIII} via a multiplication by $(-1)$ (a similar transformation relates equations~\eqref{eq:ThirdExampleConsistencyRelI} and~\eqref{eq:ThirdExampleConsistencyRelIV}). Therefore, the considered set~\eqref{eq:PhotoproductionThirdExampleSet} is {\it not} fully complete.

In the same way as for the examples discussed above, we can consider all possible selections of pairs of observables, for each of the three start-topologies shown in Figure~\ref{fig:PhotoproductionStartTopologies}, then draw the graphs that follow from them and check for completeness using Theorem~1 from section~\ref{sec:NewCriterion}. When doing this, one has $36$ possible combinations for each start-topology, resulting in a total of $108$ combinations to consider. 

%\newpage

We found that~$60$ of these $108$ combinations are fully complete according to Theorem~1, using an automated procedure\footnote{Our code is really an automated way of checking the conditions of Theorem~1 algebraically, via specific checks on the orders of the indices in the respective relative-phases. I.e., in our code the cases for the consistency-relation are not evaluated explicitly and rather the code is doing what a person would do when checking the conditions for Theorem~1 by hand. For the photoproduction problem, one actually can check all~$108$ cases by hand in an acceptable timespan and thus would not need a code. This is very different for the electroproduction problem, see section~\ref{sec:Electroproduction}.} in Mathematica~\cite{Mathematica}. From each start-topology 'I', 'II' and 'III', there originate $20$ complete sets, respectively. The~$60$ complete sets found in this way are listed in the supplemental material~\cite{Supplement}.

In the formulation of Theorem~1 from section~\ref{sec:NewCriterion}, as well as in the derivations discussed in this section up to this point, we always started with a given selection of observables, then drew the correpsonding graph and checked this graph for completeness. The reason for this is that the mapping between combinations of observables and graphs is actually {\it not} bijective. On other words, when starting from a specific set of observables, one can always arrive at a uniquely specified graph. However, when going in the reverse direction, i.e. when starting from a graph, one may find multiple sets of observables that fit this graph. For instance, the set of observables~$\left\{ \Ocal^{b}_{1-}, \Ocal^{b}_{2+}, \Ocal^{c}_{1+}, \Ocal^{c}_{1-}  \right\}$ fits the graph shown in Figure~\ref{fig:PhotoproductionFirstExampleGraph}, exactly the same graph that originates from the first example-set~\eqref{eq:PhotoproductionFirstExampleSet}. 

\begin{figure*}[htb]
 \begin{center}
\includegraphics[width = 0.985 \textwidth,trim={0.0cm 4.25cm 0.0cm 2.25cm},clip]{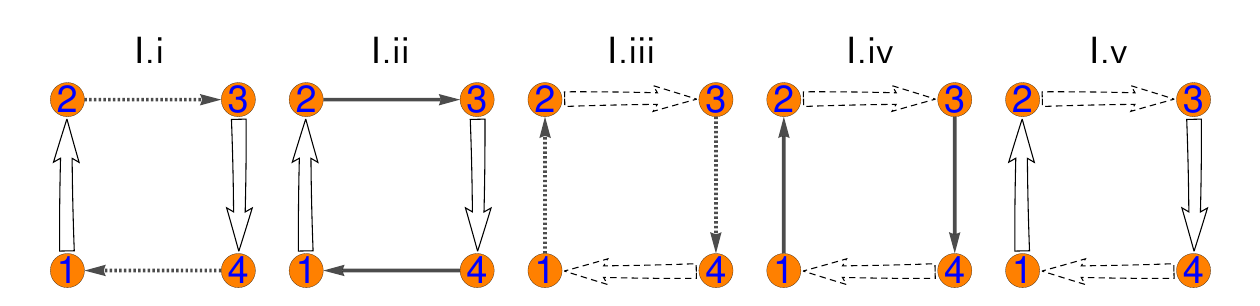} 
\end{center}
\vspace*{-5pt}
\caption{These five types of graphs have been derived from the first (i.e. box-like) topology 'I' shown in Figure~\ref{fig:PhotoproductionStartTopologies}. They correspond to different combinations of selections of type 'A' and 'B' taken from the shape-classes $b$ (relative-phases $\left\{ \phi_{14}, \phi_{23} \right\}$) and $c$ (relative-phases $\left\{ \phi_{12}, \phi_{34} \right\}$), as described in section~\ref{sec:NewCriterion}. At least one pair of double-lined arrows has to be contained in the graphs, i.e. at least one selection of type 'B' has to have been made. Graphs resulting solely from selections of type 'A' have not been shown (there exist $4$ of such graphs), since they cannot yield fully complete sets (cf. Theorem~2 from appendix~\ref{sec:ReviewMoravcsik}).}
\label{fig:PhotoproductionFiveImpliedTopologies}
\end{figure*}

\begin{figure}
 \begin{center}
\includegraphics[width = 0.485 \textwidth,trim={1.1cm 0.4cm 1.1cm 0.4cm},clip]{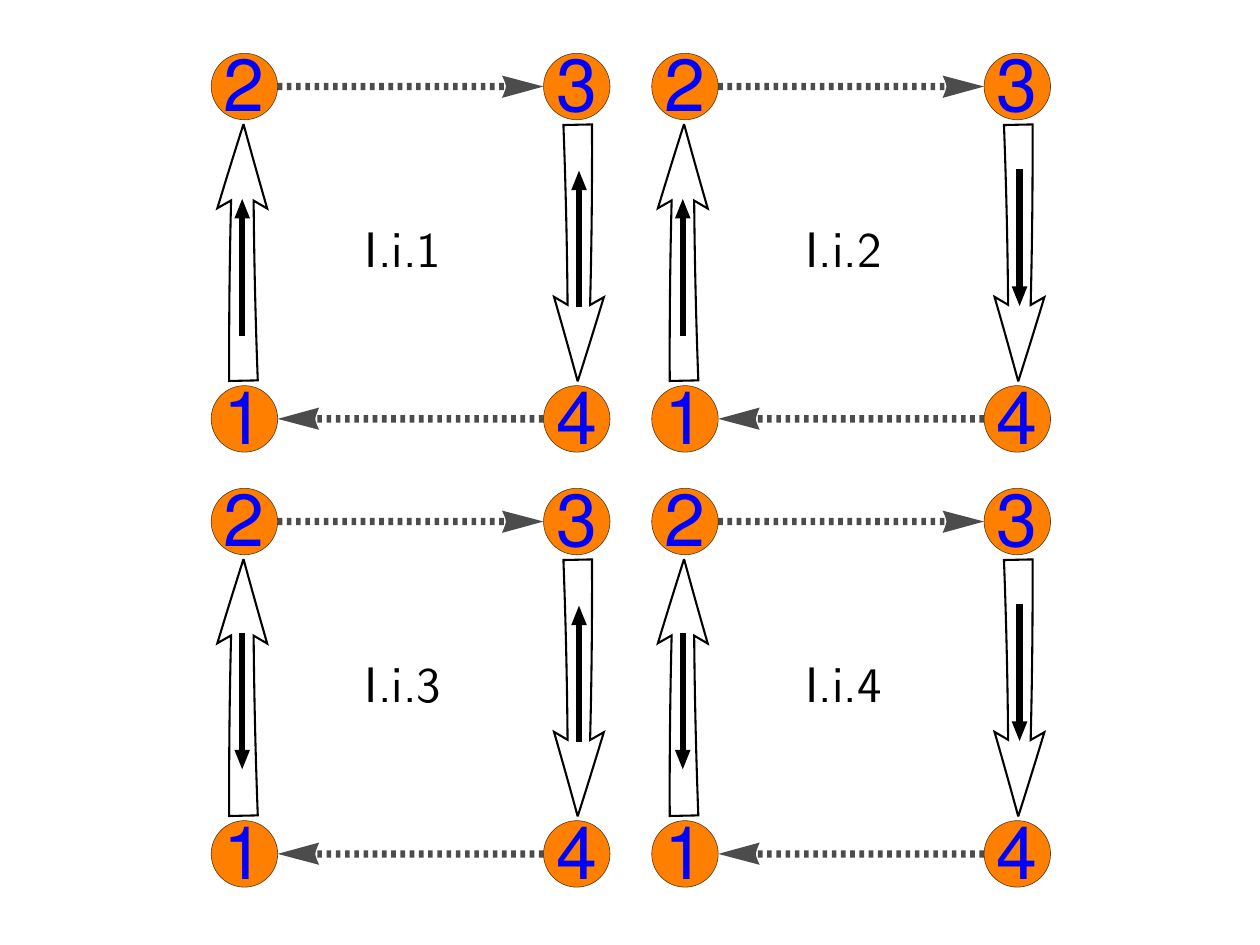} 
\end{center}
\vspace*{-5pt}
\caption{These four graphs are directly deduced from graph-type (I.i) shown in Figure~\ref{fig:PhotoproductionFiveImpliedTopologies}, by inserting all possible combinations of $\zeta$-sign arrows into the double-lined arrows. The two graphs (I.i.2) and (I.i.3) satisfy all the completeness-conditions posed by Theorem~1 from section~\ref{sec:NewCriterion}, while the graphs (I.i.1) and (I.i.4) violate these criteria.}
\label{fig:PhotoproductionFirstFourTopologies}
\end{figure}

\begin{figure*}
 \begin{center}
\includegraphics[width = 0.99 \textwidth,trim={1.0cm 0.4cm 1.0cm 0.4cm},clip]{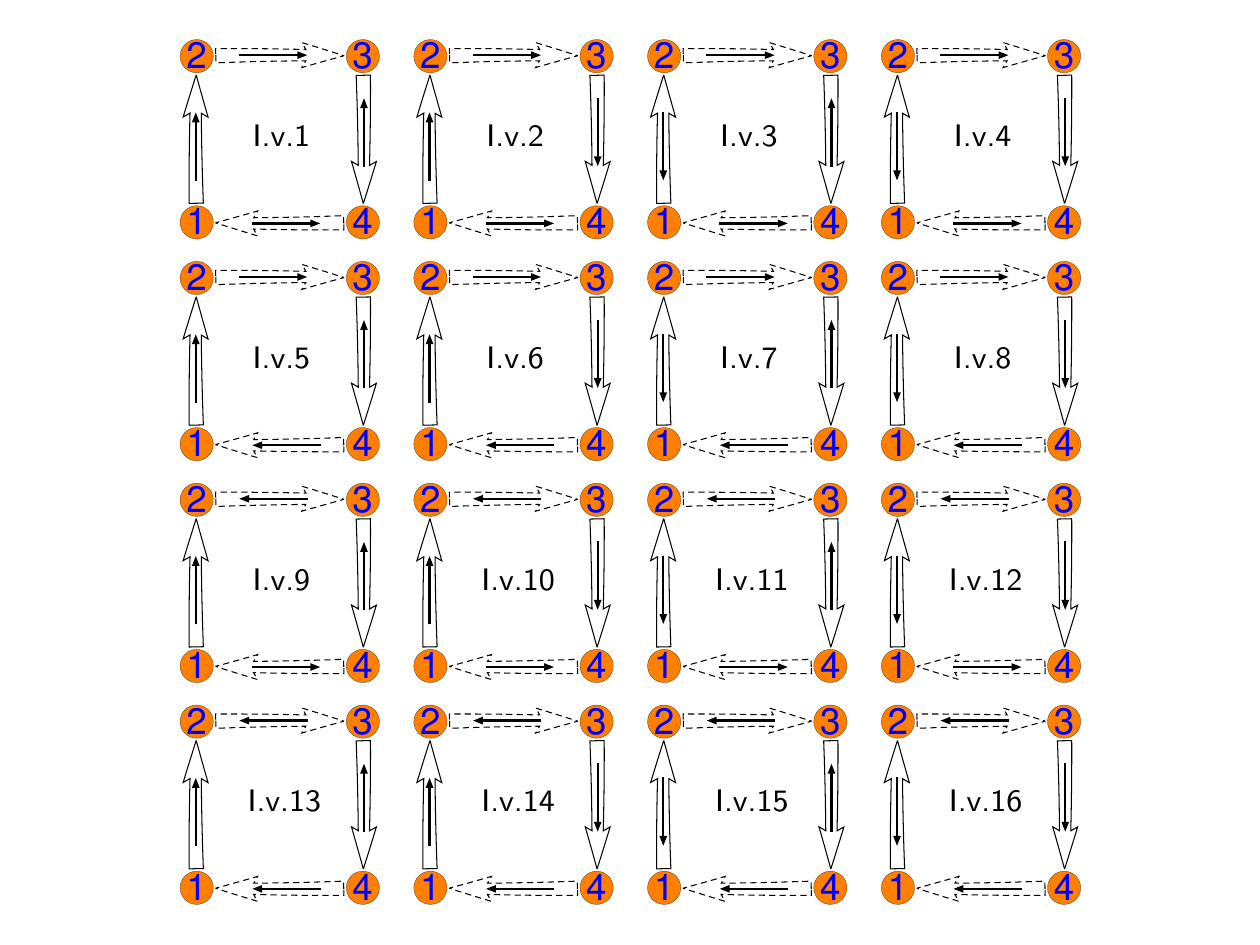} 
\end{center}
\vspace*{-5pt}
\caption{This set of $16$ graphs is directly deduced from graph-type (I.v) shown in Figure~\ref{fig:PhotoproductionFiveImpliedTopologies}, by inserting all possible combinations of $\zeta$-sign arrows into the double-lined arrows. From these $16$ graphs, only four violate the completeness-criterion posed in Theorem~1 of section~\ref{sec:NewCriterion}, namely: (I.v.1), (I.v.4), (I.v.13) and (I.v.16).}
\label{fig:PhotoproductionSixteenTopologies}
\end{figure*}

This leads one to question whether it is possible at all to find the above-mentioned~$60$ complete sets starting solely from graphs, i.e. without selecting observables first. In the following, we outline the individual steps for this alternative procedure and provide arguments for the fact that the above-mentioned non-bijectivity is actually not a problem. When deriving complete sets solely from graphs, one can proceed as follows:
\begin{itemize}
 \item[1.)] Starting from one of the basic topologies with direction shown in Figure~\ref{fig:PhotoproductionStartTopologies}, draw all possible types of graphs that result from it by applying all possible allowed combinations of pairs of single- and double-lined arrows. For the considered problem with $N = 4$ amplitudes, this would result in $9$ possible graph-types originating from one particular start-topology. However, from these~$9$ graph-types, only~$5$ contain double-lined arrows at all and thus can lead to complete sets. For the box-like topology 'I', these $5$ graph-types are shown in Figure~\ref{fig:PhotoproductionFiveImpliedTopologies}.
 \item[2.)] For each graph-type obtained in '1.)' which contains at least one pair of double-lined arrows, draw all possible combinations of $\zeta$-sign arrows into the given pairs of double-lined arrows. This implies $4$ new graphs in case the graph-type contains one pair of double-lined arrows and $16$ new graphs for the one graph-type shown in Figure~\ref{fig:PhotoproductionFiveImpliedTopologies} that has two pairs of double-lined arrows. In this step one therefore obtains $(4+4+4+4+16) = 32$ graphs from each start-topology. Thus, one obtains~$96$ graphs in total. Examples for such graphs are shown for the start-topology 'I' in Figures~\ref{fig:PhotoproductionFirstFourTopologies} and~\ref{fig:PhotoproductionSixteenTopologies}.
 \item[3.)] From the graphs determined in step '2.)', single out all graphs that satisfy the completeness-criterion posed in Theorem~1. For each graph-type containing one pair of double-lined arrows shown in Figure~\ref{fig:PhotoproductionFiveImpliedTopologies}, one obtains two complete cases of combinations of $\zeta$-signs, cf. Figure~\ref{fig:PhotoproductionFirstFourTopologies}. For each graph-type that has two pairs of double-lined arrows as shown in Figure~\ref{fig:PhotoproductionFiveImpliedTopologies}, one gets $12$ complete combinations of $\zeta$-signs, see Figure~\ref{fig:PhotoproductionSixteenTopologies}. Thus, one obtains $(2 + 2 + 2 + 2 + 12) = 20$ fully complete graphs from each start-topology and therefore $60$ fully complete graphs in total. We observe that the combinatorics match with the case discussed before, where we started from selections of observables.
 \item[4.)] For each complete graph obtained in step '3.)', find all selections of observables that fit it according to the cases 'A.1', 'A.2' and 'B.1', $\ldots$, 'B.4' listed in section~\ref{sec:NewCriterion}. Here, one runs into the problem that the graphs constructed in steps '1.)', $\ldots$, '3.)' allow for more combinations of $\zeta$-signs than exist in the cases listed in section~\ref{sec:NewCriterion}. In the cases 'B.1', $\ldots$, 'B.4' from section~\ref{sec:NewCriterion}, the possible combinations of the signs of the $\zeta$'s in the ambiguity-formulas for $(\phi_{ij},\phi_{kl})$ were: $(- \zeta, - \zeta)$ and $(- \zeta, + \zeta)$. This leads one to question to which combinations of observables the other two cases for the $\zeta$-signs, i.e. $(+ \zeta, + \zeta)$ and $(+ \zeta, - \zeta)$, correspond. \\
 As described in detail in appendix~\ref{sec:FlippedSignCases} and summarized in Table~\ref{tab:ZetaSignAssociationTable}, the 'new' cases $(+ \zeta, + \zeta)$ and $(+ \zeta, - \zeta)$ just correspond to the already known pairings of observables listed in section~\ref{sec:NewCriterion}, but with the sign of one of the two observables flipped. Therefore, these new cases introduce in principle redundant information, but they still can be very useful for the derivation of complete sets when starting solely from graphs. \\
 As an example, consider the two complete graphs (I.i.2) and (I.i.3) shown in Figure~\ref{fig:PhotoproductionFirstFourTopologies}. These graphs are related to each by a flip of the direction of both $\zeta$-sign arrows within the appearing pair of double-lined arrows. These two graphs correspond to the same observables: the graph (I.i.3) corresponds to the sets
 \begin{align}
  &\left\{ \Ocal^{b}_{1+}, \Ocal^{b}_{1-}, \Ocal^{c}_{1+}, \Ocal^{c}_{2+} \right\} \text{ and } \nonumber \\
  &\left\{ \Ocal^{b}_{1+}, \Ocal^{b}_{1-}, \Ocal^{c}_{1-}, \Ocal^{c}_{2-} \right\}    , \label{eq:GraphIi2Sets}
 \end{align}
 while the graph (I.i.2) corresponds to the sets (cf. Table~\ref{tab:ZetaSignAssociationTable})
 \begin{align}
  &\left\{ \Ocal^{b}_{1+}, \Ocal^{b}_{1-}, \Ocal^{c}_{1+}, - \Ocal^{c}_{2+} \right\} \text{ and } \nonumber \\
  &\left\{ \Ocal^{b}_{1+}, \Ocal^{b}_{1-}, \Ocal^{c}_{1-}, - \Ocal^{c}_{2-} \right\}      . \label{eq:GraphIi3Sets}
 \end{align}
 In this way, we see that graphs related to each other by the flip of {\it one} pair of $\zeta$-sign arrows generally introduce redundant information. However, this does not harm our ability to derive all~$60$ complete sets starting solely from graphs, since during the steps~'1.)' to~'3.)' outlined above, we have determined all possible fully complete graphs anyway. We just have to assign all complete sets of observables to each graph that is fully complete according to Theorem~1, using the associations shown in Table~\ref{tab:ZetaSignAssociationTable}. Then, from the resulting overall sets of observables, we have to sort out the non-redundant ones, which should then leave only the $60$ complete sets which we already determined when starting from the $108$ possible combinations of observables.
\end{itemize}

The steps '1.)' to '4.)' described above can in principle be automated on a computer.
\begin{table*}
\begin{center}
\begin{tabular}{l|cccc}
\hline 
\hline
  ($\zeta$-sign for $\phi_{ij}$, $\zeta$-sign for $\phi_{kl}$)  &   \multicolumn{4}{c}{Possible selections of observables} \\
\hline
$(- \zeta, - \zeta)$ & $\left( \Ocal^{n}_{1+}, \Ocal^{n}_{2+} \right)$, & $\left( - \Ocal^{n}_{1+}, - \Ocal^{n}_{2+} \right)$, & $\left( \Ocal^{n}_{1-}, \Ocal^{n}_{2-} \right)$, & $\left( - \Ocal^{n}_{1-}, - \Ocal^{n}_{2-} \right)$ \\
 $(- \zeta, + \zeta)$ & $\left( \Ocal^{n}_{1+}, \Ocal^{n}_{2-} \right)$,  & $\left( - \Ocal^{n}_{1+}, - \Ocal^{n}_{2-} \right)$,  & $\left( \Ocal^{n}_{1-}, \Ocal^{n}_{2+} \right)$,  & $\left( - \Ocal^{n}_{1-}, - \Ocal^{n}_{2+} \right)$  \\
  $(+ \zeta, - \zeta)$ & $\left( - \Ocal^{n}_{1+}, \Ocal^{n}_{2-} \right)$,  & $\left( \Ocal^{n}_{1+}, - \Ocal^{n}_{2-} \right)$,  & $\left( - \Ocal^{n}_{1-}, \Ocal^{n}_{2+} \right)$,  & $\left( \Ocal^{n}_{1-}, - \Ocal^{n}_{2+} \right)$  \\
 $(+ \zeta, + \zeta)$ &  $\left( - \Ocal^{n}_{1+}, \Ocal^{n}_{2+} \right)$, & $\left( \Ocal^{n}_{1+}, - \Ocal^{n}_{2+} \right)$, & $\left( - \Ocal^{n}_{1-}, \Ocal^{n}_{2-} \right)$, & $\left( \Ocal^{n}_{1-}, - \Ocal^{n}_{2-} \right)$ \\
 \hline
 \hline
\end{tabular}
\end{center}
\caption{The different cases for the signs of the $\zeta$-angles are given here, as they appear in the formulas for the discrete ambiguities for the relative-phases $\phi_{ij}$ and $\phi_{kl}$, as listed in section~\ref{sec:NewCriterion}. The possible selections of pairs of observables (including possible sign-flips for both observables), which correspond to the different sign-combinations for the $\zeta$'s, are given on the right. The cases~$(- \zeta, - \zeta)$ and~$(- \zeta, + \zeta)$ have been listed in section~\ref{sec:NewCriterion} and derived explicitly in appendix~\ref{sec:NakayamaDerivation}. The other two cases~$(+ \zeta, - \zeta)$ and~$(+ \zeta, + \zeta)$ follow from flipping the sign of one of the two observables in the respective pairs and thus give in principle redundant information. However, the information given in this Table is still useful in case one wishes to derive complete sets of observables starting solely from graphs, as described in the main text. Further details on how the results shown here were obtained are given in appendix~\ref{sec:FlippedSignCases}.}
\label{tab:ZetaSignAssociationTable}
\end{table*}

We report that all relevant complete sets found by Nakayama in the case '{\bf (2+2)}' (cf.~\cite{Nakayama:2018yzw}) have been recovered using the graphical criterion devised in Theorem~1 from section~\ref{sec:NewCriterion}. Furthermore, we also verified the completeness of the obtained sets with Mathematica~\cite{Mathematica}, using similar methods as those described in appendix~A of reference~\cite{Kroenert:2020ahf}. We refrain from listing all these complete sets here again, due to reasons of space. The sets have been collected in the supplemental material~\cite{Supplement}. The photoproduction problem has already been treated at length in the literature. Therefore, we refer to tables and lists already given in references~\cite{Chiang:1996em, Nakayama:2018yzw} for a further confirmation of our results.

It should come as no surprise that the results obtained by Nakayama were reproduced by Theorem~1 from section~\ref{sec:NewCriterion}, since this theorem is basically a graphical reformulation of derivations and criteria already contained in reference~\cite{Nakayama:2018yzw} (see in particular section~III there). However, the usefulness of the proposed graphical criterion will become apparent once we utilize it in order to derive complete sets of minimal length for the more involved problem of single-meson electroproduction, which will be the subject of the next section.

\clearpage

\section{Application to pseudoscalar meson electroproduction ($N = 6$)} \label{sec:Electroproduction}

Pseudoscalar meson electroproduction is described by $N = 6$ amplitudes $b_{1}, \ldots, b_{6}$, which are accompanied by $36$ polarization ob\-serva\-bles~\cite{Tiator:2017cde}. The expressions for the ob\-serva\-bles are collected in Table~\ref{tab:ElectroObservables}. Again, an algebra of Dirac-matrices is behind the definitions of the observables and therefore also behind their subdivision into different shape-classes~\cite{Wunderlich:2020umg}. This time, one has the $6 \times 6$ Dirac-matrices $\tilde{\Gamma}^{\alpha}$ (these are listed in appendix~B of reference~\cite{Wunderlich:2020umg}). For the ob\-serva\-bles in non-diagonal shape-classes, we again use the systematic notation $\Ocal^{n}_{\nu \pm}$ introduced by Nakayama~\cite{Nakayama:2018yzw}. However, in Table~\ref{tab:ElectroObservables}, we also give the observables in the usual physical notation, which is taken from the paper by Tiator and collaborators~\cite{Tiator:2017cde} and which states that each observable corresponds to a so-called 'response-function' $R^{\beta \alpha}_{i}$. The (physical) meaning of the indices on these response-functions is explained further in Table~\ref{tab:ElectroObservables}

The observables (and thus also the corresponding $\tilde{\Gamma}^{\alpha}$-matrices) can be grouped into $10$ overall shape-classes. Two shape-classes contain diagonal ob\-serva\-bles: one of these two classes, called 'D1', contains $4$ ob\-serva\-bles which correspond to matrices with non-vanishing entries in the first $4$ diagonal elements. The second diagonal shape-class 'D2' contains $2$ matrices with non-vanishing entries in the fifth and sixth diagonal element. The remaining $30$ ob\-serva\-bles are divided into $8$ non-diagonal shape-classes. These non-diagonal classes comprise four shape-classes of anti-diagonal structure ('AD1'$,\ldots,$'AD4'), three shape-classes of right-parallelogram type ('PR1'$,\ldots,$'PR3') and one class of left-parallelogram structure ('PL1'). All non-diagonal shape-classes each contain $4$ ob\-serva\-bles, apart from the class 'AD2' which is composed of just $2$ quantities.

%\clearpage

\begin{table*}%[h]
 \begin{center}
 \begin{tabular}{lcr}
  \hline
  \hline
  Observable & \hspace*{5pt} Relative-phases  & \hspace*{10pt} Shape-class \\
  \hline 
  $R^{00}_{T} = \frac{1}{2} \left( \left| b_{1} \right|^{2} + \left| b_{2} \right|^{2} + \left| b_{3} \right|^{2} + \left| b_{4} \right|^{2} \right)$  &   &    \\
  $- \hspace*{1pt}^{c} R^{00}_{TT} = \frac{1}{2} \left( \left| b_{1} \right|^{2} + \left| b_{2} \right|^{2} - \left| b_{3} \right|^{2} - \left| b_{4} \right|^{2} \right)$  &   &  $ \mathrm{D1}$ \\
  $- R_{T}^{0y} = \frac{1}{2} \left( - \left| b_{1} \right|^{2} + \left| b_{2} \right|^{2} + \left| b_{3} \right|^{2} - \left| b_{4} \right|^{2} \right)$ &  &    \\
  $ - R^{y' 0}_{T} = \frac{1}{2} \left( - \left| b_{1} \right|^{2} + \left| b_{2} \right|^{2} - \left| b_{3} \right|^{2} + \left| b_{4} \right|^{2} \right)$ &  &    \\
  \hline
   $\Ocal^{a}_{1+} = \left| b_{1} \right| \left| b_{3} \right| \sin \phi_{13} + \left| b_{2} \right| \left| b_{4} \right| \sin \phi_{24} = \mathrm{Im} \left[ b_{3}^{\ast} b_{1} + b_{4}^{\ast} b_{2} \right] = - \hspace*{1pt}^{s} R_{TT}^{0z}$  &  &  \\
   $\Ocal^{a}_{1-} = \left| b_{1} \right| \left| b_{3} \right| \sin \phi_{13} - \left| b_{2} \right| \left| b_{4} \right| \sin \phi_{24}  = \mathrm{Im} \left[ b_{3}^{\ast} b_{1} - b_{4}^{\ast} b_{2} \right] = R_{TT'}^{0x}$ &  $\left\{ \phi_{13}, \phi_{24}  \right\}$  & $a = \mathrm{PR1}$ \\
   $\Ocal^{a}_{2+} = \left| b_{1} \right| \left| b_{3} \right| \cos \phi_{13} + \left| b_{2} \right| \left| b_{4} \right| \cos \phi_{24}  = \mathrm{Re} \left[ b_{3}^{\ast} b_{1} + b_{4}^{\ast} b_{2} \right] = R_{TT'}^{0z}$  &  &  \\
   $\Ocal^{a}_{2-} = \left| b_{1} \right| \left| b_{3} \right| \cos \phi_{13} - \left| b_{2} \right| \left| b_{4} \right| \cos \phi_{24} = \mathrm{Re} \left[ b_{3}^{\ast} b_{1} - b_{4}^{\ast} b_{2} \right] =  \hspace*{1pt}^{s} R^{0x}_{TT}$  &   &  \\
   \hline
   $\Ocal^{b}_{1+} = \left| b_{1} \right| \left| b_{4} \right| \sin \phi_{14} + \left| b_{2} \right| \left| b_{3} \right| \sin \phi_{23} = \mathrm{Im} \left[ b_{4}^{\ast} b_{1} + b_{3}^{\ast} b_{2} \right] = - \hspace*{1pt}^{s} R^{z'0}_{TT}$  &   &  \\
   $\Ocal^{b}_{1-} = \left| b_{1} \right| \left| b_{4} \right| \sin \phi_{14} - \left| b_{2} \right| \left| b_{3} \right| \sin \phi_{23}  = \mathrm{Im} \left[ b_{4}^{\ast} b_{1} - b_{3}^{\ast} b_{2} \right] = - R^{x'0}_{TT'}$  &  $\left\{ \phi_{14}, \phi_{23}  \right\}$  & $b = \mathrm{AD1}$ \\
   $\Ocal^{b}_{2+} = \left| b_{1} \right| \left| b_{4} \right| \cos \phi_{14} + \left| b_{2} \right| \left| b_{3} \right| \cos \phi_{23}  = \mathrm{Re} \left[ b_{4}^{\ast} b_{1} + b_{3}^{\ast} b_{2} \right] = R^{z'0}_{TT'}$  &   &  \\
   $\Ocal^{b}_{2-} = \left| b_{1} \right| \left| b_{4} \right| \cos \phi_{14} - \left| b_{2} \right| \left| b_{3} \right| \cos \phi_{23} = \mathrm{Re} \left[ b_{4}^{\ast} b_{1} - b_{3}^{\ast} b_{2} \right] = - \hspace*{1pt}^{s} R^{x'0}_{TT}$  &   &  \\
   \hline
   $\Ocal^{c}_{1+} = \left| b_{1} \right| \left| b_{2} \right| \sin \phi_{12} + \left| b_{3} \right| \left| b_{4} \right| \sin \phi_{34} = \mathrm{Im} \left[ b_{2}^{\ast} b_{1} + b_{4}^{\ast} b_{3} \right] = - R^{x'z}_{T}$  &   &  \\
   $\Ocal^{c}_{1-} = \left| b_{1} \right| \left| b_{2} \right| \sin \phi_{12} - \left| b_{3} \right| \left| b_{4} \right| \sin \phi_{34}  = \mathrm{Im} \left[ b_{2}^{\ast} b_{1} - b_{4}^{\ast} b_{3} \right] = R^{z'x}_{T}$  &  $\left\{ \phi_{12}, \phi_{34}  \right\}$  & $c = \mathrm{PL1}$ \\
   $\Ocal^{c}_{2+} = \left| b_{1} \right| \left| b_{2} \right| \cos \phi_{12} + \left| b_{3} \right| \left| b_{4} \right| \cos \phi_{34}  = \mathrm{Re} \left[ b_{2}^{\ast} b_{1} + b_{4}^{\ast} b_{3} \right] = R_{T}^{z' z}$  &   &  \\
   $\Ocal^{c}_{2-} = \left| b_{1} \right| \left| b_{2} \right| \cos \phi_{12} - \left| b_{3} \right| \left| b_{4} \right| \cos \phi_{34} = \mathrm{Re} \left[ b_{2}^{\ast} b_{1} - b_{4}^{\ast} b_{3} \right] = R_{T}^{x' x}$  &   &  \\
   \hline
  $ R_{L}^{00} =\left| b_{5} \right|^{2} + \left| b_{6} \right|^{2}$  &   & $\mathrm{D2}$ \\
   $ R_{L}^{0y} = \left| b_{5} \right|^{2} - \left| b_{6} \right|^{2}$  &  &   \\
   \hline 
  $ \Ocal^{d}_{1} = 2 \left| b_{5} \right| \left| b_{6} \right| \sin \phi_{56} = 2 \mathrm{Im} \left[ b_{6}^{\ast} b_{5} \right] =  R_{L}^{z'x}$ &   $\left\{ \phi_{56}  \right\}$  & $d = \mathrm{AD2}$ \\
   $ \Ocal^{d}_{2} = 2 \left| b_{5} \right| \left| b_{6} \right| \cos \phi_{56}  = 2 \mathrm{Re} \left[ b_{6}^{\ast} b_{5} \right] = - R_{L}^{x'x}$  &  &   \\
  \hline 
   $\Ocal^{e}_{1+} = \left| b_{3} \right| \left| b_{6} \right| \sin \phi_{36} + \left| b_{4} \right| \left| b_{5} \right| \sin \phi_{45} = \mathrm{Im} \left[ b_{6}^{\ast} b_{3} + b_{5}^{\ast} b_{4} \right] = - \hspace*{1pt}^{s} R^{00}_{LT'}$   &   &  \\
   $\Ocal^{e}_{1-} = \left| b_{3} \right| \left| b_{6} \right| \sin \phi_{36} - \left| b_{4} \right| \left| b_{5} \right| \sin \phi_{45}  = \mathrm{Im} \left[ b_{6}^{\ast} b_{3} - b_{5}^{\ast} b_{4} \right] = \hspace*{1pt}^{s} R^{0y}_{LT'}$  &  $\left\{ \phi_{36}, \phi_{45}  \right\}$  & $e = \mathrm{AD3}$ \\
   $\Ocal^{e}_{2+} = \left| b_{3} \right| \left| b_{6} \right| \cos \phi_{36} + \left| b_{4} \right| \left| b_{5} \right| \cos \phi_{45}  = \mathrm{Re} \left[ b_{6}^{\ast} b_{3} + b_{5}^{\ast} b_{4} \right] = \hspace*{1pt}^{c} R^{00}_{LT}$  &   &  \\
   $\Ocal^{e}_{2-} = \left| b_{3} \right| \left| b_{6} \right| \cos \phi_{36} - \left| b_{4} \right| \left| b_{5} \right| \cos \phi_{45} = \mathrm{Re} \left[ b_{6}^{\ast} b_{3} - b_{5}^{\ast} b_{4} \right] = - \hspace*{1pt}^{c} R^{0y}_{LT}$  &    &  \\
   \hline 
   $\Ocal^{f}_{1+} = \left| b_{1} \right| \left| b_{6} \right| \sin \phi_{16} + \left| b_{2} \right| \left| b_{5} \right| \sin \phi_{25} = \mathrm{Im} \left[ b_{6}^{\ast} b_{1} + b_{5}^{\ast} b_{2} \right] = - \hspace*{1pt}^{s} R^{0z}_{LT}$  &   &  \\
   $\Ocal^{f}_{1-} = \left| b_{1} \right| \left| b_{6} \right| \sin \phi_{16} - \left| b_{2} \right| \left| b_{5} \right| \sin \phi_{25}  = \mathrm{Im} \left[ b_{6}^{\ast} b_{1} - b_{5}^{\ast} b_{2} \right] = \hspace*{1pt}^{c} R^{0x}_{LT'}$ &  $\left\{ \phi_{16}, \phi_{25}  \right\}$  & $f = \mathrm{AD4}$ \\
   $\Ocal^{f}_{2+} = \left| b_{1} \right| \left| b_{6} \right| \cos \phi_{16} + \left| b_{2} \right| \left| b_{5} \right| \cos \phi_{25}  = \mathrm{Re} \left[ b_{6}^{\ast} b_{1} + b_{5}^{\ast} b_{2} \right] = \hspace*{1pt}^{c} R^{0z}_{LT'}$  &   &  \\
   $\Ocal^{f}_{2-} = \left| b_{1} \right| \left| b_{6} \right| \cos \phi_{16} - \left| b_{2} \right| \left| b_{5} \right| \cos \phi_{25} = \mathrm{Re} \left[ b_{6}^{\ast} b_{1} - b_{5}^{\ast} b_{2} \right] =  \hspace*{1pt}^{s} R^{0x}_{LT}$  &    &  \\
   \hline 
   $\Ocal^{g}_{1+} = \left| b_{1} \right| \left| b_{5} \right| \sin \phi_{15} + \left| b_{2} \right| \left| b_{6} \right| \sin \phi_{26} = \mathrm{Im} \left[ b_{5}^{\ast} b_{1} + b_{6}^{\ast} b_{2} \right] = -  \hspace*{1pt}^{s} R^{z'0}_{LT}$ &  &  \\
   $\Ocal^{g}_{1-} = \left| b_{1} \right| \left| b_{5} \right| \sin \phi_{15} - \left| b_{2} \right| \left| b_{6} \right| \sin \phi_{26}  = \mathrm{Im} \left[ b_{5}^{\ast} b_{1} - b_{6}^{\ast} b_{2} \right] = - \hspace*{1pt}^{c} R^{x'0}_{LT'}$ &  $\left\{ \phi_{15}, \phi_{26}  \right\}$  & $g = \mathrm{PR2}$ \\
   $\Ocal^{g}_{2+} = \left| b_{1} \right| \left| b_{5} \right| \cos \phi_{15} + \left| b_{2} \right| \left| b_{6} \right| \cos \phi_{26}  = \mathrm{Re} \left[ b_{5}^{\ast} b_{1} + b_{6}^{\ast} b_{2} \right] = \hspace*{1pt}^{c} R^{z'0}_{LT'} $  &   &  \\
   $\Ocal^{g}_{2-} = \left| b_{1} \right| \left| b_{5} \right| \cos \phi_{15} - \left| b_{2} \right| \left| b_{6} \right| \cos \phi_{26} = \mathrm{Re} \left[ b_{5}^{\ast} b_{1} - b_{6}^{\ast} b_{2} \right] = - \hspace*{1pt}^{s} R^{x'0}_{LT}$  &    &  \\
   \hline 
   $\Ocal^{h}_{1+} = \left| b_{3} \right| \left| b_{5} \right| \sin \phi_{35} + \left| b_{4} \right| \left| b_{6} \right| \sin \phi_{46} = \mathrm{Im} \left[ b_{5}^{\ast} b_{3} + b_{6}^{\ast} b_{4} \right] = \hspace*{1pt}^{s} R^{x'x}_{LT'}$  &   &  \\
   $\Ocal^{h}_{1-} = \left| b_{3} \right| \left| b_{5} \right| \sin \phi_{35} - \left| b_{4} \right| \left| b_{6} \right| \sin \phi_{46}  = \mathrm{Im} \left[ b_{5}^{\ast} b_{3} - b_{6}^{\ast} b_{4} \right] = - \hspace*{1pt}^{c} R^{z'x}_{LT}$ &  $\left\{ \phi_{35}, \phi_{46}  \right\}$  & $h = \mathrm{PR3}$ \\
   $\Ocal^{h}_{2+} = \left| b_{3} \right| \left| b_{5} \right| \cos \phi_{35} + \left| b_{4} \right| \left| b_{6} \right| \cos \phi_{46}  = \mathrm{Re} \left[ b_{5}^{\ast} b_{3} + b_{6}^{\ast} b_{4} \right] = - \hspace*{1pt}^{c} R^{x'x}_{LT}$  &  &  \\
   $\Ocal^{h}_{2-} = \left| b_{3} \right| \left| b_{5} \right| \cos \phi_{35} - \left| b_{4} \right| \left| b_{6} \right| \cos \phi_{46} = \mathrm{Re} \left[ b_{5}^{\ast} b_{3} - b_{6}^{\ast} b_{4} \right] = - \hspace*{1pt}^{s} R^{z'x}_{LT'}$  &    &  \\
   \hline
   \hline
 \end{tabular}
 \end{center}
 \caption{The definitions of electroproduction ob\-serva\-bles are collected here for the diagonal ob\-serva\-bles of types $\mathrm{D1}$ and $\mathrm{D2}$, as well as for the non-diagonal shape-classes $\left\{ a,b,c,d,e,f,g,h \right\}$. The specific combinations of relative-phases belonging to each individual non-diagonal shape-class are indicated as well. The definitions and sign-conventions for the ob\-serva\-bles have been adopted from reference~\cite{Tiator:2017cde}. \\ Every ob\-serva\-ble from a non-diagonal shape-class is written in the systematic symbolic notation~$\Ocal^{n}_{\nu \pm}$ introduced by Nakayama~\cite{Nakayama:2018yzw}. Furthermore, we also give for the observables the usual physical notation, which is defined as follows~\cite{Tiator:2017cde}: every observable corresponds to a 'response-function' $R^{\beta \alpha}_{i}$. The superscript-index $\alpha$ represents the target-polarization, the index $\beta$ indicates the recoil-polarization and the sub-script $i$ represents the polarization of the virtual photon in electroproduction, which can take the following configurations: $i \in \left\{ T, L, TL, TT, TL', TT' \right\}$ (meaning purely longitudinal, purely transverse or 'mixed' interference contributions to the differential cross section). In case the letter '$s$' or '$c$' is written as an additional superscript on the left of the respective response-function, then this indicates a possible sine- or cosine-dependence of the respective contribution to the differential cross section (with the sine or cosine depending on the azimuthal angle of the produced pseudoscalar meson).}
 \label{tab:ElectroObservables}
\end{table*}

We again make the standard-assumption that all six observables $\left\{ R^{00}_{T}, \hspace*{1pt}^{c} R^{00}_{TT}, R_{T}^{0y}, R^{y' 0}_{T}, R_{L}^{00} , R_{L}^{0y}  \right\}$ from the diagonal shape-classes 'D1' and 'D2' have been already used to uniquely fix the six moduli $\left| b_{1} \right| , \ldots, \left| b_{6} \right| $. Then, one has to select six more observables from the remaining non-diagonal shape-classes, which corresponds to the determination of complete sets with minimal length $2 N = 12$. Such minimal complete sets should then be able to uniquely specify the relative-phases. This is where we again use the criterion formulated in Theorem~1 of section~\ref{sec:NewCriterion}.

\begin{figure*}
 \begin{center}
\includegraphics[width = 0.975 \textwidth,trim={0.0cm 1.8cm 0.0cm 0.4cm},clip]{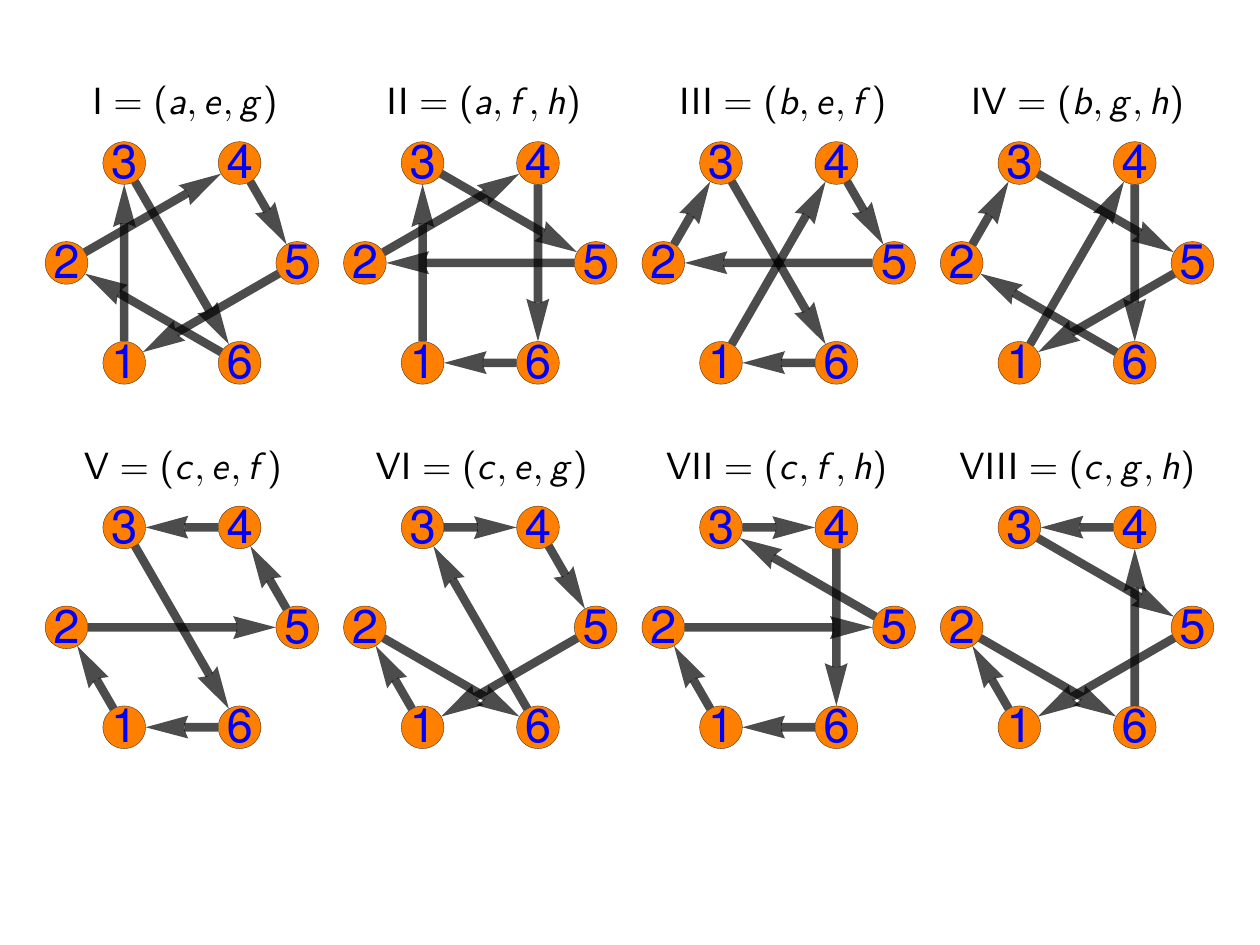} 
\end{center}
\vspace*{-5pt}
\caption{The $8$ possible start-topologies for pseudoscalar meson electroproduction ($N = 6$ amplitudes), which allow for a selection of three pairs of observables from three different shape-classes, are shown here. The {\it direction of translation} is indicated for each graph as well. This direction is intimately connected to our convention of writing the corresponding consistency relation, i.e. equations~\eqref{eq:MinimalConsistencyRelElectroproductionI} to~\eqref{eq:MinimalConsistencyRelElectroproductionVIII} (see also the comments made below equation~\eqref{eq:GeneralConsistencyRelation} in section~\ref{sec:NewCriterion}). Each of the shown topologies corresponds to the relative phases from a particular combination of three shape-classes for electroproduction (cf. Table~\ref{tab:ElectroObservables}). The combinations of shape-classes are also indicated above the graphs (cf. discussion in the main text).}
\label{fig:ElectroproductionStartTopologies}
\end{figure*}

For the problem of electroproduction ($N = 6$ amplitudes), there exist $60$ possible topologies for fully connected graphs with $6$ vertices, where every vertex has order $2$ (i.e. is touched by two edges). We refrain here from showing all these $60$ topologies, due to reasons of space. They can be found in section~VI of reference~\cite{Wunderlich:2020umg}. The additional constraints formulated for the considered graphs in the beginning of Theorem~1 from section~\ref{sec:NewCriterion} place further restrictions on the topologies. We have only to consider those topologies which correspond to three pairs of relative phases from three different shape-classes of four (cf. Table~\ref{tab:ElectroObservables}). From the above-mentioned total of $60$ possible topologies, only $8$ topologies remain that satisfy this constraint. These $8$ possibilities are shown in Figure~\ref{fig:ElectroproductionStartTopologies}. They constitute the possible start-topologies for our application of Theorem~1 to electroproduction.

%

%\clearpage

Each of the $8$ start-topologies corresponds to a particular combination of observables from three different shape-classes. Furthermore, the directions for the graphs shown in Figure~\ref{fig:ElectroproductionStartTopologies} stand in a one-to-one correspondence to our convention for writing the consistency relations (cf. the comments made below equation~\eqref{eq:GeneralConsistencyRelation} in section~\ref{sec:NewCriterion}). The consistency relations corresponding to the $8$ topologies shown in Figure~\ref{fig:ElectroproductionStartTopologies} read as follows:
\begin{align}
\text{I} = (a,e,g) \text{: } &\phi_{13} + \phi_{36} + \phi_{62} + \phi_{24} + \phi_{45} + \phi_{51} \nonumber \\
&= 0    ,  \label{eq:MinimalConsistencyRelElectroproductionI} \\
\text{II} = (a,f,h) \text{: }  &\phi_{13} + \phi_{35} + \phi_{52} + \phi_{24} + \phi_{46} + \phi_{61} \nonumber \\
&= 0    ,  \label{eq:MinimalConsistencyRelElectroproductionII} \\
\text{III} = (b,e,f) \text{: }  &\phi_{14} + \phi_{45} + \phi_{52} + \phi_{23} + \phi_{36} + \phi_{61} \nonumber \\
&= 0    ,  \label{eq:MinimalConsistencyRelElectroproductionIII} \\
\text{IV} = (b,g,h) \text{: }  &\phi_{14} + \phi_{46} + \phi_{62} + \phi_{23} + \phi_{35} + \phi_{51} \nonumber \\
&= 0    ,  \label{eq:MinimalConsistencyRelElectroproductionIV} \\
\text{V} = (c,e,f) \text{: }  &\phi_{12} + \phi_{25} + \phi_{54} + \phi_{43} + \phi_{36} + \phi_{61} \nonumber \\
&= 0    ,  \label{eq:MinimalConsistencyRelElectroproductionV} \\
\text{VI} = (c,e,g) \text{: }  &\phi_{12} + \phi_{26} + \phi_{63} + \phi_{34} + \phi_{45} + \phi_{51} \nonumber \\
&= 0    ,  \label{eq:MinimalConsistencyRelElectroproductionVI} \\
\text{VII} = (c,f,h) \text{: }  &\phi_{12} + \phi_{25} + \phi_{53} + \phi_{34} + \phi_{46} + \phi_{61} \nonumber \\
&= 0    ,  \label{eq:MinimalConsistencyRelElectroproductionVII} \\
\text{VIII} = (c,g,h) \text{: }  &\phi_{12} + \phi_{26} + \phi_{64} + \phi_{43} + \phi_{35} + \phi_{51} \nonumber \\
&= 0    .  \label{eq:MinimalConsistencyRelElectroproductionVIII}
\end{align}
We stress again the fact that the sign-conventions fixed by these directions are of vital importance for the applicability of Theorem~1.

\begin{figure}
 \begin{center}
\includegraphics[width = 0.475 \textwidth,trim={0.0cm 0.5cm 0.0cm 0.5cm},clip]{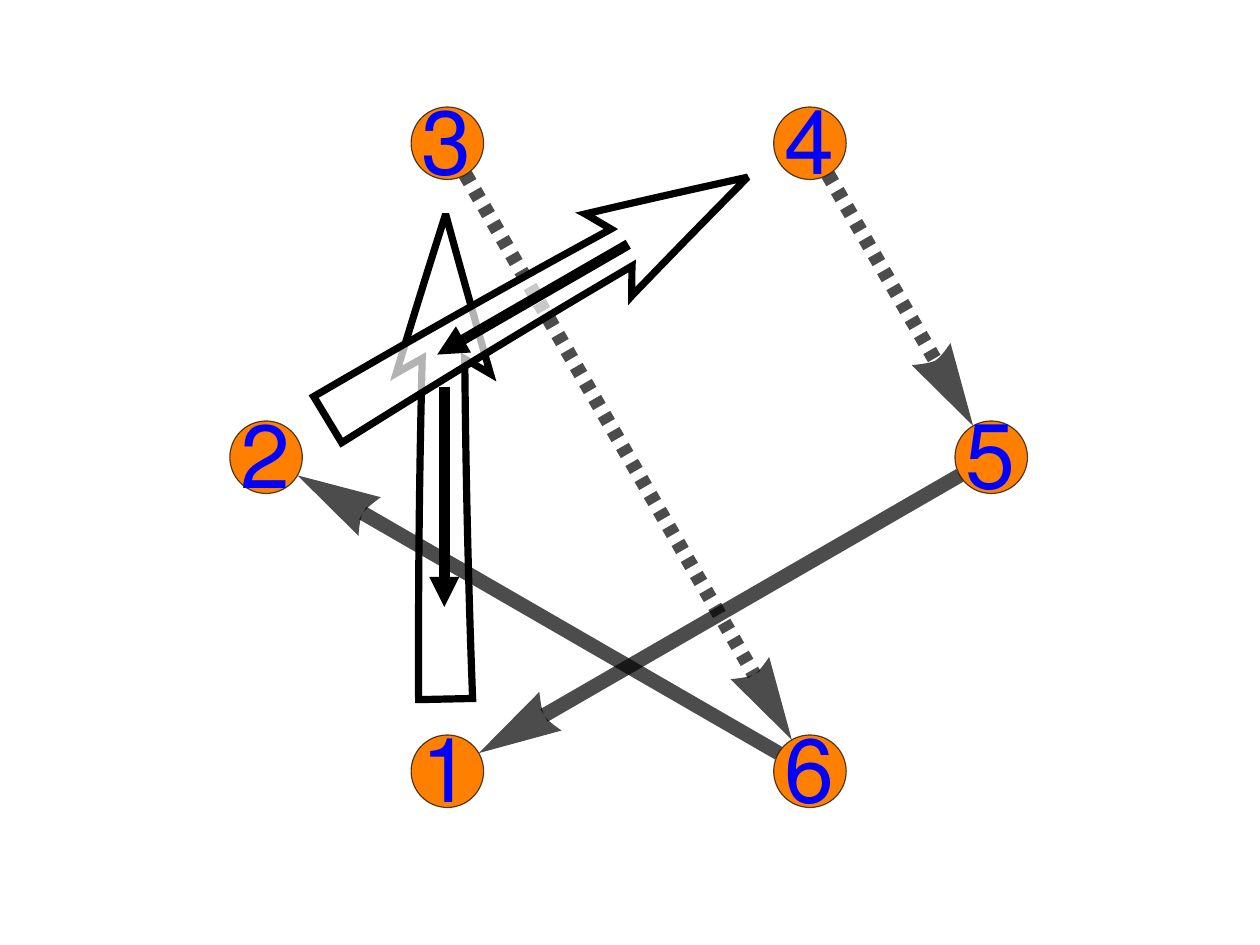} 
\end{center}
\vspace*{-5pt}
\caption{The first example for a graph in electroproduction ($N = 6$ amplitudes) is shown. This graph is fully complete according to Theorem~1 from section~\ref{sec:NewCriterion}. It can be inferred from the selection of observables~\eqref{eq:ElectroproductionFirstExampleSet}. The dashed single-lined arrows indicate the selection of type 'A.1' for the shape-class $e$, while the solid single-lined arrows represent the selection of type 'A.2' for the shape-class~$g$. The solid double-lined arrows indicate the fact that a selection of type 'B' has been applied for the two observables from the shape-class $a$. The $\zeta$-sign arrows have been drawn into the solid double-lined arrows according to the selection~\eqref{eq:ElectroproductionFirstExampleSet}.}
\label{fig:ElectroproductionFirstExampleGraph}
\end{figure}

As a first example for a selection of $6$ observables from the non-diagonal shape-classes (i.e. of three pairs of observables), we consider the following set:
\begin{equation}
  \left\{ \Ocal^{a}_{1+}, \Ocal^{a}_{2+}, \Ocal^{e}_{1+}, \Ocal^{e}_{1-}, \Ocal^{g}_{2+}, \Ocal^{g}_{2-} \right\}  , \label{eq:ElectroproductionFirstExampleSet}
\end{equation}
which implies the graph shown in Figure~\ref{fig:ElectroproductionFirstExampleGraph}.
This specific graph fulfills the completeness-criterion posed in Theorem~1 and therefore the set~\eqref{eq:ElectroproductionFirstExampleSet} is in fact complete. When combined with the $6$ observables from the diagonal shape-classes 'D1' and 'D2', the set~\eqref{eq:ElectroproductionFirstExampleSet} thus forms a complete set of minimal length $2 N = 12$. We refrain here from writing all the cases for the consistency-relation~\eqref{eq:MinimalConsistencyRelElectroproductionI} explicitly and instead again mention the fact that the graph shown in Figure~\ref{fig:ElectroproductionFirstExampleGraph} constitues a useful summary of all these cases.

\begin{figure}
 \begin{center}
\includegraphics[width = 0.475 \textwidth,trim={0.0cm 0.5cm 0.0cm 0.5cm},clip]{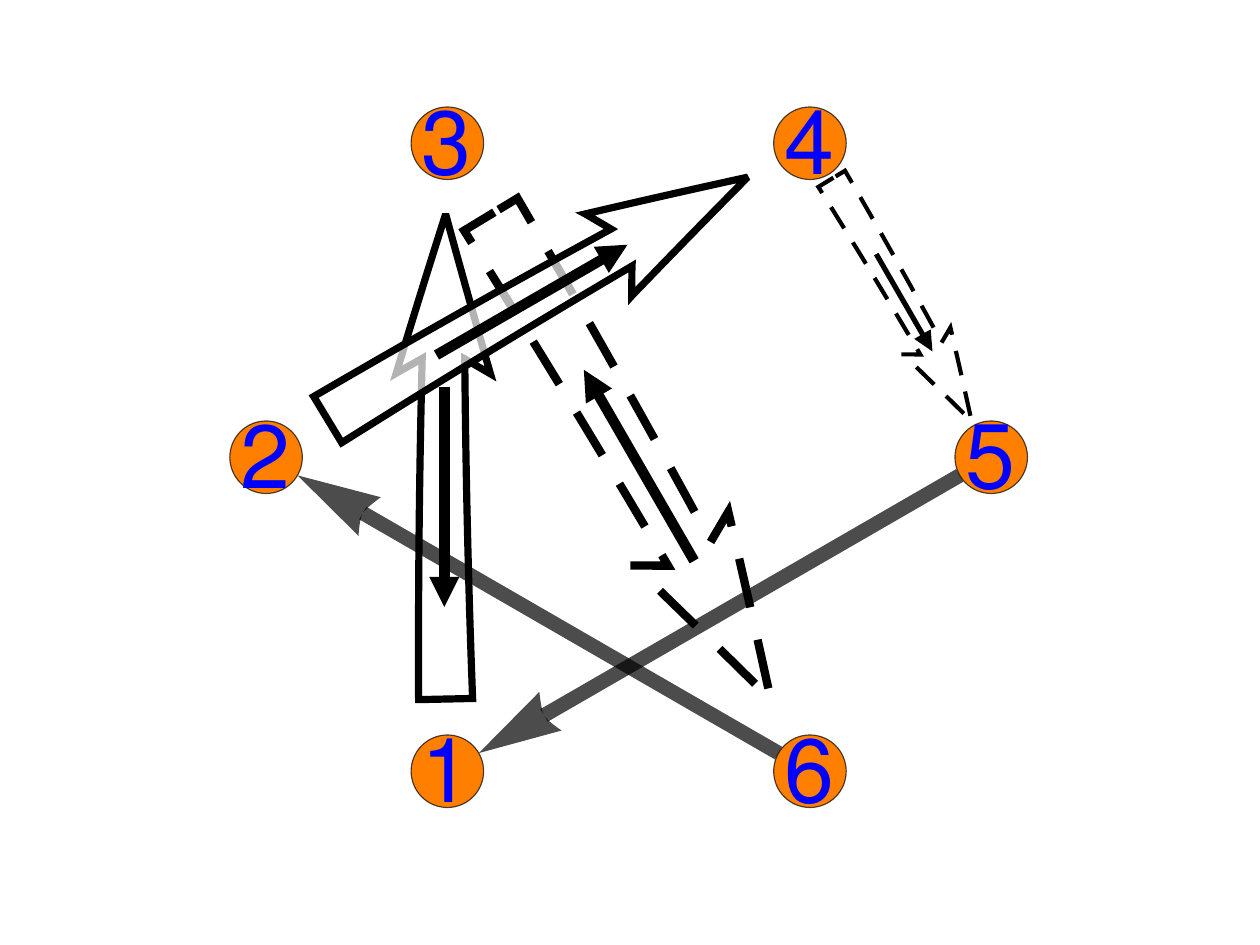} 
\end{center}
\vspace*{-5pt}
\caption{The second example-graph for electroproduction is shown. This graph violates the completeness-criterion posed in Theorem~1 from section~\ref{sec:NewCriterion}. It can be inferred from the selection of observables~\eqref{eq:ElectroproductionSecondExampleSet}. The solid single-lined arrows indicate the selection of type 'A.2' for the shape-class~$g$. The solid double-lined arrows indicate the fact that a selection of type 'B' has been applied for the two observables from the shape-class $a$, while the dashed double-lined arrows represent the same fact for the relative-phases belonging to shape-class~$e$. The $\zeta$-sign arrows have been drawn into the double-lined arrows according to the selection~\eqref{eq:ElectroproductionSecondExampleSet}.}
\label{fig:ElectroproductionSecondExampleGraph}
\end{figure}

As a second example-set, we consider the following selection of $6$ observables
\begin{equation}
  \left\{ \Ocal^{a}_{1+}, \Ocal^{a}_{2-}, \Ocal^{e}_{1-}, \Ocal^{e}_{2+}, \Ocal^{g}_{2+}, \Ocal^{g}_{2-} \right\}  , \label{eq:ElectroproductionSecondExampleSet}
\end{equation}
which implies the graph shown in Figure~\ref{fig:ElectroproductionSecondExampleGraph}. This graph violates the completeness-criterion from Theorem~1 and therefore the set~\eqref{eq:ElectroproductionSecondExampleSet} is not complete.

The search for all complete sets using Theorem~1 can again proceed via considerations of all relevant combinations of observables (cf. the discussion in section~\ref{sec:Photoproduction}). For each of the~$8$ possible start-topologies shown in Figure~\ref{fig:ElectroproductionStartTopologies}, one then has to consider~$216$ combinations, which are made up of all the possibilities to select three pairs of observables using the cases 'A.1', $\ldots$, 'B.4' outlined in section~\ref{sec:NewCriterion}. Thus, there exists a total of $(8*216) = 1728$ combinations that have to be considered. Using the Mathematica-routines already mentioned in section~\ref{sec:Photoproduction}, we found~$1216$ combinations from these $1728$ different possibilities to be fully complete. These complete sets are composed of~$152$ sets for each of the $8$ start-topologies shown in Figure~\ref{fig:ElectroproductionStartTopologies}. One example-set for each start-topology is given in Table~\ref{tab:MinimalCompleteExampleSetsElectroproduction}. The full list of~$1216$ complete sets is given in the supplemental material~\cite{Supplement}.

\begin{table}%[h]
\begin{tabular}{l|cccccc}
%\hline
\hline
\hline
Set-Nr. & \multicolumn{6}{c}{Observables} \\
\hline   
$(a, e, g): 42$ & $\hspace*{1pt}^{s} R_{TT}^{0z}$ & $R_{TT'}^{0z}$ & $\hspace*{1pt}^{s} R_{LT'}^{00}$ & $\hspace*{1pt}^{s} R_{LT'}^{0y}$ & $\hspace*{1pt}^{c} R_{LT'}^{z'0}$ & $\hspace*{1pt}^{s} R_{LT}^{x'0}$ \\
& $\Ocal^{a}_{1+}$ & $\Ocal^{a}_{2+}$ & $\Ocal^{e}_{1+}$ & $\Ocal^{e}_{1-}$ & $\Ocal^{g}_{2+}$ & $\Ocal^{g}_{2-}$ \\  
$ (a, f, h): 10$ & $\hspace*{1pt}^{s} R_{TT}^{0z}$ & $R_{TT'}^{0x}$ & $\hspace*{1pt}^{s} R_{LT}^{0z}$ & $\hspace*{1pt}^{c} R_{LT'}^{0z}$ & $\hspace*{1pt}^{c} R_{LT}^{z'x}$ & $\hspace*{1pt}^{s} R_{LT'}^{z'x}$ \\   & $\Ocal^{a}_{1+}$ & $\Ocal^{a}_{1-}$ & $\Ocal^{f}_{1+}$ & $\Ocal^{f}_{2+}$ & $\Ocal^{h}_{1-}$ & $\Ocal^{h}_{2-}$ \\
$ (b, e, f): 100$ & $R_{TT'}^{x'0}$ & $R_{TT'}^{z'0}$ & $\hspace*{1pt}^{c} R_{LT}^{00}$ & $\hspace*{1pt}^{c} R_{LT}^{0y}$ & $\hspace*{1pt}^{c} R_{LT'}^{0x}$ & $\hspace*{1pt}^{s} R_{LT}^{0x}$ \\   & $\Ocal^{b}_{1-}$ & $\Ocal^{b}_{2+}$ & $\Ocal^{e}_{2+}$ & $\Ocal^{e}_{2-}$ & $\Ocal^{f}_{1-}$ & $\Ocal^{f}_{2-}$ \\
$ (b, g, h): 150$ & $R_{TT'}^{x'0}$ & $\hspace*{1pt}^{s} R_{TT}^{x'0}$ & $\hspace*{1pt}^{c} R_{LT'}^{x'0}$ & $\hspace*{1pt}^{s} R_{LT}^{x'0}$ & $\hspace*{1pt}^{s} R_{LT'}^{x'x}$ & $\hspace*{1pt}^{s} R_{LT'}^{z'x}$ \\   & $\Ocal^{b}_{1-}$ & $\Ocal^{b}_{2-}$ & $\Ocal^{g}_{1-}$ & $\Ocal^{g}_{2-}$ & $\Ocal^{h}_{1+}$ & $\Ocal^{h}_{2-}$ \\
$ (c, e, f): 30$ & $R_{T}^{z'z}$ & $R_{T}^{x'x}$ & $\hspace*{1pt}^{s} R_{LT'}^{00}$ & $\hspace*{1pt}^{c} R_{LT}^{0y}$ & $\hspace*{1pt}^{s} R_{LT}^{0z}$ & $\hspace*{1pt}^{s} R_{LT}^{0x}$ \\   & $\Ocal^{c}_{2+}$ & $\Ocal^{c}_{2-}$ & $\Ocal^{e}_{1+}$ & $\Ocal^{e}_{2-}$ & $\Ocal^{f}_{1+}$ & $\Ocal^{f}_{2-}$ \\
 $(c, e, g): 80$ & $R_{T}^{x'z}$ & $R_{T}^{x'x}$ & $\hspace*{1pt}^{c} R_{LT}^{00}$ & $\hspace*{1pt}^{c} R_{LT}^{0y}$ & $\hspace*{1pt}^{c} R_{LT'}^{x'0}$ & $\hspace*{1pt}^{c} R_{LT'}^{z'0}$ \\   & $\Ocal^{c}_{1+}$ & $\Ocal^{c}_{2-}$ & $\Ocal^{e}_{2+}$ & $\Ocal^{e}_{2-}$ & $\Ocal^{g}_{1-}$ & $\Ocal^{g}_{2+}$ \\
 $(c, f, h): 1$ & $R_{T}^{x'z}$ & $R_{T}^{z'x}$ & $\hspace*{1pt}^{s} R_{LT}^{0z}$ & $\hspace*{1pt}^{c} R_{LT'}^{0x}$ & $\hspace*{1pt}^{s} R_{LT'}^{x'x}$ & $\hspace*{1pt}^{s} R_{LT'}^{z'x}$ \\   & $\Ocal^{c}_{1+}$ & $\Ocal^{c}_{1-}$ & $\Ocal^{f}_{1+}$ & $\Ocal^{f}_{1-}$ & $\Ocal^{h}_{1+}$ & $\Ocal^{h}_{2-}$ \\
 $(c, g, h): 50$ & $R_{T}^{x'z}$ & $R_{T}^{z'z}$ & $\hspace*{1pt}^{s} R_{LT}^{z'0}$ & $\hspace*{1pt}^{s} R_{LT}^{x'0}$ & $\hspace*{1pt}^{s} R_{LT'}^{x'x}$ & $\hspace*{1pt}^{s} R_{LT'}^{z'x}$ \\   & $\Ocal^{c}_{1+}$ & $\Ocal^{c}_{2+}$ & $\Ocal^{g}_{1+}$ & $\Ocal^{g}_{2-}$ & $\Ocal^{h}_{1+}$ & $\Ocal^{h}_{2-}$ \\ 
\hline
\hline
\end{tabular}
\caption{Here we collect $8$ selected examples for the minimal complete sets composed of $12$ observables. One example has been chosen for each of the $8$ different start-topologies shown in Figure~\ref{fig:ElectroproductionStartTopologies}. The $6$ ob\-serva\-bles given here in each case have to be combined with the $6$ diagonal ob\-serva\-bles $\left\{ R^{00}_{T},\hspace*{1pt}^{c}  R^{00}_{TT}, R_{T}^{0y}, R^{y' 0}_{T}, R_{L}^{00} , R_{L}^{0y}  \right\}$ in order to form a complete set of $12$. Every example is given in the 'response-function' notation $R^{\beta \alpha}_{i}$ and also in Nakayama's~\cite{Nakayama:2018yzw} systematic notation~$\Ocal^{n}_{\nu \pm}$~(cf. Table~\ref{tab:ElectroObservables}). The labelling-scheme for the set-number contains the combination of shape-classes for each of the $8$ start-topologies, as well as the number that the respective set has been given in the full lists contained in the supplemental material~\cite{Supplement}. This Table contains the first example-set~\eqref{eq:ElectroproductionFirstExampleSet} (i.e. the graph shown in Figure~\ref{fig:ElectroproductionFirstExampleGraph}) discussed in the main text, which is here the example for the shape-class combination '$(a,e,g)$'. The supplemental material~\cite{Supplement} contains all the~$1216$ minimal complete sets derived for electroproduction in this work. These~$1216$ minimal complete sets can be further subdivided into~$152$ complete sets for each of the~$8$ relevant shape-class combinations, i.e. for each of the~$8$ relevant start-topologies.}
\label{tab:MinimalCompleteExampleSetsElectroproduction}
\end{table}

\begin{figure*}
 \begin{center}
\includegraphics[width = 0.985 \textwidth,trim={0.25cm 0.0cm 0.25cm 0.0cm},clip]{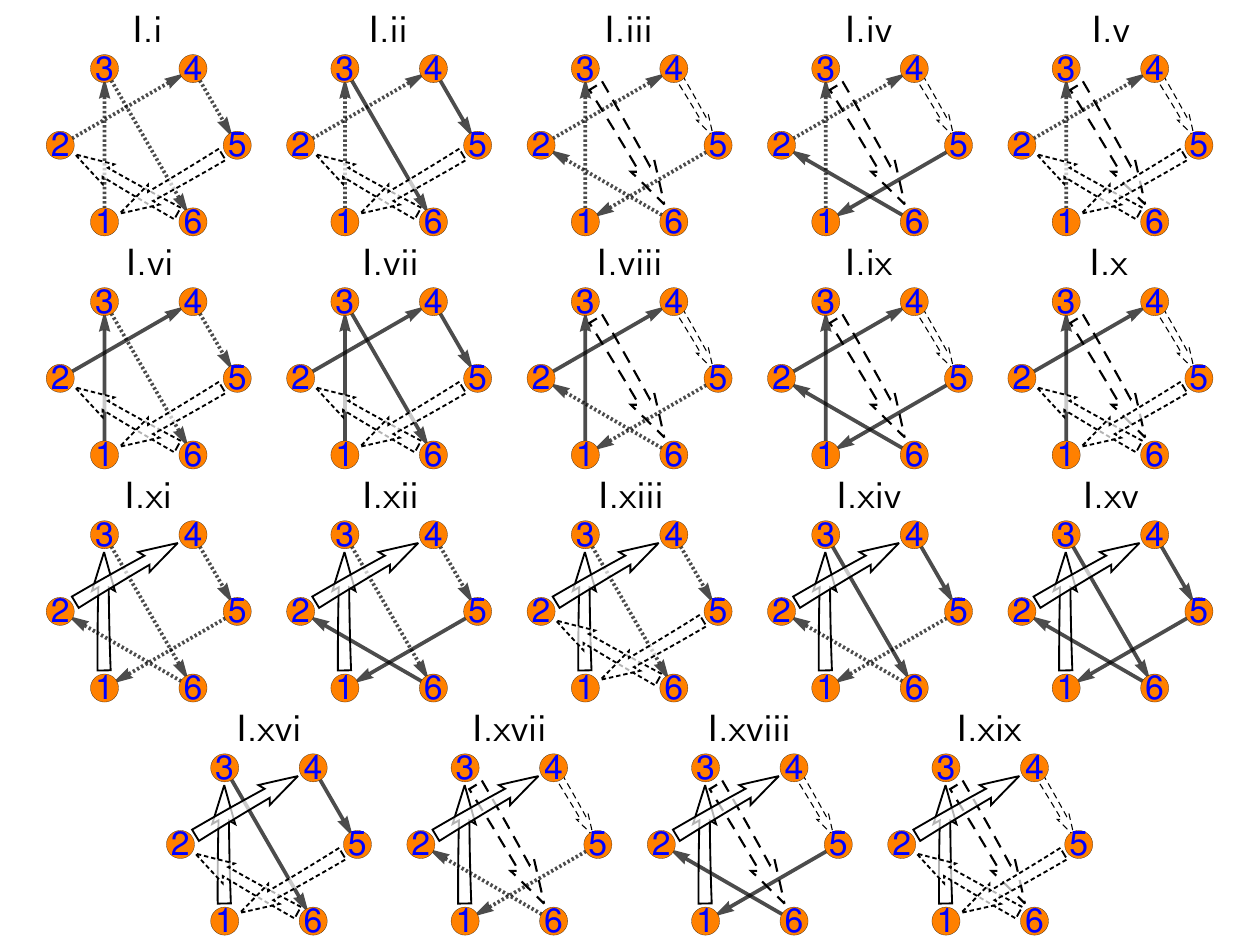} 
\end{center}
\vspace*{-5pt}
\caption{These nineteen types of graphs arise from the first topology with direction shown in Figure~\ref{fig:ElectroproductionStartTopologies}. They correspond to different combinations of selections of type 'A' and 'B' (cf. section~\ref{sec:NewCriterion}) for the three pairs of relative-phases belonging to the shape-classes $(a,e,g)$ (i.e. pairs of relative-phases~$\left\{ \phi_{13}, \phi_{24} \right\}$,~$\left\{ \phi_{36}, \phi_{45} \right\}$ and~$\left\{ \phi_{15}, \phi_{26} \right\}$, respectively), with at least one pair of double-lined arrows, i.e. at least one selection of type 'B'. The solid double-lined arrows mark relative-phases belonging to the shape-class~$a$, dashed double-lined arrows mark relative-phases from the shape-class~$e$ and the dotted double-lined arrows refer to the shape-class~$g$. Graphs resulting solely from selections of type 'A' are not shown in this plot (there exist $8$ such graphs), since they cannot yield fully complete sets according to Theorem~2 from appendix~\ref{sec:ReviewMoravcsik}.}
\label{fig:ElectroproductionNineteenImpliedTopologies}
\end{figure*}

Conversely, one can also derive the~$1216$ complete sets starting solely from considerations of graphs, similar to the steps~'1.)' to~'4.)' described at the end of section~\ref{sec:Photoproduction}. From each of the~$8$ start-topologies shown in Figure~\ref{fig:ElectroproductionStartTopologies}, one can derive~$19$ different types of graphs that contain at least one pair of double-lined arrows. These~$19$ graph-types are plotted for the first start-topology 'I' in Figure~\ref{fig:ElectroproductionNineteenImpliedTopologies}. Then, one has to draw all possible combinations of $\zeta$-sign arrows into the double-lined arrows in the relevant graph-types. For each graph-type with one pair of double-lined arrows, one thus obtains $4$ graphs with $\zeta$-sign arrows, each graph-type with two pairs of double-lined arrows implies~$16$ graphs with $\zeta$-sign arrows and for the one possible graph-type (one per start-topology) which contains three pairs of double-lined arrows, one gets~$64$ possible graphs with $\zeta$-sign arrows. Therefore, for each start-topology, $(12*4+6*16+1*64) = 208$ graphs have to be considered (cf. Figure~\ref{fig:ElectroproductionNineteenImpliedTopologies}). From these~$208$ graphs,~$152$ turn out to fulfill the completeness-criteria\footnote{In more details: each graph-type with one pair of double-lined arrows implies~$2$ complete graphs, each graph-type with two pairs of double-lined arrows implies~$12$ complete graphs and the one graph-type (one per start-topology) with three pairs of double-lined arrows implies~$56$ complete graphs. Thus, one gets~$(12*2 + 6*12 + 1*56) = 152$ complete graphs from each of the~$8$ possible start-topologies (see also Figure~\ref{fig:ElectroproductionNineteenImpliedTopologies}).} posed in Theorem~1 from section~\ref{sec:NewCriterion}. For all start-topologies, this leads to~$(8*152) = 1216$ complete graphs. Again, the combinatorics in the purely graphical approach match exactly the number of complete sets which has been determined by starting from all possible combinations of observables, i.e. in the first approach outlined above.

In the following, we discuss the general structure of the~$1216$ derived complete sets in a bit more detail and also compare them to complete sets for electroproduction already discussed in the literature~\cite{Tiator:2017cde,Wunderlich:2020umg}. Due to the basic structure of the~$8$ start-topologies shown in Figure~\ref{fig:ElectroproductionStartTopologies}, one always obtains a combination of two observables from one of the shape-classes with purely transverse photon-polarization~$\left\{ a,b,c \right\}$ with four observables from two of the shape-classes with mixed transverse-longitudinal photon polarization~$\left\{ e,f,g,h \right\}$ (cf. also Table~\ref{tab:MinimalCompleteExampleSetsElectroproduction}). These observables of course always have to be combined with the~$6$ 'diagonal' observables~$\left\{ R^{00}_{T},\hspace*{1pt}^{c}  R^{00}_{TT}, R_{T}^{0y}, R^{y' 0}_{T}, R_{L}^{00} , R_{L}^{0y}  \right\}$, where the latter~$6$ quantities are composed of both purely transverse and purely longitudinal observables. Furthermore, for each of the shape-class combinations corresponding to one of the~$8$ start-topologies shown in Figure~\ref{fig:ElectroproductionStartTopologies}, there always occurs at least one shape-class that contains only observables with recoil-polarization, i.e. one of the shape-classes~$\left\{ b,c,g,h \right\}$. This means that just as in the case of photoproduction~\cite{Chiang:1996em,Nakayama:2018yzw}, double-polarization observables with recoil-polarization cannot be avoided for a minimal complete set in electroproduction, at least within the context of the search-strategy employed in the present work. This fact has also been pointed out in the work on electroproduction by Tiator and collaborators~\cite{Tiator:2017cde}. Furthermore, the unavoidability of double-polarization observables with recoil-polarization has also turned out to be true for the Moravcsik-complete sets\footnote{We denote complete sets of observables derived using Theorem~2 from appendix~\ref{sec:ReviewMoravcsik} as 'Moravcsik-complete sets', cf. reference~\cite{Wunderlich:2020umg}.} consisting of~$13$ as well as~$14$ observables, which have been derived and listed for electroproduction in reference~\cite{Wunderlich:2020umg}.

Another interesting property of the minimal complete sets given for electroproduction in Table~\ref{tab:MinimalCompleteExampleSetsElectroproduction} as well as the supplemental material~\cite{Supplement} is that they contain no observables from the purely longitudinal shape-class 'AD2', i.e. none of the observables~$\left\{ R_{L}^{z' x}, R_{L}^{x' x} \right\}$. This fact is a consequence of the shapes of the $8$ start-topologies, since the relative-phase~$\phi_{56}$ is never present in any of them. This is very different from the Moravcsik-complete sets with~$13$ observables derived in reference~\cite{Wunderlich:2020umg}, since one observable from the pair~$\left\{ R_{L}^{z' x}, R_{L}^{x' x} \right\}$ is contained in all of them.

Furthermore, it is interesting to analyze the~$1216$ minimal complete sets regarding their recoil-polarization content. For each of the shape-class combinations~$(a,e,g)$, $(a,f,h)$, $(b,e,f)$ and~$(c,e,f)$, or equivalently for each of the four corresponding start-topologies (cf. Figure~\ref{fig:ElectroproductionStartTopologies}), one obtains a complete set that contains exactly two recoil-polarization observables, apart from the observable $R_{T}^{y' 0}$ which is contained in the~$6$ 'diagonal' observables and thus always measured, of course. For each of the two shape-class combinations~$(c,e,g)$ and $(c,f,h)$, one gets a complete set with four recoil-polarization observables (apart from $R_{T}^{y' 0}$) and for each of the combinations~$(b,g,h)$ and $(c,g,h)$, one gets a complete set with the maximal recoil-polarization content of six recoil-polarization observables. The statements made here are reflected in the~$8$ example-sets shown in Table~\ref{tab:MinimalCompleteExampleSetsElectroproduction}.

Further interesting facts arise as soon as we compare the~$1216$ minimal complete sets derived in this work to the~$96$ possible Moravcsik-complete sets composed of~$14$ observables, which have been derived and listed in reference~\cite{Wunderlich:2020umg}. For the Moravcsik-complete sets with~$14$ observables, also some examples exist that contain only $2$ observables with recoil-polarization (apart from $R_{T}^{y' 0}$). Furthermore, interestingly the combinations of shape-classes that lie at the heart of the Moravcsik-complete sets with~$14$ observables are exactly the same combinations as for the minimal complete sets derived in this work (cf. lists in appendix~D of reference~\cite{Wunderlich:2020umg}). In other words, the Moravcsik-complete sets of~$14$ are derived from the exact same graph-topologies shown in Figure~\ref{fig:ElectroproductionStartTopologies}, but using Theorem~2 from appendix~\ref{sec:ReviewMoravcsik} instead of Theorem~1 from section~\ref{sec:NewCriterion}. Furthermore, we suspect that many of the~$1216$ minimal complete sets with $12$ observables, which have been derived in this work, are in fact subsets of the Moravcsik-complete sets of~$14$ from reference~\cite{Wunderlich:2020umg}.
In section~VI from reference~\cite{Wunderlich:2020umg}, this has been illustrated explicitly for the minimal complete set
\begin{equation}
  \left\{ \Ocal^{c}_{1+}, \Ocal^{c}_{2-}, \Ocal^{g}_{2+}, \Ocal^{g}_{2-} , \Ocal^{h}_{2+},   \Ocal^{h}_{2-}   \right\}    , \label{eq:ExampleSetFromMoravcsikPaper}
\end{equation}
which is numerated as the set '$(c,g,h) : 68$' in the lists of the supplemental material~\cite{Supplement}, which result from the calculations performed in the present work. Furthermore, it has been demonstrated explicitly in reference~\cite{Wunderlich:2020umg} how this minimal complete set with~$12$ observables can be deduced from a Moravcsik-complete set of~$14$ via a mathematical reduction-procedure. We assume that similar facts are true for many more cases, but have not checked all cases explicitly in the course of this work.

When comparing to the statements made on minimal complete sets for electroproduction in the work by Tiator and collaborators~\cite{Tiator:2017cde}, we have to state that our findings corroborate their statements. Still, our work complements reference~\cite{Tiator:2017cde} by giving an explicit graphical construction-procedure for minimal complete sets, which has implied an extensive list of~$1216$ such sets~\cite{Supplement}. Such a graphical procedure was not given in reference~\cite{Tiator:2017cde} and neither has been a list of complete sets. However, as a final remark, we would like to re-cite here a useful alternative construction-procedure for complete sets in electroproduction, which has in fact been proposed in reference~\cite{Tiator:2017cde}. This alternative procedure consists of combining a complete photoproduction-set with one full shape-class of electroproduction observables from the four possibilities~$\left\{ e,f,g,h \right\}$. We see that the definitions of the shape-classes 'D1', $a$, $b$ and $c$ are algebraically identical to the definitions of the photoproduction observables (compare Tables~\ref{tab:PhotoObservables} and~\ref{tab:ElectroObservables}). Thus, one can take for instance any of the complete sets derived in section~\ref{sec:Photoproduction} and combine it for example with the full shape-class~$e$. This yields then a set of $(8 + 4) = 12$ observables. The four complex amplitudes~$b_{1}, \ldots, b_{4}$ are determined uniquely up to one overall phase from the complete photoproduction-set. The $2$ moduli and the $2$ relative-phases which are missing from just the $8$ observables in the complete photoproduction-set are then uniquely fixed via the $4$ quantities from the shape-class $e$, as has been already pointed out in reference~\cite{Tiator:2017cde}. Therefore, this method of combining complete photoproduction-sets with a full shape-class from~$\left\{ e,f,g,h \right\}$ represents a powerful and elegant alternative scheme for the derivation of complete sets for electroproduction, which is complementary to the approach followed in the present work. 

This concludes our discussion of the case '{\bf (2+2+2)}' for pseudoscalar meson electroproduction. By this we mean complete sets emerging from selections of three pairs of observables from three different shape-classes, based on the $8$ start-topologies shown in Figure~\ref{fig:ElectroproductionStartTopologies}. Although a quite extensive list of~$1216$ complete sets of minimal length was found this way, we acknowledge that many more such sets exist for electroproduction (cf. statements made in reference~\cite{Tiator:2017cde}). We continue our discussion with comments on a possible generalization of the graphical criterion to problems with~$N > 6$ amplitudes.

%\clearpage

\section{Generalization to problems involving $N > 6$ amplitudes} \label{sec:Generalization}

In order to provide an idea on how the new graphical criterion proposed in this work can be generalized to processes with larger numbers of $N > 6$ amplitudes, we consider here the next more complicated case of two-meson photoproduction, which is generally described by $N = 8$ amplitudes~\cite{Roberts:2004mn,Kroenert:2020ahf}. We do not list here the definitions of all the $N^{2} = 64$ polarization observables for two-meson photoproduction, due to reasons of space. Their definitions, as well as an explanation of their physical meaning in terms of actual measurements, can be found in references~\cite{Roberts:2004mn,Kroenert:2020ahf}. It is clear that a minimal complete set for two-meson photoproduction has to contain at least $2 N = 16$ observables (cf. remarks in the introduction, section~\ref{sec:Intro}, as well as reference~\cite{Kroenert:2020ahf}). \\
The general structure encountered in this case is as follows: the $64$ observables contain one shape-class with $8$ diagonal observables, which are capable of uniquely fixing the moduli $ \left| b_{1} \right|, \ldots, \left| b_{8} \right|$ of the transversity amplitudes. The remaining $56$ observables can be grouped into $7$ distinct shape-classes, containing $8$ observables each, which all have the same repeating mathematical structure. For each of the non-diagonal shape-classes $n = 1, \ldots, 7$, the $8$ observables have the following generic form~\cite{Kroenert:2020ahf}
\begin{widetext}
 \begin{align}
 \Ocal^{n}_{s1} &= \left| b_{i} \right| \left| b_{j} \right| \sin \phi_{ij} + \left| b_{k} \right| \left| b_{l} \right| \sin \phi_{kl} +  \left| b_{m} \right| \left| b_{p} \right| \sin \phi_{mp} + \left| b_{q} \right| \left| b_{r} \right| \sin \phi_{qr}   , \label{eq:TwoMesonShapeClassObsI} \\
 \Ocal^{n}_{s2} &= \left| b_{i} \right| \left| b_{j} \right| \sin \phi_{ij} + \left| b_{k} \right| \left| b_{l} \right| \sin \phi_{kl} -  \left| b_{m} \right| \left| b_{p} \right| \sin \phi_{mp} - \left| b_{q} \right| \left| b_{r} \right| \sin \phi_{qr}   , \label{eq:TwoMesonShapeClassObsII} \\
 \Ocal^{n}_{s3} &=  \left| b_{i} \right| \left| b_{j} \right| \sin \phi_{ij} - \left| b_{k} \right| \left| b_{l} \right| \sin \phi_{kl} +  \left| b_{m} \right| \left| b_{p} \right| \sin \phi_{mp} - \left| b_{q} \right| \left| b_{r} \right| \sin \phi_{qr}  , \label{eq:TwoMesonShapeClassObsIII} \\
 \Ocal^{n}_{s4} &=  \left| b_{i} \right| \left| b_{j} \right| \sin \phi_{ij} - \left| b_{k} \right| \left| b_{l} \right| \sin \phi_{kl} -  \left| b_{m} \right| \left| b_{p} \right| \sin \phi_{mp} + \left| b_{q} \right| \left| b_{r} \right| \sin \phi_{qr}  , \label{eq:TwoMesonShapeClassObsIV} \\
 \Ocal^{n}_{c1} &= \left| b_{i} \right| \left| b_{j} \right| \cos \phi_{ij} + \left| b_{k} \right| \left| b_{l} \right| \cos \phi_{kl} +  \left| b_{m} \right| \left| b_{p} \right| \cos \phi_{mp} + \left| b_{q} \right| \left| b_{r} \right| \cos \phi_{qr}  , \label{eq:TwoMesonShapeClassObsV} \\
 \Ocal^{n}_{c2} &= \left| b_{i} \right| \left| b_{j} \right| \cos \phi_{ij} + \left| b_{k} \right| \left| b_{l} \right| \cos \phi_{kl} -  \left| b_{m} \right| \left| b_{p} \right| \cos \phi_{mp} - \left| b_{q} \right| \left| b_{r} \right| \cos \phi_{qr}    , \label{eq:TwoMesonShapeClassObsVI} \\
 \Ocal^{n}_{c3} &=  \left| b_{i} \right| \left| b_{j} \right| \cos \phi_{ij} - \left| b_{k} \right| \left| b_{l} \right| \cos \phi_{kl} +  \left| b_{m} \right| \left| b_{p} \right| \cos \phi_{mp} - \left| b_{q} \right| \left| b_{r} \right| \cos \phi_{qr}   , \label{eq:TwoMesonShapeClassObsVII} \\
 \Ocal^{n}_{c4} &=  \left| b_{i} \right| \left| b_{j} \right| \cos \phi_{ij} - \left| b_{k} \right| \left| b_{l} \right| \cos \phi_{kl} -  \left| b_{m} \right| \left| b_{p} \right| \cos \phi_{mp} + \left| b_{q} \right| \left| b_{r} \right| \cos \phi_{qr}   , \label{eq:TwoMesonShapeClassObsVIII}
\end{align}
\end{widetext}
where the indices $i,j,k,l,m,p,q,r \in 1,\ldots,8$ have to be all pairwise distinct. The $7$ non-diagonal shape-classes are otherwise only distinguished in the combinations of indices $i,j, \ldots,r$ which appear in the above-given definitions. Every shape-class composed of $8$ observables, which has the structure given above, is in one-to-one correspondence to the particular combination of four relative phases $\left\{ \phi_{ij}, \phi_{kl}, \phi_{mp}, \phi_{qr} \right\}$. We mention the fact that an algebra of $8 \times 8$ $\tilde{\Gamma}^{\alpha}$-matrices is behind the shape-class structure shown in equations~\eqref{eq:TwoMesonShapeClassObsI} to~\eqref{eq:TwoMesonShapeClassObsVIII}, as is described in more detail in reference~\cite{Kroenert:2020ahf}. \\

When confronted with a more elaborate structure such as the one given in equations~\eqref{eq:TwoMesonShapeClassObsI} to~\eqref{eq:TwoMesonShapeClassObsVIII}, one can at first make an attempt to reduce the problem to already known cases (cf. Theorems~1 and~2 in section~\ref{sec:NewCriterion} and appendix~\ref{sec:ReviewMoravcsik}). Such a reduction can be achieved in the following two ways:

\begin{itemize}
 \item[(i)] {\it Full decoupling:} \\
 Considering the definitions of the $8$ observables in the two-meson photoproduction shape-class~\eqref{eq:TwoMesonShapeClassObsI} to~\eqref{eq:TwoMesonShapeClassObsVIII}, one can see quickly that the real- and imaginary parts of the bilinear amplitude products, or equivalently the cosines and sines of the corresponding relative phases, can be isolated by defining certain linear combinations of observables. For instance, the sines of the relative phases $\phi_{ij}, \ldots, \phi_{qr}$ can be isolated via evaluation of the combinations: 
 \begin{align}
  \left| b_{i} \right| \left| b_{j} \right| \sin \phi_{ij} &=  \frac{1}{4} \left( \Ocal^{n}_{s1} + \Ocal^{n}_{s2} + \Ocal^{n}_{s3} + \Ocal^{n}_{s4} \right)  , \label{eq:SinFullDecouplingFirstEq}  \\
  \left| b_{k} \right| \left| b_{l} \right| \sin \phi_{kl} &=  \frac{1}{4} \left( \Ocal^{n}_{s1} + \Ocal^{n}_{s2} - \Ocal^{n}_{s3} - \Ocal^{n}_{s4} \right)  , \label{eq:SinFullDecouplingSecondEq}  \\
  \left| b_{m} \right| \left| b_{p} \right| \sin \phi_{mp} &=  \frac{1}{4} \left( \Ocal^{n}_{s1} - \Ocal^{n}_{s2} + \Ocal^{n}_{s3} - \Ocal^{n}_{s4} \right)  , \label{eq:SinFullDecouplingThirdEq}  \\
  \left| b_{q} \right| \left| b_{r} \right| \sin \phi_{qr} &=  \frac{1}{4} \left( \Ocal^{n}_{s1} - \Ocal^{n}_{s2} - \Ocal^{n}_{s3} + \Ocal^{n}_{s4} \right)  . \label{eq:SinFullDecouplingFourthEq} 
 \end{align}
 In exactly the same way, one can isolate the cosines of the relative-phases, by defining analogous linear-combinations for the observables $\Ocal^{n}_{c1}, \ldots, \Ocal^{n}_{c4}$. We say that the bilinear combinations appearing in the original shape-class have been {\it fully decoupled}. In other words, one has reduced the ambiguity-problem provided by the bilinear-forms defined in terms of $8 \times 8$ $\tilde{\Gamma}^{\alpha}$-matrices~\cite{Kroenert:2020ahf} for two-meson photoproduction to the ambiguities implied by certain $2$-dimensional sub-algebras of said $\tilde{\Gamma}^{\alpha}$-matrices. \\
 It is now possible to apply Moravcsik's theorem in the modified form (Theorem~2 in appendix~\ref{sec:ReviewMoravcsik}) to the fully decoupled shape-class. This has been done in reference~\cite{Kroenert:2020ahf}, where complete sets containing at least $24$ observables were found using this method.
 \item[(ii)] {\it Partial decoupling:} \\
 Instead of trying to isolate the real- and imaginary parts of bilinear products alone, one can try to only isolate sub shape-classes with four elements, of the same structure as the one given in equations~\eqref{eq:NonTrivialShapeClassObsI} to~\eqref{eq:NonTrivialShapeClassObsIV} of section~\ref{sec:NewCriterion}, which are contained in the $8$ observables given above (i.e. in equations~\eqref{eq:TwoMesonShapeClassObsI} to~\eqref{eq:TwoMesonShapeClassObsVIII}). For instance, we can find such a sub-class, indicated by the symbol $\OcalTilde^{n,a}$, i.e. with an additional '$a$' in the super-script, via the following linear combinations of pairs of observables, which contain only the two relative phases $\phi_{ij}$ and $\phi_{kl}$ (cf. reference~\cite{Kroenert:2020ahf}):
\begin{align}
 \OcalTilde^{n,a}_{1+} &= \frac{1}{2} \left( \Ocal^{n}_{s1} + \Ocal^{n}_{s2} \right)    \nonumber \\
 &=  \left| b_{i} \right| \left| b_{j} \right| \sin \phi_{ij} + \left| b_{k} \right| \left| b_{l} \right| \sin \phi_{kl} , \label{eq:PartiallyDecoupledClassAEquationI}  \\
 \OcalTilde^{n,a}_{1-} &= \frac{1}{2} \left( \Ocal^{n}_{s3} + \Ocal^{n}_{s4} \right) \nonumber \\
 &=  \left| b_{i} \right| \left| b_{j} \right| \sin \phi_{ij} - \left| b_{k} \right| \left| b_{l} \right| \sin \phi_{kl} , \label{eq:PartiallyDecoupledClassAEquationII}  \\
 \OcalTilde^{n,a}_{2+} &= \frac{1}{2} \left( \Ocal^{n}_{c1} + \Ocal^{n}_{c2} \right) \nonumber \\
 &=  \left| b_{i} \right| \left| b_{j} \right| \cos \phi_{ij} + \left| b_{k} \right| \left| b_{l} \right| \cos \phi_{kl} , \label{eq:PartiallyDecoupledClassAEquationIII}  \\
 \OcalTilde^{n,a}_{2-} &= \frac{1}{2} \left( \Ocal^{n}_{c3} + \Ocal^{n}_{c4} \right) \nonumber \\
 &=  \left| b_{i} \right| \left| b_{j} \right| \cos \phi_{ij} - \left| b_{k} \right| \left| b_{l} \right| \cos \phi_{kl} . \label{eq:PartiallyDecoupledClassAEquationIV} 
\end{align}
In the same way, one can isolate a sub-class $\OcalTilde^{n,b}$ for the remaining two relative-phases $\phi_{mp}$ and $\phi_{qr}$:
\begin{align}
 \OcalTilde^{n,b}_{1+} &= \frac{1}{2} \left( \Ocal^{n}_{s1} - \Ocal^{n}_{s2} \right) \nonumber \\
 &=  \left| b_{m} \right| \left| b_{p} \right| \sin \phi_{mp} + \left| b_{q} \right| \left| b_{r} \right| \sin \phi_{qr} , \label{eq:PartiallyDecoupledClassBEquationI}  \\
 \OcalTilde^{n,b}_{1-} &= \frac{1}{2} \left( \Ocal^{n}_{s3} - \Ocal^{n}_{s4} \right) \nonumber \\
 &=  \left| b_{m} \right| \left| b_{p} \right| \sin \phi_{mp} - \left| b_{q} \right| \left| b_{r} \right| \sin \phi_{qr} , \label{eq:PartiallyDecoupledClassBEquationII}  \\
 \OcalTilde^{n,b}_{2+} &= \frac{1}{2} \left( \Ocal^{n}_{c1} - \Ocal^{n}_{c2} \right) \nonumber \\
 &=  \left| b_{m} \right| \left| b_{p} \right| \cos \phi_{mp} + \left| b_{q} \right| \left| b_{r} \right| \cos \phi_{qr} , \label{eq:PartiallyDecoupledClassBEquationIII}  \\
 \OcalTilde^{n,b}_{2-} &= \frac{1}{2} \left( \Ocal^{n}_{c3} - \Ocal^{n}_{c4} \right) \nonumber \\
 &=  \left| b_{m} \right| \left| b_{p} \right| \cos \phi_{mp} - \left| b_{q} \right| \left| b_{r} \right| \cos \phi_{qr} . \label{eq:PartiallyDecoupledClassBEquationIV} 
\end{align}
This pair of disjoint sub-classes is not the only such pair which can be formed from the $8$ observables~\eqref{eq:TwoMesonShapeClassObsI} to~\eqref{eq:TwoMesonShapeClassObsVIII}. One can actually split the full shape-class into three possible pairs of disjoint sub-classes, for which the above-given two ('$n,a$' and '$n,b$') are just the first example. The second possibility would be given by disjoint shape-classes '$n,c$' and '$n,d$' with associated pairs of relative-phases $\left\{ \phi_{ij}, \phi_{mp} \right\}$ and $\left\{ \phi_{kl}, \phi_{qr} \right\}$, respectively. The precise definitions of the classes '$n,c$' and '$n,d$' then proceed analogously to the equations~\eqref{eq:PartiallyDecoupledClassAEquationI} to~\eqref{eq:PartiallyDecoupledClassBEquationIV}. Furthermore, one can also form a third combination of disjoint shape-classes, we call them '$n,e$' and '$n,f$', which correspond to the pairs of relative phases $\left\{ \phi_{ij}, \phi_{qr} \right\}$ and $\left\{ \phi_{kl}, \phi_{mp} \right\}$, respectively. This exhausts all the possibilities to achieve a partial decoupling of the full shape-class shown in equations~\eqref{eq:TwoMesonShapeClassObsI} to~\eqref{eq:TwoMesonShapeClassObsVIII}. We see that some freedom on how to achieve this decoupling indeed exists. \\
To the partially decoupled shape-classes with four elements, such as those described above, our new graphical criterion (Theorem~$1$ from section~\ref{sec:NewCriterion}) can be directly applied. Thus, one would search for {\bf (2+2+2+2)}-combinations selected from four of the partially decoupled shape-classes, i.e. four shape-classes in the $\OcalTilde$-basis. However, due to basic topological reasons, it is only possible to derive complete sets with at least $20$ observables in this way, when considering observables in the $\Ocal$-basis. This is true due to the fact that at least three shape-classes in the $\Ocal$-basis have to be combined in order to get a connected graph when combining the indices from all their relative-phases~\cite{Kroenert:2020ahf}. This means that the above-mentioned {\bf (2+2+2+2)}-combinations in the $\OcalTilde$-basis have to correspond to at least~$12$ observables in the $\Ocal$-basis. Together with the~$8$ 'diagonal' observables for two-meson photoproduction, this implies~$20$ observables in total. \\
Lastly, we mention the fact that the observables in the partially decoupled shape-classes (i.e. in the $\OcalTilde$-basis) can be used for an explicit algebraic derivation of minimal complete sets of~$16$ observables in the $\Ocal$-basis, as has been done in reference~\cite{Kroenert:2020ahf}. However, in this derivation, a selection-pattern was used which does not correspond to Theorem~1 from section~\ref{sec:NewCriterion}.
\end{itemize}

The approaches (i) and (ii) described above clearly exhaust all the possibilities of directly applying the criteria stated in section~\ref{sec:NewCriterion} and appendix~\ref{sec:ReviewMoravcsik} to the problem of two-meson photoproduction. In case one wishes to make statements about more general combinations of observables, one has no other choice but to execute a new dedicated derivation of the general structure of the phase-ambiguities allowed by the definitions~\eqref{eq:TwoMesonShapeClassObsI} to~\eqref{eq:TwoMesonShapeClassObsVIII}.

\begin{itemize}
 \item[(iii)] {\it Full derivation of new phase-ambiguity structure:} \\
 A full dedicated derivation of the phase-ambiguities implied by more general selections from a combination of shape-classes such as the one defined in equations~\eqref{eq:TwoMesonShapeClassObsI} to~\eqref{eq:TwoMesonShapeClassObsVIII} has not been performed in the course of this work. Therefore, in the following we can only speculate how such a derivation might look like. \\
 The most nearby option would be to try patterns of {\bf (2+2+2+2)}-combinations in the $\Ocal$-basis, i.e. selections of $8$ non-diagonal observables with $2$ observables picked from each individual shape-class. These selections do not fit the patterns outlined in point (ii) above. Therefore, some algebraic ingenuity is needed in the derivation and it is likely that trying to exactly replicate the derivation shown in detail in appendix~\ref{sec:NakayamaDerivation} might not be enough. Since the derivations from appendix~\ref{sec:NakayamaDerivation} are already quite involved, we only expect the new calculations needed for the larger shape-classes to be more complicated. However, once the derivation of the discrete phase-ambiguities implied by, for instance, the {\bf (2+2+2+2)}-combinations mentioned above has been completed, we expect that again a graphical criterion similar to the one formulated in Theorem~1 from section~\ref{sec:NewCriterion} can be used in order to derive complete sets. These complete sets then have the minimal length of $2 N = 16$ observables.
\end{itemize}

The discussion in this section has illustrated the problems which are encountered when trying to generalize the criterion posed in Theorem~1 from section~\ref{sec:NewCriterion} to more involved problems with~$N > 6$ amplitudes. Theorem~1 has been able to directly yield minimal complete sets of length~$2N$ for photoproduction and electroproduction, i.e. for $N \leq 6$. Therefore, in these cases it has outperformed the modified form of Moravcsik's theorem (Theorem~2 from appendix~\ref{sec:ReviewMoravcsik}). However, the price one has to pay for this achievement is that a simple selection of minimal complete sets for $N > 6$ is not so easily possible. Instead, the advantage of deriving minimal complete sets is paid with new and quite involved algebraic derivations, which have to be performed for more complicated amplitude-extraction problems, with larger and more involved shape-classes.

\vspace*{5pt}

\section{Conclusions and Outlook} \label{sec:ConclusionsAndOutlook}

We have introduced a generalization of the graphs originally introduced by Moravcsik~\cite{Moravcsik:1984uf}, which has lead to a new graphical criterion that allows for the determination of minimal complete sets of length $2 N$, for an amplitude-extraction problem with $N \leq 6$ complex amplitudes, where furthermore~$N$ has to be {\it even}. The new method rests heavily on the known discrete phase-ambiguities implied by the selection of any pair of observables from a non-diagonal shape-class composed of four quantities. In order to achieve the selection of minimal complete sets for $N \leq 6$, the considered graphs must be provided with additional directional information. Our new criterion has been applied to single-meson photoproduction ($N = 4$ amplitudes) as well as electroproduction ($N = 6$ amplitudes), with success. In particular, we were able to determine for the first time an extensive list of complete sets with minimal length $2 N = 12$ for the case of electroproduction.

However, the generalization of our new criterion to problems involving a larger number of $N > 6$ amplitudes is difficult, due to the fact that one has to perform new and more involved algebraic derivations for the ambiguity structure implied by the then appearing larger shape-classes, i.e. with more than four observables in each class. It is possible to decouple such a problem, in order to reduce it to the already known case of the above-mentioned shape-classes of four. However, the complete sets of observables determined in this way are generally not of minimal length $2N$ any more. \\ Once the algebraic derivation of the full ambiguity structure allowed by such larger shape-classes has been performed successfully, the selection of complete sets according to a graphical criterion similar to the one proposed in this work is straightforward. The derivation of the ambiguity structure is actually the only significant hurdle in the treatment of the more complicated problems. Once this hurdle is taken, graphical criteria can be applied with full effect. 

This work can be extended into multiple directions. The obvious first choice would be to work out the new graphical criteria for more complicated reactions with $N > 6$ amplitudes, such as two-meson photoproduction ($N = 8$) or even vector-meson photoproduction ($N = 12$), and to try to extract minimal complete sets with length~$2N$ for these reactions. For this, some algebraic ingenuity is needed to treat the larger shape-classes. Another possible direction of research would be to try to establish a closer contact to mathematicians, in order to see whether they have deeper insights into the discussed matters. When Moravcsik published his paper in 1985~\cite{Moravcsik:1984uf}, he called the mathematical theory for the ambiguities of a set of bilinear algebraic equations for several unknowns to be 'nonexistent'~(see the second-last paragraph in the introduction of reference~\cite{Moravcsik:1984uf}). However, this may have changed in the most recent years and it could be that mathematicians use the objects and approaches introduced in this work, but in different guises. Nevertheless, we have to admit that also the present work seems to lead to a kind of more general theory for the unique solvability of systems of bilinear equations.

\begin{acknowledgments}
   The work of Y.W. was supported by the {\it Transdisciplinary Research Area - Building Blocks of Matter and Fundamental Interactions (TRA Matter)} during the completion of this manuscript.
   The author would like to thank Philipp Kroenert for useful comments on the manuscript and for helpful remarks regarding the software needed for the creation of the presented figures. %In addition, he wishes to thank~$\ldots$ for a careful reading of the manuscript.
\end{acknowledgments}

\appendix

\section{Review of Moravcsik's theorem} \label{sec:ReviewMoravcsik}

This appendix provides a review of Moravcsik's theorem~\cite{Moravcsik:1984uf} in a slightly modified form, which resulted from a recent reexamination~\cite{Wunderlich:2020umg}. The review is included both to keep the present work self-contained and also because Moravcsik's theorem serves as a useful reference point to the new graphical criterion, which is developed in section~\ref{sec:NewCriterion} of the main text.

Consider an amplitude-extraction problem formulated in terms of $N$ complex transversity-amplitudes\footnote{One can also formulate all statements made in this work for helicity- instead of transversity amplitudes. However, in the transversity-basis, the observables which can uniquely fix amplitude-moduli are generally most easily measured~\cite{Chiang:1996em,Tiator:2017cde,Nakayama:2018yzw} (cf. discussions on photoproduction~\ref{sec:Photoproduction} and electroproduction~\ref{sec:Electroproduction}). Therefore, we have decided to choose the latter basis.} $b_{1},\ldots,b_{N}$. For such a problem, one can consider the $N^{2}$ bilinear amplitude-products
\begin{equation}
 b_{j}^{\ast} b_{i}  , \text{ for } i,j=1,\ldots,N. \label{eq:GenericBilinearProducts}
\end{equation}
Due to the bilinear structure of the products~\eqref{eq:GenericBilinearProducts}, as well as the fact that polarization observables are most generally linear combinations of such products, the amplitudes generally can only be determined up to one unknown overall phase~\cite{Chiang:1996em, MyDiploma, MyPhD, Nakayama:2018yzw}, which can depend on all kinematic variables describing the considered process. Therefore, the maximal amount of information which can be extracted is contained in the moduli and relative-phases of the $N$ amplitudes.

An important initial standard-assumption is that all the $N$ moduli
\begin{equation}
  \left| b_{1} \right|,  \left| b_{2} \right|, \ldots,  \left| b_{N} \right|  , \label{eq:Moduli}
\end{equation}
have already been determined from a suitable subset composed of $N$ 'diagonal' ob\-serva\-bles. This assumption makes the algebraic analysis of complete experiments easier (cf. \cite{Moravcsik:1984uf, Chiang:1996em, Nakayama:2018yzw, Wunderlich:2020umg}) and therefore we shall always adopt it in this paper.

When introducing polar coordinates (i.e. modulus and phase) for each amplitude, the real parts of the generally complex bilinear products become
\begin{equation}
 %\mathrm{Re} \left[ b_{j}^{\ast} b_{i} \right] = \left| b_{i} \right|\left| b_{j} \right|  \mathrm{Re} \left[ e^{i \phi_{ij}} \right] = \left| b_{i} \right|\left| b_{j} \right|  \cos \phi_{ij}  . \label{eq:ReBicomEq}
 \mathrm{Re} \left[ b_{j}^{\ast} b_{i} \right] = \left| b_{i} \right|\left| b_{j} \right|  \cos \phi_{ij}  . \label{eq:ReBicomEq}
\end{equation}
The real parts thus fix their corresponding relative-phase $\phi_{ij} := \phi_{i} - \phi_{j}$ up to the discrete ambiguity~\cite{Moravcsik:1984uf,Nakayama:2018yzw}
\begin{equation}
 \phi_{ij}^{\lambda} = \phi_{ij}^{\pm} = \begin{cases}  + \alpha_{ij}, \\ - \alpha_{ij} ,   \end{cases}  \label{eq:CosTypeAmbiguity}
\end{equation}
where $\alpha_{ij}$ can be extracted uniquely from the value of $\mathrm{Re} \left[ b_{j}^{\ast} b_{i} \right]$, and on the interval $\alpha_{ij} \in \left[ 0, \pi \right]$. We call a discrete ambiguity of the form~\eqref{eq:CosTypeAmbiguity} a 'cosine-type' ambiguity~\cite{Wunderlich:2020umg}.

The imaginary part of a bilinear product is written as
\begin{equation}
 %\mathrm{Im} \left[ b_{j}^{\ast} b_{i} \right] = \left| b_{i} \right|\left| b_{j} \right|  \mathrm{Im} \left[ e^{i \phi_{ij}} \right] = \left| b_{i} \right|\left| b_{j} \right|  \sin \phi_{ij}  , \label{eq:ImBicomEq}
 \mathrm{Im} \left[ b_{j}^{\ast} b_{i} \right] = \left| b_{i} \right|\left| b_{j} \right|  \sin \phi_{ij}  , \label{eq:ImBicomEq}
\end{equation}
and it yields the corresponding relative-phase $\phi_{ij}$ up to the discrete phase-ambiguity~\cite{Moravcsik:1984uf,Nakayama:2018yzw}
\begin{equation}
 \phi_{ij}^{\lambda} = \phi_{ij}^{\pm} = \begin{cases}  + \alpha_{ij}, \\ \pi - \alpha_{ij} ,   \end{cases} , \label{eq:SinTypeAmbiguity}
\end{equation}
where $\alpha_{ij}$ can be extracted uniquely from the quantity $\mathrm{Im} \left[ b_{j}^{\ast} b_{i} \right]$, and on the interval $\alpha_{ij} \in \left[ - \pi / 2, \pi / 2 \right]$. We refer to a discrete ambiguity of the form~\eqref{eq:SinTypeAmbiguity} as a 'sine-type' ambiguity~\cite{Wunderlich:2020umg}.

The original theorem by Moravcsik is formulated as a 'geometrical analog'~\cite{Moravcsik:1984uf}. In this analog, every amplitude is represented by a point and each bilinear amplitude-product is thus given as a line connecting the points that correspond to the amplitudes among which the product is taken (cf. equation~\eqref{eq:GenericBilinearProducts}). Furthermore, a solid line represents the real part $\mathrm{Re} \left[ b_{j}^{\ast} b_{i} \right]$ and a broken (dashed) line denotes the imaginary part $\mathrm{Im} \left[ b_{j}^{\ast} b_{i} \right]$.

Moravcsik's theorem has been recently reexamined~\cite{Wunderlich:2020umg}, which has lead to a slight modification of the original statement. The modified form of Moravcsik's theorem has then been applied to single-meson photo- and electroproduction~\cite{Wunderlich:2020umg}, and also very recently to two-meson photoproduction~\cite{Kroenert:2020ahf}. It reads~\cite{Wunderlich:2020umg}: \\

\textbf{\underline{Theorem 2 (Modified form of Moravcsik's}} \\ \hspace*{69.5pt}  \textbf{\underline{theorem):}} \\

Consider the following 'most economical'~\cite{Moravcsik:1984uf} situation in the geometrical analog: a large open chain which contains \textit{all} amplitude points, and thus consists of $N - 1$ lines for a problem with $N$ amplitudes. This open chain is now turned into a \textit{fully} complete set, by adding one additional connecting line which turns it into a closed loop of $N$ lines, which has to contain all amplitude points exactly once. Furthermore, in such a closed loop every amplitude point is touched by exactly $2$ link-lines. In other words, we have generated a fully connected graph with $N$ vertices and $N$ edges, where all vertices have to have exactly the order $2$.

%Then, for the created loop to represent a fully complete set of bilinear products, i.e. one which does \textit{not} allow any residual discrete phase-ambiguities, we find the following single criterion, which is a bit different and seemingly simpler than in Moravcsik's case:
Such a connected graph (or closed loop) corresponds to a unique solution for the amplitude-extraction problem, which does not permit any residual discrete ambiguities, in case it satisfies the following criterion:
\begin{itemize}
 \item[(C2)] The connected graph has to contain an odd number of dashed lines $n_{\text{d}} \geq 1$. \\
 In particular, the graph does not have to contain any solid lines at all. For an odd number of links $N$, the connected graph with $n_{\text{d}} = N$ therefore still represents a fully complete set. \\
 It is irrelevant which of the bilinear amplitude-products are represented by the dashed lines, as long as the overall number of dashed lines is odd.
\end{itemize}

\begin{figure*}
 \begin{center}
\includegraphics[width = 0.31 \textwidth,trim={0 0 0 0.7cm},clip]{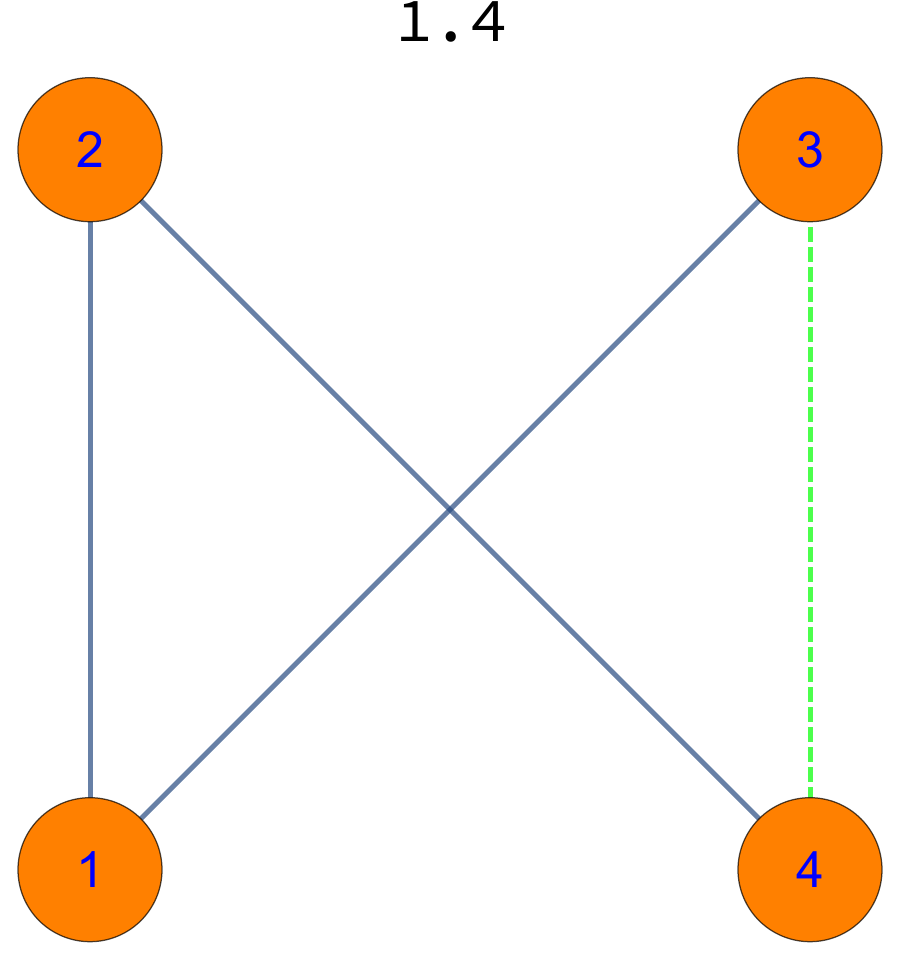} \hspace*{5pt}
\includegraphics[width = 0.31 \textwidth,trim={0 0 0 0.7cm},clip]{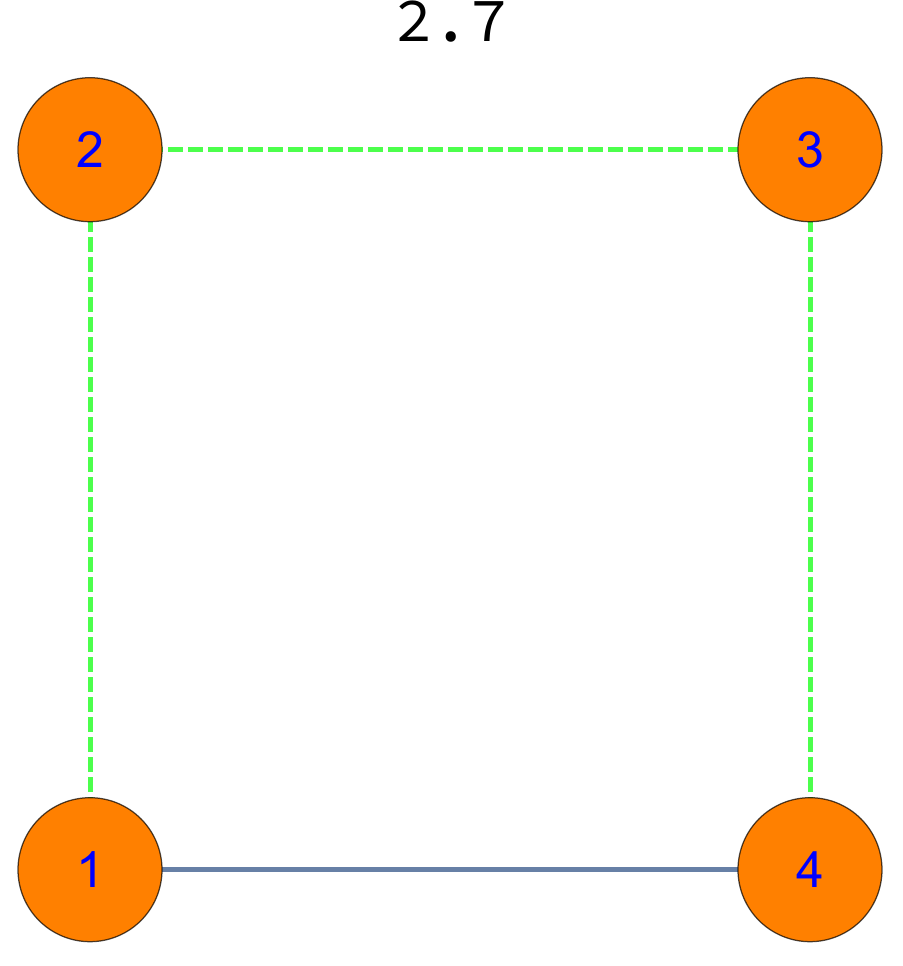} \hspace*{5pt}
\includegraphics[width = 0.31 \textwidth,trim={0 0 0 0.7cm},clip]{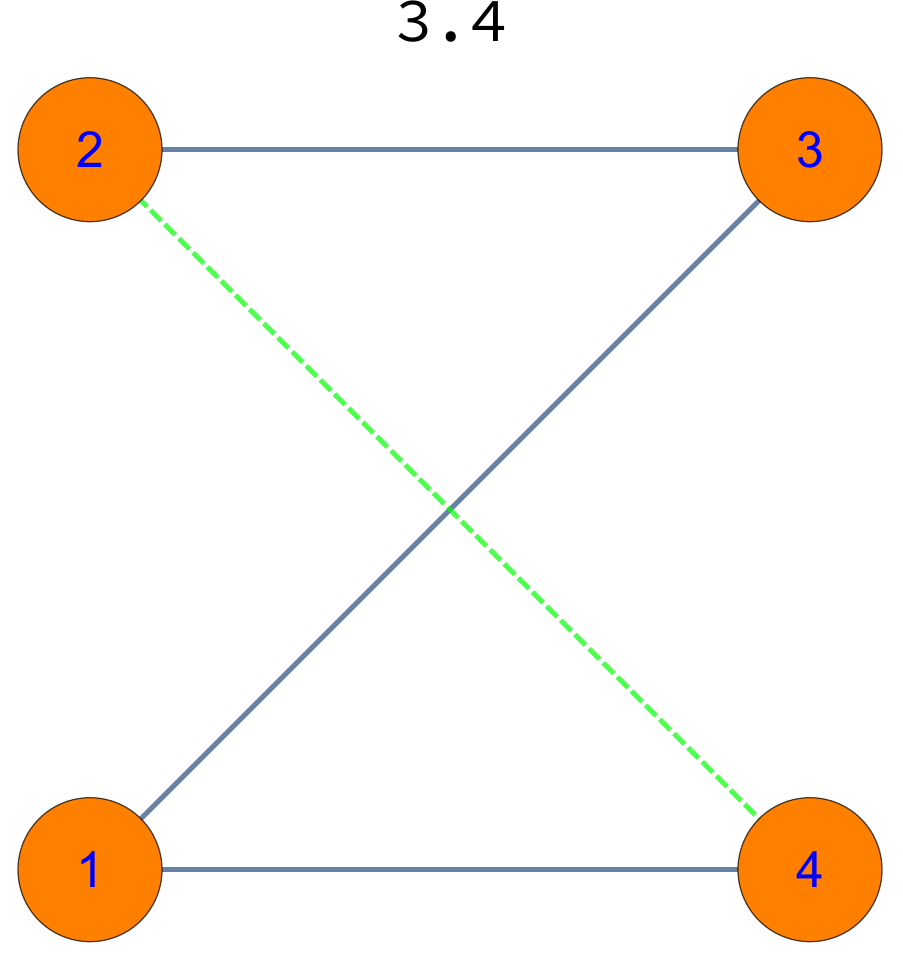}
\end{center}
\vspace*{-5pt}
\caption{The diagrams show three connected graphs which meet all the criteria posed by Theorem~2. These graphs with $4$ vertices correspond to the well-known example-case of pseudoscalar meson photoproduction, which is a problem with $N = 4$ amplitudes. Green dashed lines denote the imaginary part of a bilinear amplitude-product, while the real part of such a product is represented by a blue solid line. These graphs have been taken over identically from reference~\cite{Wunderlich:2020umg}. (Color online)}
\label{fig:PhotoproductionExampleLoops}
\end{figure*}

A detailed proof of this theorem can be found in appendix~A of reference~\cite{Wunderlich:2020umg}. As an illustration of the somewhat abstract criterion formulated in Theorem~2, we show three fully complete example-graphs for pseudoscalar meson photoproduction ($N = 4$ amplitudes) in Figure~\ref{fig:PhotoproductionExampleLoops}. In the following, we comment on some features of the theorem which are important for the present work.

The requirement of a connected topology for the graph considered in Theorem~2 is crucial, due to the fact that it directly forbids combinations of relative-phases corresponding to multiple disconnected sub-sets of amplitudes in the complex plane, where the relative-phases among all amplitudes within one sub-set are uniquely fixed, but the multiple sub-sets are still allowed to rotate freely relative to each other. The latter case occurs when at least one relative-phase connecting at least two such sub-sets is missing, and it leads to so-called {\it continuous ambiguities}. The connectedness-criterion for the graph directly removes such continuous ambiguities. 

Furthermore, in case the connectedness-criterion is fulfilled, one can establish a so-called {\it consistency relation}~\cite{Nakayama:2018yzw, Wunderlich:2020umg} among all the occurring relative-phases. The generic form as well as the importance of such consistency relations is elaborated in more detail in the beginning of section~\ref{sec:NewCriterion} of the main text.

We note that there can exist certain singular surfaces in the parameter space composed of the relative-phases, on which Theorem~2 no longer holds. Such singular configurations have been mentioned in section~III and appendix~A of reference~\cite{Wunderlich:2020umg}. However, we will disregard such cases in the following discussion, since the measure of such singular surfaces is always negligible, when compared to the full parameter space.

The modified form of Moravcsik's theorem (i.e. Theorem~2) has turned out to be very useful, since it leads to a fully automated procedure for the construction of complete sets for in principle {\it any} amplitude-extraction problem with a general number of $N$ complex amplitudes. However, when applied to realistic processes with $N \geq 4$ amplitudes, this theorem has lead to sets which are slightly over-complete when compared to complete sets of minimal length $2 N$ (see in particular section~VII of reference~\cite{Wunderlich:2020umg}, as well as the discussion in reference~\cite{Kroenert:2020ahf}). This happens essentially due to fact that isolated real- and imaginary parts of bilinear amplitude-products (equations~\eqref{eq:ReBicomEq} and~\eqref{eq:ImBicomEq}) enter the statement of Theorem~2, while the observables encountered in processes with $N \geq 4$ amplitudes are generally linear combinations of such bilinear products. The new graphical criterion, which is developed in section~\ref{sec:NewCriterion} of the main text, represents an attempt to improve this situation by allowing for a direct selection of fully complete sets with minimal length $2 N$.

\section{Derivation of the phase-ambiguities for the simplest non-trivial shape-class (eqs~\eqref{eq:NonTrivialShapeClassObsI} -~\eqref{eq:NonTrivialShapeClassObsIV})} \label{sec:NakayamaDerivation}

In the following, we repeat the derivations from section~III of Nakayama's work~\cite{Nakayama:2018yzw}. First, we consider the example-case 'B.1' (selection of~$\left( \Ocal^{n}_{1+}, \Ocal^{n}_{2+} \right)$) mentioned in section~\ref{sec:NewCriterion}. We define $B_{ij} := \left| b_{i} \right| \left| b_{j} \right|$ and consider the following pair of observables:
%
%\begin{widetext}
\begin{align}
 \Ocal^{n}_{1+} &= B_{ij} \sin \phi_{ij} + B_{kl} \sin \phi_{kl}  , \label{eq:ExampleDerivationObsI} \\
 \Ocal^{n}_{2+}  &=  B_{ij} \cos \phi_{ij} + B_{kl} \cos \phi_{kl}  . \label{eq:ExampleDerivationObsII} 
\end{align}
%\end{widetext}
%
When combining both of these definitions, a basic addition-theorem for the cosine leads to the following expression:
\begin{equation}
 \left( \Ocal^{n}_{1+} \right)^{2} + \left( \Ocal^{n}_{2+} \right)^{2} = B_{ij}^{2} + B_{kl}^{2} + 2 B_{ij} B_{kl} \cos \left( \phi_{ij} - \phi_{kl} \right)  .  \label{eq:ExampleDerivationStepI}
\end{equation}
We define the length 
\begin{equation}
N \equiv N^{n}_{1+,2+} := \sqrt{\left( \Ocal^{n}_{1+} \right)^{2} + \left( \Ocal^{n}_{2+} \right)^{2}}.  \label{eq:DefLengthN} 
\end{equation}
Then, we formally introduce the transitional angle $\zeta \equiv \zeta^{n}_{1+,2+}$ via:
\begin{equation}
   \cos \zeta \equiv \frac{\Ocal^{n}_{1+}}{N} \text{, } \sin \zeta \equiv \frac{\Ocal^{n}_{2+}}{N}   . \label{eq:ZetaAngleFormalDefinition}
\end{equation}
Compare this definition to the comments made below equation~\eqref{eq:TwoFoldPhaseAmbiguityII1}, as well as to the graphical representation shown in Figure~\ref{fig:ZetaAngleDefinition}, in section~\ref{sec:NewCriterion}. One should always keep in mind that the general angle $\zeta^{n}_{\nu \pm, \nu' \pm}$ and the length $N^{n}_{\nu \pm, \nu' \pm}$ both depend on the considered pair of observables.

With these definitions, the observables given above can be re-expressed as
%
%\begin{widetext}
\begin{align}
 N \cos \zeta &= B_{ij} \sin \phi_{ij} + B_{kl} \sin \phi_{kl}  , \label{eq:ExampleDerivationObsReExpressedI} \\
 N \sin \zeta  &=  B_{ij} \cos \phi_{ij} + B_{kl} \cos \phi_{kl}  . \label{eq:ExampleDerivationObsReExpressedII} 
\end{align}
%\end{widetext}
%
Next, we multiply the first of these equations by $\sin \phi_{kl}$ and the second one by $\cos \phi_{kl}$ and then add both equations, which leads to:
\begin{equation}
\cos \left( \phi_{ij} - \phi_{kl} \right)  = \frac{- B_{kl} + N \sin \left( \zeta + \phi_{kl} \right)}{B_{ij}}  .  \label{eq:ExampleDerivationStepII}
\end{equation}
Inserting this result in equation~\eqref{eq:ExampleDerivationStepI}, one obtains:
\begin{equation}
  \sin \left( \zeta + \phi_{kl} \right) = \frac{N^{2} - B_{ij}^{2} + B_{kl}^{2}}{2 N B_{kl}}  . \label{eq:ExampleDerivationStepIII}
\end{equation}
Applying the $\arcsin$-function to this equation, one can derive the phase-ambiguity
\begin{equation}
  \phi_{kl} = \begin{cases} - \zeta + \alpha_{kl}  , \\ - \zeta - \alpha_{kl} + \pi ,  \end{cases}    \label{eq:PhaseAmbExampleDerivationI}
\end{equation}
where the right-hand-side of equation~\eqref{eq:ExampleDerivationStepIII} uniquely fixes $\alpha_{kl}$ on the interval $[- \pi / 2, \pi / 2]$.

In exactly the same way that has lead to equation~\eqref{eq:ExampleDerivationStepIII}, one can prove the following constraint:
\begin{equation}
  \sin \left( \zeta + \phi_{ij} \right) = \frac{ B_{ij}^{2} - B_{kl}^{2} + N^{2}}{2 N B_{ij}}  . \label{eq:ExampleDerivationStepIV}
\end{equation}
This constraint leads to the following discrete phase-ambiguity
\begin{equation}
  \phi_{ij} = \begin{cases} - \zeta + \alpha_{ij}  , \\ - \zeta - \alpha_{ij} + \pi ,  \end{cases}    \label{eq:PhaseAmbExampleDerivationII}
\end{equation}
where the right-hand-side of equation~\eqref{eq:ExampleDerivationStepIV} uniquely specifies $\alpha_{ij}$ on the interval $[- \pi / 2, \pi / 2]$.

Up to now, the discrete phase-ambiguity given by equations~\eqref{eq:PhaseAmbExampleDerivationI} and~\eqref{eq:PhaseAmbExampleDerivationII} looks like a four-fold one, but this is not true due to the fact that the relative-phases $\phi_{ij}$ and $\phi_{kl}$ are {\it not} independent. Rather, equation~\eqref{eq:ExampleDerivationStepI} fixes a constraint for $\cos \left( \phi_{ij} - \phi_{kl} \right)$. In other words, one has $\pm \tilde{\alpha} \equiv \phi_{ij} - \phi_{kl}$, where $\tilde{\alpha}$ is uniquely specified on the interval $[0,\pi]$. The ambiguities~\eqref{eq:PhaseAmbExampleDerivationI} and~\eqref{eq:PhaseAmbExampleDerivationII} leave the following four possibilities for the difference of the relative phases:
\begin{equation}
  \phi_{ij} - \phi_{kl} = \begin{cases} \alpha_{ij} - \alpha_{kl} , \\  \alpha_{ij} + \alpha_{kl}  - \pi , \\ - \alpha_{ij} - \alpha_{kl} + \pi , \\ - \alpha_{ij} + \alpha_{kl} . \end{cases}  \label{eq:PhaseDifferenceFourCases}
\end{equation}
After taking out the indeterminacy of the sign of $\pm \tilde{\alpha}$, which is true due to the symmetry of the cosine, one is left with the following set of two non-redundant cases for the quantity $\tilde{\alpha}$:
\begin{equation}
  \tilde{\alpha} = \begin{cases}  \alpha_{ij} - \alpha_{kl}  ,   \\  \alpha_{ij} + \alpha_{kl}  - \pi  .  \end{cases}      \label{eq:AlphaTildeTwoNonRedundantCases}
\end{equation}
We evaluate the cosine of the second possibility for $\tilde{\alpha}$, as follows:

\clearpage

\begin{widetext}
\begin{align}
 \cos \left( \phi_{ij} - \phi_{kl} \right) &= \cos \left( \alpha_{ij} + \alpha_{kl} - \pi \right)  = - \cos \alpha_{ij} \cos \alpha_{kl} +  \sin \alpha_{ij} \sin  \alpha_{kl}  \nonumber \\
 &= - \sqrt{\left( 1 - \sin^{2} \alpha_{ij} \right) \left( 1 - \sin^{2} \alpha_{kl} \right)}  + \sin \alpha_{ij} \sin  \alpha_{kl} \nonumber \\
 &= - \sqrt{\left( 1 - \left[ \frac{ B_{ij}^{2} - B_{kl}^{2} + N^{2}}{2 N B_{ij}}  \right]^{2} \right) \left( 1 - \left[ \frac{N^{2} - B_{ij}^{2} + B_{kl}^{2}}{2 N B_{kl}} \right]^{2} \right)} +  \frac{ B_{ij}^{2} - B_{kl}^{2} + N^{2}}{2 N B_{ij}}  \frac{N^{2} - B_{ij}^{2} + B_{kl}^{2}}{2 N B_{kl}}  \nonumber \\
 &=  \frac{N^{2} - B_{ij}^{2} - B_{kl}^{2}}{2 B_{ij} B_{kl}}  . \label{eq:CosineConstraintCheck}  
\end{align}
\end{widetext}
The calculation needed to get from the third to the fourth step in equation~\eqref{eq:CosineConstraintCheck} is quite involved. The numerator of the square-root term becomes the absolute value $\left| B_{ij}^{4} + \left( B_{kl}^{2} - N^{2} \right)^{2} - 2 B_{ij}^{2} \left( B_{kl}^{2} + N^{2} \right) \right|$. Once we insert the equation for $N^{2}$, i.e. eq.~\eqref{eq:ExampleDerivationStepI}, into the term within the absolute value, we see that this term equals~$- 4 B_{ij}^{2} B_{kl}^{2} \sin^{2} \left( \phi_{ij} - \phi_{kl} \right)$, which is definitely a negative number. Thus, evaluating the absolute value amounts to flipping the sign of this term. Keeping track of this sign-change, it is then possible to derive the correct result~\eqref{eq:CosineConstraintCheck}.

We observe that the second of the two non-redundant possibilities~\eqref{eq:AlphaTildeTwoNonRedundantCases} satisfies the correct constraint~\eqref{eq:ExampleDerivationStepI} for $\cos \left( \phi_{ij} - \phi_{kl} \right) $. In exactly the same way as done above, one can check that the first possibility in~\eqref{eq:AlphaTildeTwoNonRedundantCases} leads to an equation which is different from~\eqref{eq:ExampleDerivationStepI}. Thus, we have derived that the difference of the relative-phases $\phi_{ij}$ and $\phi_{kl}$ always has to satisfy:
\begin{equation}
 \phi_{ij} - \phi_{kl} = \pm \left( \alpha_{ij} + \alpha_{kl} - \pi \right)  . \label{eq:PhaseDifferenceCorrectConstraint}
\end{equation}
The fact that this relation has to hold reduces the four-fold discrete ambiguity derived initially to a two-fold one. The remaining constraint~\eqref{eq:PhaseDifferenceCorrectConstraint} is only generally fulfilled by the following two cases:
\begin{equation}
 \begin{cases}  \phi_{ij} = - \zeta + \alpha_{ij}  ,  \\ \phi_{kl} = - \zeta - \alpha_{kl} + \pi   , \end{cases}  \text{or } \begin{cases}  \phi_{ij} = - \zeta - \alpha_{ij} + \pi ,  \\ \phi_{kl} = - \zeta + \alpha_{kl} . \end{cases} \label{eq:ExampleIPhaseAmbiguityEndResult}
\end{equation}
Consider next the case 'B.4' described in section~\ref{sec:NewCriterion}, i.e. the combination of observables~$\left( \Ocal^{n}_{1-}, \Ocal^{n}_{2-} \right)$. From their definitions
\begin{align}
 \Ocal^{n}_{1-} &= B_{ij} \sin \phi_{ij} - B_{kl} \sin \phi_{kl}  , \label{eq:Example2DerivationObsI} \\
 \Ocal^{n}_{2-}  &=  B_{ij} \cos \phi_{ij} - B_{kl} \cos \phi_{kl}  , \label{eq:Example2DerivationObsII} 
\end{align}
we see that the entire derivation given above for the case 'B.1' can be mimicked, with the only change needed being a flip of the sign of $B_{kl}$ in every intermediate step. Thus, we can infer the following two constraints:
\begin{align}
 \sin \left( \zeta + \phi_{kl} \right) &= \frac{- N^{2} + B_{ij}^{2} - B_{kl}^{2}}{2 N B_{kl}}  , \label{eq:Example2DerivationStepIII} \\ \nonumber \\
 \sin \left( \zeta + \phi_{ij} \right) &= \frac{N^{2} + B_{ij}^{2} - B_{kl}^{2}}{2 N B_{ij}}  . \label{eq:Example2DerivationStepIV}
\end{align}
Applying the $\arcsin$-function yields here again an apparent four-fold discrete phase-ambiguity, with formally the same expressions as given in equations~\eqref{eq:PhaseAmbExampleDerivationI} and~\eqref{eq:PhaseAmbExampleDerivationII}. However, now the derivation~\eqref{eq:CosineConstraintCheck} changes. We again consider the two non-redundant cases for the $\tilde{\alpha}$-variable given in equation~\eqref{eq:AlphaTildeTwoNonRedundantCases}. However, now some quite involved algebra shows that:
\begin{widetext}
\begin{align}
 \cos \left( \phi_{ij} - \phi_{kl} \right) &= \cos \left( \alpha_{ij} - \alpha_{kl} \right) =  \cos \alpha_{ij} \cos \alpha_{kl} +  \sin \alpha_{ij} \sin  \alpha_{kl}  \nonumber \\
 &=   \sqrt{\left( 1 - \sin^{2} \alpha_{ij} \right) \left( 1 - \sin^{2} \alpha_{kl} \right)}  + \sin \alpha_{ij} \sin  \alpha_{kl} \nonumber \\
 &=    \sqrt{\left( 1 - \left[ \frac{ B_{ij}^{2} - B_{kl}^{2} + N^{2}}{2 N B_{ij}}  \right]^{2} \right) \left( 1 - \left[ \frac{- N^{2} + B_{ij}^{2} - B_{kl}^{2}}{2 N B_{kl}} \right]^{2} \right)} +  \frac{ B_{ij}^{2} - B_{kl}^{2} + N^{2}}{2 N B_{ij}}  \frac{- N^{2} + B_{ij}^{2} - B_{kl}^{2}}{2 N B_{kl}}  \nonumber \\
 &=  \frac{- N^{2} + B_{ij}^{2} + B_{kl}^{2}}{2 B_{ij} B_{kl}}  . \label{eq:CosineConstraintCheckExampleII}  
\end{align}
\end{widetext}
%

%\clearpage

The second possibility mentioned in equation~\eqref{eq:AlphaTildeTwoNonRedundantCases} leads to the wrong constraint for $\cos \left( \phi_{ij} - \phi_{kl} \right)$. Therefore, in the case 'B.4', the following constraint has to hold:
\begin{equation}
 \phi_{ij} - \phi_{kl} = \pm \left( \alpha_{ij} - \alpha_{kl} \right)  . \label{eq:PhaseDifferenceCorrectConstraintExampleII}
\end{equation}
This constraint leaves only the following two-fold phase-ambiguity for the case 'B.4':
\begin{equation}
 \begin{cases}  \phi_{ij} = - \zeta + \alpha_{ij}  ,  \\ \phi_{kl} = - \zeta + \alpha_{kl}  , \end{cases}  \text{or } \begin{cases}  \phi_{ij} = - \zeta - \alpha_{ij} + \pi ,  \\ \phi_{kl} = - \zeta - \alpha_{kl} + \pi . \end{cases} \label{eq:ExampleIIPhaseAmbiguityEndResult}
\end{equation}
Now, we treat the case 'B.2' from section~\ref{sec:NewCriterion}, i.e. we consider the following pair of observables
\begin{align}
 \Ocal^{n}_{1+} &= B_{ij} \sin \phi_{ij} + B_{kl} \sin \phi_{kl}  , \label{eq:Example3DerivationObsI} \\
 \Ocal^{n}_{2-}  &=  B_{ij} \cos \phi_{ij} - B_{kl} \cos \phi_{kl}  . \label{eq:Example3DerivationObsII} 
\end{align}
The derivation for this case corresponds to the expressions derived for 'B.1' above, with both the signs of $B_{kl}$ and $\phi_{kl}$ flipped. The expression~\eqref{eq:ExampleDerivationStepI} changes as follows:
\begin{equation}
 \left( \Ocal^{n}_{1+} \right)^{2} + \left( \Ocal^{n}_{2-} \right)^{2} = B_{ij}^{2} + B_{kl}^{2} - 2 B_{ij} B_{kl} \cos \left( \phi_{ij} + \phi_{kl} \right)  .  \label{eq:Example3DerivationStepI}
\end{equation}
One has to keep in mind that now the constraint for the cosine holds for $\cos \left( \phi_{ij} + \phi_{kl} \right)$. Following the derivation further, one obtains the following pair of constraints:
\begin{align}
 \sin \left( \zeta - \phi_{kl} \right) &= \frac{- N^{2} + B_{ij}^{2} - B_{kl}^{2}}{2 N B_{kl}}  , \label{eq:Example3DerivationStepIII} \\
 \sin \left( \zeta + \phi_{ij} \right) &= \frac{N^{2} + B_{ij}^{2} - B_{kl}^{2}}{2 N B_{ij}}  . \label{eq:Example3DerivationStepIV}
\end{align}
These two constraints fix the relative phases $\phi_{ij}$ and $\phi_{kl}$ up to the following discrete ambiguities
\begin{equation}
  \phi_{kl} = \begin{cases} \zeta - \alpha_{kl}  , \\ \zeta + \alpha_{kl} - \pi ,  \end{cases} \text{and } \phi_{ij} = \begin{cases} - \zeta + \alpha_{ij}  , \\ - \zeta - \alpha_{ij} + \pi .  \end{cases}  \label{eq:PhaseAmbsExample3Derivation}
\end{equation}
For the sum of both relative phases, one thus obtains the four possible cases
\begin{equation}
  \phi_{ij} + \phi_{kl} = \begin{cases} \alpha_{ij} - \alpha_{kl} , \\  \alpha_{ij} + \alpha_{kl}  - \pi , \\ - \alpha_{ij} - \alpha_{kl} + \pi , \\ - \alpha_{ij} + \alpha_{kl} . \end{cases}  \label{eq:PhaseSumFourCasesExample3}
\end{equation}
One again has to single out the non-redundant cases
\begin{equation}
  \tilde{\alpha} = \begin{cases}  \alpha_{ij} - \alpha_{kl}  ,   \\  \alpha_{ij} + \alpha_{kl}  - \pi  .  \end{cases}      \label{eq:AlphaTildeTwoNonRedundantCasesExample3}
\end{equation}
In a calculation which is formally quite similar to~\eqref{eq:CosineConstraintCheckExampleII}, one can check that the correct constraint for $\cos \left( \phi_{ij} + \phi_{kl} \right)$ is satisfied by:
\begin{equation}
 \phi_{ij} + \phi_{kl} = \pm \left( \alpha_{ij} - \alpha_{kl} \right)  . \label{eq:PhaseDifferenceCorrectConstraintExampleIII}
\end{equation}
This relation for the sum of the relative-phases singles out the following two-fold discrete phase-ambiguity for the case 'B.2':
\begin{equation}
 \begin{cases}  \phi_{ij} = - \zeta + \alpha_{ij}  ,  \\ \phi_{kl} =  \zeta - \alpha_{kl}  , \end{cases}  \text{or } \begin{cases}  \phi_{ij} = - \zeta - \alpha_{ij} + \pi ,  \\ \phi_{kl} = \zeta + \alpha_{kl} - \pi . \end{cases} \label{eq:ExampleIIIPhaseAmbiguityEndResult}
\end{equation}
Finally, we consider the case 'B.3' from section~\ref{sec:NewCriterion}, i.e. the selection of the following pair of observables:
\begin{align}
 \Ocal^{n}_{1-} &= B_{ij} \sin \phi_{ij} - B_{kl} \sin \phi_{kl}  , \label{eq:Example4DerivationObsI} \\
 \Ocal^{n}_{2+}  &=  B_{ij} \cos \phi_{ij} + B_{kl} \cos \phi_{kl}  . \label{eq:Example4DerivationObsII} 
\end{align}
Here, one can follow the steps in the derivation for the case 'B.1', which has been described above, and only has to flip the sign of the relative phase~$\phi_{kl}$. Thus, equation~\eqref{eq:ExampleDerivationStepI} turns into:
\begin{equation}
 \left( \Ocal^{n}_{1-} \right)^{2} + \left( \Ocal^{n}_{2+} \right)^{2} = B_{ij}^{2} + B_{kl}^{2} + 2 B_{ij} B_{kl} \cos \left( \phi_{ij} + \phi_{kl} \right)  ,  \label{eq:Example4DerivationStepI}
\end{equation}
The constraint for the cosine is again valid for the sum of both relative-phases $\phi_{ij}$ and $\phi_{kl}$.

For this case 'B.3', one can derive the following set of constraints:
\begin{align}
 \sin \left( \zeta - \phi_{kl} \right) &= \frac{N^{2} - B_{ij}^{2} + B_{kl}^{2}}{2 N B_{kl}}  , \label{eq:Example4DerivationStepIII} \\
 \sin \left( \zeta + \phi_{ij} \right) &= \frac{N^{2} + B_{ij}^{2} - B_{kl}^{2}}{2 N B_{ij}}  . \label{eq:Example4DerivationStepIV}
\end{align}
The two relative-phases $\phi_{ij}$ and $\phi_{kl}$ are thus fixed up to the following discrete ambiguities
\begin{equation}
  \phi_{kl} = \begin{cases} \zeta - \alpha_{kl}  , \\ \zeta + \alpha_{kl} - \pi ,  \end{cases} \text{and } \phi_{ij} = \begin{cases} - \zeta + \alpha_{ij}  , \\ - \zeta - \alpha_{ij} + \pi .  \end{cases}  \label{eq:PhaseAmbsExample4Derivation}
\end{equation}
The thus implied possible cases for the sum $\phi_{ij} + \phi_{kl}$ are formally the same as in equation~\eqref{eq:PhaseSumFourCasesExample3} and the non-redundant cases for the quantity~$\tilde{\alpha}$ are formally the same as in equation~\eqref{eq:AlphaTildeTwoNonRedundantCasesExample3}. 

One has again to be careful with the constraint for $\cos \left( \phi_{ij} + \phi_{kl} \right)$. In a calculation which is formally similar to the steps taken in~\eqref{eq:CosineConstraintCheck}, one can deduce that for the case 'B.3', the following constraint has to hold:
\begin{equation}
 \phi_{ij} + \phi_{kl} = \pm \left( \alpha_{ij} + \alpha_{kl} - \pi \right)  . \label{eq:PhaseDifferenceCorrectConstraintExampleIV}
\end{equation}
This relation singles out the following two-fold discrete phase-ambiguity for the case 'B.3':
\begin{equation}
 \begin{cases}  \phi_{ij} = - \zeta + \alpha_{ij}  ,  \\ \phi_{kl} = \zeta + \alpha_{kl} - \pi  , \end{cases}  \text{or } \begin{cases}  \phi_{ij} = - \zeta - \alpha_{ij} + \pi ,  \\ \phi_{kl} =  \zeta - \alpha_{kl}  . \end{cases} \label{eq:ExampleIVPhaseAmbiguityEndResult}
\end{equation}

\clearpage

\section{Discrete phase-ambiguities of the simplest non-trivial shape-class (eqs~\eqref{eq:NonTrivialShapeClassObsI} -~\eqref{eq:NonTrivialShapeClassObsIV}) for combinations of observables with flipped signs} \label{sec:FlippedSignCases}

In this appendix, we derive the discrete phase-ambiguities for observables selected from the shape-class given in equations~\eqref{eq:NonTrivialShapeClassObsI} -~\eqref{eq:NonTrivialShapeClassObsIV} in section~\ref{sec:NewCriterion}, provided that at least one of the observables has a flipped sign. We will see that this leads to more general combinations of signs for the $\zeta$-angles in the ambiguity-formulas (more general compared to those listed in section~\ref{sec:NewCriterion}). Although this derivation yields in principle redundant information, the obtained results are still useful in case one wishes to derive complete sets starting solely from considerations of graphs, as discussed in more detail in section~\ref{sec:Photoproduction}. The results derived in the following have been collected in Table~\ref{tab:ZetaSignAssociationTable} of the main text.

The first and simplest way of deriving the phase-ambiguities for combinations of observables with flipped signs proceeds via consideration of equation~\eqref{eq:ZetaAngleFormalDefinition} from appendix~\ref{sec:NakayamaDerivation}. Suppose we reverse the sign of the second observable in this formula, i.e. we consider the observables~$\left(  \Ocal^{n}_{1+}, - \Ocal^{n}_{2+} \right)$. Then, equation~\eqref{eq:ZetaAngleFormalDefinition} becomes:
\begin{equation}
   \cos \zeta \equiv \frac{\Ocal^{n}_{1+}}{N} \text{, } \sin \zeta \equiv \frac{- \Ocal^{n}_{2+}}{N}   . \label{eq:ZetaAngleFormalDefinitionFlippedSign}
\end{equation}
Suppose now that another angle, called '$\tilde{\zeta}$', exists for which the equation retains its original form:
\begin{equation}
   \cos \tilde{\zeta} \equiv \frac{\Ocal^{n}_{1+}}{N} \text{, } \sin \tilde{\zeta} \equiv \frac{ \Ocal^{n}_{2+}}{N}   . \label{eq:ZetaAngleFormalDefinitionFlippedSignII}
\end{equation}
Then, we have
\begin{equation}
 \cos \tilde{\zeta} = \cos \zeta \text{, and } \sin \tilde{\zeta} = - \sin \zeta ,  \label{eq:ZetaTildeFormulasI}
\end{equation}
or
\begin{equation}
 \tilde{\zeta} = - \zeta   . \label{eq:ZetaTildeFormulaII}
\end{equation}
We see that the $\zeta$-angle has reversed its sign when going from~$\left(  \Ocal^{n}_{1+}, \Ocal^{n}_{2+} \right)$ to~$\left(  \Ocal^{n}_{1+}, - \Ocal^{n}_{2+} \right)$. This can also be seen via consideration of Figure~\ref{fig:ZetaAngleDefinition}.

Alternatively, consider now the combination~$\left( - \Ocal^{n}_{1+}, \Ocal^{n}_{2+} \right)$, which leads to the following equations:
\begin{equation}
   \cos \zeta \equiv \frac{- \Ocal^{n}_{1+}}{N} \text{, } \sin \zeta \equiv \frac{\Ocal^{n}_{2+}}{N}   . \label{eq:ZetaAngleFormalDefinitionFlippedSignIII}
\end{equation}
In the same way as before, one derives for this case:
\begin{equation}
 \tilde{\zeta} = \pi - \zeta   . \label{eq:ZetaTildeFormulaIII}
\end{equation}
An additional summand of~$\pi$ appears now in the ambiguity-formulas, which is however irrelevant. The important point is that also in this case, the sign of the $\zeta$-angle has been reversed. In exactly the same way, one can show that for the combination~$\left( - \Ocal^{n}_{1+}, - \Ocal^{n}_{2+} \right)$, the $\zeta$-angle retains its original sign.

The second possible way of deriving the ambiguities for flipped signs of observables proceeds by going through the steps for case 'B.1' as discussed in appendix~\ref{sec:NakayamaDerivation}. However, this time we flip the sign of the first observable:~$\left( - \Ocal^{n}_{1+}, \Ocal^{n}_{2+} \right)$. Considering the definitions in equations~\eqref{eq:ExampleDerivationObsI} and~\eqref{eq:ExampleDerivationObsII}, we observe that flipping the sign of $\Ocal^{n}_{1+}$ is tantamount to reversing the signs of both relative-phases $\phi_{ij}$ and $\phi_{kl}$. Following the derivation through, we obtain the following constraints:
\begin{align}
  \sin \left( \zeta - \phi_{kl} \right) = \frac{N^{2} - B_{ij}^{2} + B_{kl}^{2}}{2 N B_{kl}}  . \label{eq:FlippedExampleDerivationStepIII} \\
  \sin \left( \zeta - \phi_{ij} \right) = \frac{ B_{ij}^{2} - B_{kl}^{2} + N^{2}}{2 N B_{ij}}  . \label{eq:FlippedExampleDerivationStepIV}
\end{align}
These two constraints directly imply the following discrete phase-ambiguities:
\begin{equation}
  \phi_{kl} = \begin{cases} \zeta - \alpha_{kl}  , \\ \zeta + \alpha_{kl} - \pi ,  \end{cases} \text{and } \phi_{ij} = \begin{cases}  \zeta - \alpha_{ij}  , \\  \zeta + \alpha_{ij} - \pi .  \end{cases}  \label{eq:PhaseAmbsFlippedExampleIDerivation}
\end{equation}
We see that from then on the $\zeta$'s appear with reversed signs in all formulas compared to the original derivation for case 'B.1', given in appendix~\ref{sec:NakayamaDerivation} (cf. equations~\eqref{eq:PhaseAmbExampleDerivationI} and~\eqref{eq:PhaseAmbExampleDerivationII}). The same is true when one considers the combination with the sign of the second observable flipped:~$\left( \Ocal^{n}_{1+}, - \Ocal^{n}_{2+} \right)$. The latter case can be obtained from the original derivation for 'B.1' by reversing the signs of all the quantities $\phi_{ij}$, $\phi_{kl}$, $B_{ij}$ and $B_{kl}$ (cf. equations~\eqref{eq:ExampleDerivationObsI} and~\eqref{eq:ExampleDerivationObsII}). However, in case one would consider the combination~$\left( - \Ocal^{n}_{1+}, - \Ocal^{n}_{2+} \right)$, all one would have to do is reverse the signs of the modulus-factors $B_{ij}$ and $B_{kl}$, which would leave the signs of the $\zeta$-angles in the ambiguity-formulas untouched.

By repeating the arguments given above for the case 'B.4' considered in appendix~\ref{sec:NakayamaDerivation}, one can derive the remaining associations between the $\zeta$-sign combinations '$(- \zeta, - \zeta)$' and '$(+ \zeta, + \zeta)$' and specific pairs of observables, which are given in Table~\ref{tab:ZetaSignAssociationTable} of the main text.

Finally, in order to obtain all the associations of pairs of observables to the $\zeta$-sign combinations '$(- \zeta, + \zeta)$' and '$(+ \zeta, - \zeta)$', one has to repeat the steps described above for the cases 'B.2' and 'B.3' discussed in appendix~\ref{sec:NakayamaDerivation}. In this way, the full set of associations given in Table~\ref{tab:ZetaSignAssociationTable} is obtained.

\end{document}